\documentclass[times]{fldauth}

\usepackage{moreverb}
\usepackage{graphicx}
\usepackage[colorlinks,bookmarksopen,bookmarksnumbered,citecolor=red,urlcolor=red]{hyperref}

\newcommand\BibTeX{{\rmfamily B\kern-.05em \textsc{i\kern-.025em b}\kern-.08em
T\kern-.1667em\lower.7ex\hbox{E}\kern-.125emX}}

\newcommand{\Q}{\mathbf{Q}}
\newcommand{\F}{\mathbf{F}}
\renewcommand{\S}{\mathbf{S}}
\newcommand{\K}{\mathbf{K}}
\newcommand{\M}{\mathbf{M}}
\renewcommand{\P}{\mathbf{P}}
\renewcommand{\v}{\mathbf{v}}
\newcommand{\w}{\mathbf{w}}
\newcommand{\x}{\mathbf{x}}

\newcommand{\q}{\mathbf{q}}
\newcommand{\f}{\mathbf{f}}
\newcommand{\g}{\mathbf{g}}

\def\volumeyear{2012}

\begin{document}

\runningheads{W. Boscheri et al.}{Unstructured Lagrangian WENO Schemes for Hydrodynamics and MHD}

\title{High Order Lagrangian ADER-WENO Schemes on Unstructured Meshes -- Application of Several Node Solvers to Hydrodynamics and Magnetohydrodynamics}

\author{W. Boscheri, M. Dumbser\corrauth, D.S. Balsara}

\address{Department of Civil, Environmental and Mechanical Engineering, University of Trento, Via Mesiano, 77, I-38123 Trento \\ 
         Physics Department, University of Notre Dame du Lac, 225 Nieuwland Science Hall, Notre Dame, IN 46556, USA}

\corraddr{walter.boscheri@unitn.it, michael.dumbser@ing.unitn.it, dinshaw.balsara@nd.edu}

\begin{abstract}
In this paper we present a class of high order accurate cell-centered Arbitrary--Eulerian--Lagrangian (ALE) one--step ADER-WENO finite volume schemes for the solution of nonlinear hyperbolic conservation 
laws on two--dimensional unstructured triangular meshes along the principles laid out in \cite{BoscheriDumbserLag}. High order of accuracy in space is achieved by a WENO reconstruction algorithm, while a local 
space--time Galerkin predictor allows the schemes to be high order accurate also in time by using an element-local weak formulation of the governing PDE on moving meshes. 
The mesh motion can be computed by choosing among three different node solvers, which are for the first time compared with each other in this article: 
the node velocity may be obtained i) either as an arithmetic average among the states surrounding the node, as suggested by Cheng and Shu \cite{chengshu1}, 
or, ii) as a solution of multiple one--dimensional half-Riemann problems around a vertex, as suggested by Maire \cite{Maire2009}, 
or, iii) by solving approximately a multidimensional Riemann problem around each vertex of the mesh using the genuinely multidimensional HLL Riemann solver recently proposed by Balsara et al. \cite{BalsaraMultiDRS}. 
Once the vertex velocity and thus the new node location has been determined by the node solver, the local mesh motion is then constructed by \textit{straight} edges connecting the vertex positions at the old time level 
$t^n$ with the new ones at the next time level $t^{n+1}$. If necessary, a rezoning step can be introduced here to overcome mesh tangling or highly deformed elements. The final ALE finite volume scheme is based directly 
on a space-time conservation formulation of the governing PDE system, which therefore makes an additional remapping stage unneccesary, since the ALE fluxes already properly take into account the rezoned geometry. 
In this sense, our scheme falls into the category of \textit{direct} ALE methods. Furthermore, the geometric conservation law is satisfied by the scheme by construction. 

We apply the high order algorithm presented in this paper to the Euler equations of compressible gas dynamics as well as to the ideal classical and relativistic MHD equations. We show numerical 
convergence results up to fifth order of accuracy in space and time together with some classical numerical test problems for each hyperbolic system under consideration.   
\end{abstract}

\maketitle

\vspace{-6pt}
\section{Introduction}
\label{sec.intro}
\vspace{-2pt}

In this article we use the family of high order accurate Arbitrary--Eulerian--Lagrangian (ALE) one--step WENO finite volume schemes presented in \cite{BoscheriDumbserLag} 
for the solution of hyperbolic systems of conservation laws in two space dimensions that can be cast in the following general form:  
\begin{equation}
\label{PDE}
  \frac{\partial \Q}{\partial t} + \nabla \cdot \F(\Q) = \S(\Q), \qquad (x,y) \in \Omega(t) \subset \mathbb{R}^2, \quad t \in \mathbb{R}_0^+, \quad \Q \in \Omega_{\Q} \subset \mathbb{R}^\nu,     
\end{equation} 
where $\Q=(q_1,q_2,...,q_\nu)$ is the vector of conserved variables defined in the space of the admissible states $\Omega_{\Q} \subset \mathbb{R}^\nu$, $\F(\Q)=\left( \f(\Q),\g(\Q) \right)$ is the nonlinear flux tensor and $\S(\Q)$ 
represents a nonlinear algebraic source term. 

A lot of research for the development of Lagrangian numerical schemes has been carried out in the last decades, because material interfaces can be precisely identified and located by using a Lagrangian formulation, where the computational mesh moves with the local fluid velocity. Lagrangian algorithms have been proposed in literature either starting directly from the conservative quantities such as mass, momentum and total energy \cite{Maire2007,Smith1999}, or from the nonconservative form of the governing equations, as proposed in \cite{Benson1992,Caramana1998,Neumann1950}. Furthermore a classification is usually done considering the location of the physical variables on the mesh: in the \textit{staggered mesh} approach the velocity is defined at the cell interfaces and the other variables at the cell barycenter, see e.g. \cite{StagLag}, whereas in the \textit{cell-centered}  approach all variables are defined at the cell barycenter, see e.g. \cite{Despres2009,Depres2012,ShashkovCellCentered,MaireCyl1,Maire2007,Maire2008,Maire2009b,Maire2010}.

The first Lagrangian schemes considered the Euler equations for compressible gas dynamics and the use of Godunov--type finite volume schemes, as done by Munz in \cite{munz94} and also by Carr\'e et al. \cite{Despres2009}, who  proposed a Godunov--type cell-centered Lagrangian algorithm on general multi-dimensional unstructured meshes. The physical system of equations was coupled with the equations for the evolution of the geometry by Despr\'es and Mazeran \cite{DepresMazeran2003,Despres2005}, hence solving a weakly hyperbolic system of conservation laws where they used a node-based finite volume solver. 

First and second order accurate cell-centered Lagrangian schemes in two- and three- space dimensions have also been introduced by Maire \cite{Maire2009,Maire2010,Maire2011}, who considers general polygonal grids and computes the time derivatives of the fluxes with a node-centered solver which can be interpreted as a multi-dimensional extension of the Generalized Riemann problem methodology introduced by Ben-Artzi et al. \cite{Artzi,ArtziWarnecke}, Le Floch et al. \cite{Raviart.GRP.1,Raviart.GRP.2} and Titarev and Toro \cite{Toro:2006a,toro3,toro10}. The node solver developed by Maire in \cite{Maire2009} and the group of Despr\'es \cite{Despres2009} is the second one used in this paper,  where it is also extended in a simplified way to the equations of magnetohydrodynamics. 

All the schemes mentioned above are at most second order accurate in space and in time. However, the order of accuracy of Lagrangian schemes may be improved by introducing a high order essentially non-oscillatory (ENO)  reconstruction, as first done by Cheng and Shu \cite{chengshu1,chengshu2,chengshu3,chengshu4}, who developed a class of cell centered Lagrangian finite volume schemes for gas dynamics where high order of accuracy in time was guaranteed either by the use of a Runge-Kutta or by a Lax-Wendroff-type time stepping. The node solver used in their approach is a simple arithmetic average of the corner-extrapolated values in the cells adjacent to a mesh vertex.  This is also the first node solver considered in this article, where the mesh velocity at each node is given by a weighted average among the velocities of the Voronoi neighbor elements surrounding the node. High order  Lagrangian schemes have also been proposed in the finite element framework by Scovazzi et al. \cite{scovazzi1,scovazzi2}. Recent work has been carried out by Dumbser et al. \cite{Dumbser2012}, who 
developed a new class of high order accurate Lagrangian one--step ADER-WENO finite volume schemes for hyperbolic conservation laws with stiff source terms. High order of accuracy in space with essentially non-oscillatory 
behaviour at  discontinuities and sharp gradients is obtained by a high order WENO reconstruction algorithm \cite{Dumbser2007204,Dumbser2007693}, while high order time accuracy is reached by using the local space--time 
Galerkin predictor method proposed in \cite{DumbserEnauxToro,HidalgoDumbser}. Purely Lagrangian and ALE methods with remapping for multi-phase and multi-material flows have been very recently  presented in \cite{ShashkovMultiMat1,ShashkovMultiMat2,ShashkovMultiMat3,ShashkovMultiMat4} and a high order extension of cell-centered Lagrangian WENO finite volume schemes to non-conservative hyperbolic systems with 
stiff source terms for compressible multi-phase flows has been recently considered in \cite{DumbserBoscheriLagNC}. The algorithm presented in \cite{DumbserBoscheriLagNC} has been applied to the seven-equation Baer--Nunziato 
model of compressible multi-phase flows, which is based on a diffuse interface approach. Independently, Lagrangian ADER-WENO schemes have also been investigated by Cheng and Toro in one space dimension, see \cite{chengtoro}. 

A fully Lagrangian approach is furthermore adopted by meshless particle schemes, such as the smooth particle hydrodynamics (SPH) method \cite{Monaghan1994,SPHLagrange,SPHWeirFlow,SPH3D,Dambreak3D}, which is suitable to follow strictly the fluid motion when the computational domain is very distorted and irregular. We can also recall a particular class of methods, namely the semi-Lagrangian schemes, which adopt a fixed mesh as in a classical Eulerian approach, but the Lagrangian trajectories of the fluid are followed backward in time in order to compute the numerical solution at the the new time level \cite{Casulli1990,CasulliCheng1992,LentineEtAl2011,HuangQiul2011,QuiShu2011,CIR,BoscheriDumbser}. Those schemes are mainly used for the solution of transport equations \cite{ALE2000Belgium,ALE1996FV}. Moreover the class of Arbitrary Lagrangian Eulerian (ALE) schemes \cite{Hirt1974,Peery2000,Smith1999,Feistauer1,Feistauer2,Feistauer3,Feistauer4} should be mentioned, where the mesh velocity can be chosen arbitrarily and does not necessarily have 
to coincide with the local fluid velocity. In the present paper, we actually follow the ALE philosophy, since it allows for a greater flexibility compared to purely Lagrangian or Eulerian methods and since it contains both as
special cases.  

Another important branch of research in finite volume methods has put many efforts to introduce multidimensional effects into Riemann solvers \cite{AbgrallmultiD1,AbgrallmultiD2,AbgrallmultiD3}. The aim was the formulation of genuinely multidimensional Riemann solvers for the solution of hyperbolic conservation laws of the form \eqref{PDE}. In a series of very recent papers \cite{BalsaraHLLE,BalsaraHLLC,BalsaraMultiDRS}, Balsara et al. presented  multidimensional HLL and HLLC Riemann solvers for hydrodynamics and magnetohydrodynamics on both structured and unstructured meshes. In \cite{BalsaraMultiDRS} the multidimensional Riemann solver is designed to work also on moving meshes, incorporating the mesh velocity in the signal speeds for the Riemann problem. Here, we use the above--mentioned strategy as the third node solver in a cell-centered Lagrangian framework, where the node velocity can be  directly extracted from the so-called strongly interacting state produced by the multi-dimensional HLL Riemann solver. 

A possible alternative to the use of multi-dimensional Riemann solvers could also be the finite volume evolution Galerkin method presented, for example, in \cite{EvolutionGalerkin1,EvolutionGalerkin2}, which would constitute a  
fourth possible node solver for the use in cell-centered Lagrangian schemes. 

The structure of this article is as follows. In Section \ref{sec.scheme} we  describe the numerical scheme, including the details of the three different node solvers that are used and compared with each other in this paper.  Numerical convergence studies as well as some classical test problems for hydrodynamics, classical and relativistic magnetohydrodynamics are reported in Section \ref{sec.validation}, while conclusions and an outlook to future  research and  developments is given in Section \ref{sec.concl}. 

\vspace{-6pt}
\section{Numerical Scheme}
\label{sec.scheme}
\vspace{-2pt}

In this paper we consider general nonlinear hyperbolic systems which can be cast in the form \eqref{PDE}. Let $\mathbf{x}=(x,y)$ be the spatial position vector and let $t$ represent the time. The computational domain $\Omega(t)$ depends on time, since in the ALE framework we are dealing with moving meshes. We use a total number $N_E$ of triangular shaped control volumes $T^n_i$ to discretize the domain at the current time level $t^n$, and the union of all elements is called the 
\textit{current triangulation} $\mathcal{T}^n_{\Omega}$ of the domain $\Omega(t^n)=\Omega^n$, which is given by 
\begin{equation}
\mathcal{T}^n_{\Omega} = \bigcup \limits_{i=1}^{N_E}{T^n_i}.
\label{trian}
\end{equation}

Due to the Lagrangian approach, the mesh is moving and deforming in time. Thus, it might be convenient to map the physical element $T^n_i$ to the \textit{local} reference element $T_e$ defined in the reference system $\xi-\eta$ with the vector of generic spatial coordinates defined as $\boldsymbol{\xi} = (\xi, \eta)$. The spatial reference element is the unit triangle composed of the nodes $\boldsymbol{\xi}^e_{1}=(\xi^e_{1},\eta^e_{1})=(0,0)$, $\boldsymbol{\xi}^e_{2}=(\xi^e_{2},\eta^e_{2})=(1,0)$ and $\boldsymbol{\xi}^e_{3}=(\xi^e_{2},\eta_{2})=(0,1)$. The mapping from the reference coordinate system $\xi-\eta$ to the physical coordinate system $x-y$ reads 
\begin{equation} 
 \mathbf{x} = \mathbf{X}^n_{1,i} + \left( \mathbf{X}^n_{2,i} - \mathbf{X}^n_{1,i} \right) \xi + \left( \mathbf{X}^n_{3,i} - \mathbf{X}^n_{1,i} \right) \eta,    
 \label{xietaTransf} 
\end{equation} 
where the physical coordinates at time $t^n$ of the $k$-th vertex of the triangle $T^n_i$ are given by $\mathbf{X}^n_{k,i} = (X^n_{k,i},Y^n_{k,i})$.

Our algorithm is based on the finite volume framework, where data are stored in each cell as piecewise constant cell averages 
\begin{equation}
  \Q_i^n = \frac{1}{|T_i^n|} \int_{T^n_i} \Q(x,y,t^n) dV,     
 \label{eqn.cellaverage}
\end{equation}  
where $|T_i^n|$ denotes the volume of element $T_i^n$ at the current time $t^n$. To achieve higher order of accuracy in space a WENO reconstruction is performed (see Section \ref{sec.WENOrec}), which produces piecewise high order polynomials $\mathbf{w}_h(x,y,t^n)$ starting from the given cell averages \eqref{eqn.cellaverage}. High order of accuracy in time will be obtained using the local continuous Galerkin predictor illustrated in Section \ref{sec.localCG}.

\subsection{Polynomial WENO Reconstruction on Unstructured Meshes}
\label{sec.WENOrec}

The first step in developing high order finite volume schemes is the spatial reconstruction of the state variables within each control volume. As done in \cite{BoscheriDumbserLag,DumbserBoscheriLagNC}, we use the WENO reconstruction  algorithm in the \textit{polynomial} formulation presented in \cite{DumbserEnauxToro,Dumbser2007204,Dumbser2007693,MixedWENO2D,MixedWENO3D}, although the original development of the WENO scheme was carried out in a \textit{pointwise} manner by Shu et al. \cite{JiangShu1996,balsarashu,HuShuVortex1999,ZhangShu3D}.  The details of the algorithm can be found in the above-mentioned references, hence in this Section we summarize only the main features of the scheme. Alternative high order accurate nonlinear reconstruction operators on unstructured meshes have been presented in \cite{MOOD1,MOOD2,MOOD3,AboiyarIske}, which can be used equally well. 

As explained in \cite{Dumbser2007693}, the reconstruction step is performed in the {reference system} $(\xi,\eta)$ according to the mapping \eqref{xietaTransf} and the starting point of the reconstruction is given by the 
known cell averages defined by \eqref{eqn.cellaverage}. Once the degree $M$ of the reconstruction polynomial has been chosen and fixed, one has to construct a set of reconstruction stencils  $S_i^s$ 
\begin{equation}
\mathcal{S}_i^s = \bigcup \limits_{j=1}^{n_e} T^n_{m(j)}, 
\label{stencil}
\end{equation}
where $1\leq j \leq n_e$ is a local index which counts the elements belonging 
to the stencil, while $m(j)$ maps the local index to the global element numbers used in the triangulation \eqref{trian}. The total number of elements $n_e$ should theoretically be equal to the smallest number $\mathcal{M} = (M+1)(M+2)/2$ needed to reach the formal order of accuracy $M+1$, but in \cite{StencilRec1990,Olliver2002,KaeserIske2005} it has been shown that this choice is not appropriate for unstructured meshes. Therefore $n_e$ is typically  taken to be $n_e = 2 \mathcal{M}$ in two space dimensions, which leads to an overdetermined system of reconstruction equations that is solved with a constrained least-squares algorithm, see \cite{Dumbser2007204,Dumbser2007693}. 
In order to circumvent Godunov's theorem, which states that no monotone and better than first order accurate linear schemes can exist for hyperbolic advection processes, a non--linearity must be introduced into the algorithm. 
Thus, we will compute the final reconstruction polynomial as a nonlinear combination between more than one reconstruction polynomials, each of those defined on a different reconstruction stencil $\mathcal{S}_i^s$. According to  \cite{Dumbser2007693,Dumbser2007204}, we always will use seven stencils in two space dimensions for each element $T^n_i$, namely one central reconstruction stencil given by $s=0$, three  primary sector stencils $s \in \{1,2,3\}$ and  three reverse sector stencils $s \in \{4,5,6 \}$, as also suggested by K\"aser and Iske \cite{KaeserIske2005}. The stencils are constructed by a recursive procedure that progressively adds new neighbors to the stencil  
until the prescribed number $n_e$ is reached. For all the stencils the first element is always given by the central element $T^n_i$.

The reconstruction polynomial $\w^s_h$ for triangle $T_i^n$ is then written for each stencil $\mathcal{S}_i^s$ in terms of some spatial basis functions $\psi_l(\xi,\eta)$ and $\mathcal{M}$ unknown degrees of freedom $\hat \w^{n,s}_{l,i}$. As basis functions $\psi_l(\xi,\eta)$ we use the orthogonal Dubiner--type basis on the reference triangle $T_e$, see \cite{Dubiner,orth-basis,CBS-book}. The expression for the reconstruction polynomial reads 
\begin{equation}
\label{eqn.recpolydef} 
\w^s_h(x,y,t^n) = \sum \limits_{l=1}^\mathcal{M} \psi_l(\xi,\eta) \hat \w^{n,s}_{l,i} := \psi_l(\xi,\eta) \hat \w^{n,s}_{l,i},   
\end{equation}
with the mapping to the reference coordinate system given by the transformation \eqref{xietaTransf}. In the rest of the paper we will use classical tensor index notation with the Einstein summation
convention, which implies summation over two equal indices. 

The reconstruction is required to be conservative w.r.t. the known cell averages $\Q^n_j$ for each element $T^n_j \in \mathcal{S}_i^s$, hence
\begin{equation}
\label{intConsRec}
\frac{1}{|T^n_j|} \int \limits_{T^n_j} \psi_l(\xi,\eta) \hat \w^{n,s}_{l,i} dV = \Q^n_j, \qquad \forall T^n_j \in \mathcal{S}_i^s     
\end{equation}
where $|T^n_j|$ is the volume of element $T^n_j$ at time $t^n$. Gaussian quadrature formulae of suitable order are used to evaluate the integrals in \eqref{intConsRec}, see \cite{stroud} for details. The system \eqref{intConsRec}  constitutes an overdetermined linear algebraic system, since the number of elements in the stencil is \textit{larger} than the one of the unknown polynomial coefficients ($n_e > \mathcal{M}$), and it can be efficiently solved  by using either a constrained least--squares technique \cite{Dumbser2007693} or using a more sophisticated singular value decomposition (SVD) algorithm. The reconstruction matrix, which is given by the multidimensional integrals in  system \eqref{intConsRec}, depends on the geometry of the triangles on which the polynomial is integrated. Hence, in a Lagrangian scheme, where the mesh geometry is continuously changing, system \eqref{intConsRec} can not be  inverted once at the beginning of the simulation for all elements during a preprocessing step, as done in the Eulerian framework  \cite{Dumbser2007693,Dumbser2007204}.  As a consequence, within a high order Lagrangian WENO finite volume scheme, the small linear systems \eqref{intConsRec} must be solved again and again at the beginning of each time step, which, however, can be done in a reasonably efficient manner by using optimized LINPACK library  functions. To keep the scheme reasonably simple and cost efficient, the choice of the stencils $\mathcal{S}_i^s$ remains \textit{fixed} for all times.   

The final WENO reconstruction polynomial is computed by weighting the above--defined stencil polynomials in a nonlinear way, where the nonlinearity is introduced in the WENO weights $\omega_s$
\begin{equation}
\tilde{\omega}_s = \frac{\lambda_s}{\left(\sigma_s + \epsilon \right)^r}, \qquad 
\omega_s = \frac{\tilde{\omega}_s}{\sum_q \tilde{\omega}_q},  
\end{equation}
through the oscillation indicators $\sigma_s$, which are computed on the reference element as 
\begin{equation}
\sigma_s = \Sigma_{lm} \hat w^{n,s}_{l,i} \hat w^{n,s}_{m,i},
\end{equation}
with 
\begin{equation}
\Sigma_{lm} = \sum \limits_{ \alpha + \beta \leq M}  \, \, \int \limits_{T_e} \frac{\partial^{\alpha+\beta} \psi_l(\xi,\eta)}{\partial \xi^\alpha \partial \eta^\beta} \cdot 
                                                                         \frac{\partial^{\alpha+\beta} \psi_m(\xi,\eta)}{\partial \xi^\alpha \partial \eta^\beta} d\xi d\eta.
\end{equation}

Here, we use $\epsilon=10^{-14}$, $r=8$, $\lambda_s=1$ for the one--sided stencils ($s>0$) and $\lambda_0=10^5$ for the central stencil, according to \cite{DumbserEnauxToro,Dumbser2007204}. 
The final nonlinear WENO reconstruction polynomial and its coefficients are then given by 
\begin{equation}
\label{eqn.weno} 
 \w_h(x,y,t^n) = \sum \limits_{l=1}^{\mathcal{M}} \psi_l(\xi,\eta) \hat \w^{n}_{l,i}, \qquad \textnormal{ with } \qquad  
 \hat \w^{n}_{l,i} = \sum_s \omega_s \hat \w^{n,s}_{l,i}.   
\end{equation}

\subsection{Local Space--Time Galerkin Predictor on Moving Meshes}
\label{sec.localCG}

High order of accuracy in time is achieved adopting the Lagrangian version of the local space--time continuous Galerkin method introduced for the Eulerian framework in \cite{Dumbser20088209} and subsequently extended to the Lagrangian case in \cite{BoscheriDumbserLag}. It consists in an \textit{element--local} space--time evolution of the high order spatial polynomials $\w_h$ obtained from the WENO reconstruction algorithm presented in the previous  Section \ref{sec.WENOrec}. Thus, for each element $T_i(t)$ the solution is evolved within one time step $[t^n;t^{n+1}]$. 

Let $\mathbf{x}=(x,y)$ and $\boldsymbol{\xi}=(\xi,\eta)$ be the purely spatial coordinate vectors in physical and reference coordinates, respectively. Introducing the physical space--time coordinate vector  $\mathbf{\tilde{x}}=(x,y,t)$ and the space--time reference coordinate vector $\boldsymbol{\tilde{\xi}}=(\xi,\eta,\tau)$, as well as the following mapping in time   
\begin{equation}
t = t_n + \tau \, \Delta t, \qquad  \tau = \frac{t - t^n}{\Delta t}, \qquad \Rightarrow \qquad \widehat{t}_l = t_n + \tau_l \, \Delta t, 
\label{timeTransf}
\end{equation} 
the governing PDE (\ref{PDE}) can be reformulated in the local reference system as
\begin{equation}
\frac{\partial \Q}{\partial \tau}\tau_t + \frac{\partial \Q}{\partial \xi}\xi_t + \frac{\partial \Q}{\partial \eta}\eta_t + \frac{\partial \f}{\partial \tau}\tau_x + \frac{\partial \f}{\partial \xi}\xi_x + \frac{\partial \f}{\partial \eta}\eta_x + \frac{\partial \g}{\partial \tau}\tau_y + \frac{\partial \g}{\partial \xi}\xi_y + \frac{\partial \g}{\partial \eta}\eta_y = \mathbf{S}(\Q),  
\label{PDEweak}
\end{equation}
where the Jacobian of the spatial and temporal transformation and its inverse read
\begin{equation}
J_{st} = \frac{\partial \mathbf{\tilde{x}}}{\partial \boldsymbol{\tilde{\xi}}} = \left( \begin{array}{ccc} x_{\xi} & x_{\eta} & x_{\tau} \\ y_{\xi} & y_{\eta} & y_{\tau} \\ 0 & 0 & \Delta_t \\ \end{array} \right), \quad J_{st}^{-1} = \frac{\partial \boldsymbol{\tilde{\xi}}}{\partial \mathbf{\tilde{x}}} = \left( \begin{array}{ccc} \xi_{x} & \xi_{y} & \xi_{t} \\ \eta_{x} & \eta_{y} & \eta_{t} \\ 0 & 0 & \frac{1}{\Delta t} \\ \end{array} \right), 
\label{Jac}
\end{equation}
since $\tau_x = \tau_y = 0$ and $\tau_t = \frac{1}{\Delta t}$, according to the definition (\ref{timeTransf}). Using the inverse of the Jacobian matrix, Eqn. \eqref{PDEweak} can then be simplified to 
\begin{equation}
\frac{\partial \Q}{\partial \tau} + \Delta t \left( \frac{\partial \Q}{\partial \xi}\xi_t + \frac{\partial \Q}{\partial \eta}\eta_t + \frac{\partial \f}{\partial \xi}\xi_x + \frac{\partial \f}{\partial \eta}\eta_x + \frac{\partial \g}{\partial \xi}\xi_y + \frac{\partial \g}{\partial \eta}\eta_y  \right) = \Delta t \mathbf{S}(\Q).
\label{PDECG}
\end{equation}
To simplify the notation let us also define the term $\P$ as 
\begin{equation}
 \P := \S(\Q) - \left( \frac{\partial \Q}{\partial \xi}\xi_t + \frac{\partial \Q}{\partial \eta}\eta_t + \frac{\partial \f}{\partial \xi}\xi_x + \frac{\partial \f}{\partial \eta}\eta_x + \frac{\partial \g}{\partial \xi}\xi_y + \frac{\partial \g}{\partial \eta}\eta_y  \right), 
 \label{eqn.pdef}
\end{equation} 
which collects all terms apart from the derivative of $\Q$ with respect to $\tau$, hence one obtains 
\begin{equation}
\frac{\partial \Q}{\partial \tau} = \P. 
\label{PCG}
\end{equation}
The discrete representation for the space--time solution $\Q$ and for the term $\P$ are denoted by $\q_h$ and $\P_h$, respectively, and are given by a nodal finite element ansatz in space-time that reads 
\begin{eqnarray}
\q_h=\q_h(\xi,\eta,\tau) = \theta_{l}(\xi,\eta,\tau) \widehat{\q}_{l,i}, \qquad & \P_h=\P_h(\xi,\eta,\tau) = \theta_{l}(\xi,\eta,\tau) \hat{\P}_{l,i}, 
\label{thetaSol}
\end{eqnarray}
where $\theta_l=\theta_l(\boldsymbol{\tilde{\xi}})=\theta_l(\xi,\eta,\tau)$ is a space--time basis function defined by the Lagrange interpolation polynomials passing through a set of space--time nodes  $\boldsymbol{\tilde{\xi}}_m=(\xi_m,\eta_m,\tau_m)$, explicitly defined in \cite{Dumbser20088209}. With the nodal approach one has $\widehat{\P}_{l,i}= \P(\mathbf{\tilde{x}}_{l,i})$, i.e. the degrees of
freedom of $\P_h$ are evaluated at each space-time node, see \cite{BoscheriDumbserLag} for more details. The same interpolation can be done for the discrete flux tensor $\F_h$  and for the discrete source 
term $\S_h$. 
We adopt an \textit{isoparametric} approach, where the \textit{same} basis functions $\theta_l$ are used to represent also the mapping between the physical space--time coordinate vector $\mathbf{\tilde{x}}$ and 
the reference space--time coordinate vector $\boldsymbol{\tilde{\xi}}$, hence yielding
\begin{equation}
 x(\xi,\eta,\tau) = \theta_l(\xi,\eta,\tau) \widehat{x}_{l,i}, \qquad y(\xi,\eta,\tau) = \theta_l(\xi,\eta,\tau) \widehat{y}_{l,i}, \qquad t(\xi,\eta,\tau) = \theta_l(\xi,\eta,\tau) \widehat{t}_l. 
 \label{eqn.isoparametric} 
\end{equation} 
The degrees of freedom $\widehat{\mathbf{x}}_{l,i} = (\widehat{x}_{l,i},\widehat{y}_{l,i})$ are partially unknown and denote the vector of physical coordinates in space of the moving space--time control volume, 
while the degrees of freedom of the physical time $\widehat{t}_l$ are \textit{known} from \eqref{timeTransf} and they are defined at each space--time node $\tilde{\x}_{l,i} = (\widehat{x}_{l,i}, \widehat{y}_{l,i}, \widehat{t}_l)$.   

The weak formulation \eqref{PDECG} of the governing PDE is integrated over the unit reference space--time element $T_e \times [0,1]$. For the sake of simplicity let us introduce the integral operator 
\begin{equation}
 \left\langle f,g \right\rangle = \int \limits_{0}^{1} \int \limits_{T_e} f(\xi,\eta,\tau)g(\xi,\eta,\tau) d\xi d\eta d\tau,  
\label{intOperators}
\end{equation}
which denotes the scalar product of two functions $f$ and $g$ over the space-time reference element $T_e\times \left[0,1\right]$. 

Using the discrete representations given in \eqref{thetaSol}, multiplication of Eqn. \eqref{PDECG} with the same space--time basis functions $\theta_k(\xi,\eta,\tau)$ and integration over the space--time reference element $T_e \times [0,1]$ leads to the following system
\begin{equation}
\left\langle \theta_k,\frac{\partial \theta_l}{\partial \tau} \right\rangle \widehat{\q}_{l,i} = \Delta  t 
 \left\langle \theta_k,\theta_l \right\rangle \widehat{\P}_{l,i}. 
\label{LagrSTPDECG}
\end{equation}   
Eqn. \eqref{LagrSTPDECG} can be shortened adopting a more compact matrix notation as 
\begin{equation}
\K_{\tau}\widehat{\q}_{l,i} = \Delta t \M \widehat{\P}_{l,i}, 
\label{LagrSTPDECGmatrix}
\end{equation}
where the following universal matrices have been defined on the space-time reference element:
\begin{equation}
\K_{\tau} = \left\langle \theta_k,\frac{\partial \theta_l}{\partial \tau} \right\rangle, \qquad \textnormal{and} \qquad \M = \left\langle \theta_k,\theta_l \right\rangle. 
\label{Ktau}
\end{equation}

We split the total vector of the degrees of freedom of the numerical solution $\widehat{\q}_{l,i}$ into two parts, namely the part $\widehat{\q}_{l,i}^{0}$ containing the degrees of freedom that are known from the initial condition $\w_h$ and the part $\widehat{\q}_{l,i}^{1}$ composed by the unknown degrees of freedom for $\tau>0$, hence obtaining $\widehat{\q}_{l,i}=(\widehat{\q}_{l,i}^{0},\widehat{\q}_{l,i}^{1})$. As done in \cite{Dumbser20088209}, 
the known degrees of freedom $\widehat{\q}_{l,i}^{0}$ are shifted onto the right-hand side of \eqref{LagrSTPDECGmatrix}, so that the nonlinear algebraic equation system \eqref{PCG} can be solved by an iterative procedure, 
i.e. \begin{equation}
  \K_{\tau} \widehat{\q}_{l,i}^{r+1} = \Delta t \M \widehat{\P}_{l,i}^r,  
\label{CGfinal}
\end{equation}
where the superscript $r$ denotes the iteration number. As initial guess ($r=0$) one can simply take the reconstruction polynomial $\w_h$ at the initial time level or a more accurate expression based on a second order MUSCL--type scheme (see \cite{HidalgoDumbser} for details).

Together with the governing PDE \eqref{PDE} we also have to evolve \textit{locally} the vertex coordinates of the local space--time element, since the mesh is moving in time. Introducing the local mesh velocity $\mathbf{V}=\mathbf{V}(x,y,t)=(U,V)$, the evolution of the local geometry is then described by the following ODE system:
\begin{equation}
\frac{d \mathbf{x}}{dt} = \mathbf{V}(x,y,t),
\label{ODEmesh}
\end{equation}
where the velocity is also expressed using a nodal approach as
\begin{equation}
\mathbf{V}_h=\mathbf{V}_h(\xi,\eta,\tau) = \theta_{l}(\xi,\eta,\tau) \widehat{\mathbf{V}}_{l,i}, \quad  \widehat{\mathbf{V}}_{l,i} = \mathbf{V}(\mathbf{\tilde{x}}_{l,i}).
\label{Vdof}
\end{equation}
We use again the local space--time Galerkin method to solve the system \eqref{ODEmesh}, as proposed in \cite{BoscheriDumbserLag}: 
\begin{equation}
\left\langle \theta_k,\frac{\partial \theta_l}{\partial \tau} \right\rangle \widehat{\mathbf{x}}_{l,i} = \Delta t \left\langle \theta_k,\theta_l \right\rangle \widehat{\mathbf{V}}_{l,i},
\label{VCG}
\end{equation}
which yields the following iteration scheme for the unknown coordinate vector $\widehat{\mathbf{x}}_l$: 
\begin{equation}
\K_{\tau} \widehat{\mathbf{x}}^{r+1}_{l,i} = \Delta t \M \widehat{\mathbf{V}}^r_{l,i}.
\label{newVertPos}
\end{equation}
The physical triangle $T_i^n$ at time $t^n$ is known, therefore the initial condition of the ODE system is given by the nodal degrees of freedom $\widehat{\x}_l$ at relative time $\tau=0$.

The physical system of the PDE \eqref{CGfinal} and the geometrical system of the ODE \eqref{newVertPos} are evolved \textit{together} within the local space--time predictor step and the iterative procedure  
stops when the residuals of the two systems are less than a prescribed tolerance $tol$ (typically $tol\approx 10^{-12}$). Once we have carried out the above procedure for all the elements of the computational 
domain, we end with  an \textit{element--local predictor} for the numerical solution $\q_h$, for the fluxes $\mathbf{F}_h=(\f_h,\g_h)$, for the source term $\S_h$ and also for the mesh velocity $\mathbf{V}_h$.  

We point out that our scheme is assumed to be an \textit{arbitrary Lagrangian-Eulerian} scheme (ALE), where the local mesh velocity can be chosen \textit{independently} from the local fluid velocity. If the mesh velocity is set to zero, i.e. $\mathbf{V}=0$, then the scheme reduces to a purely Eulerian approach, whereas we obtain a Lagrangian method if $\mathbf{V}$ coincides with the local fluid velocity. Any other choice regarding $\mathbf{V}$ is also  possible. 

Since each element has evolved its vertex coordinates \textit{locally}, at the new time level $t^{n+1}$ the geometry may be discontinuous. Therefore we have to update the mesh \textit{globally} in such a way that each vertex location is uniquely defined and this can be done with a suitable node solver algorithm.

\subsection{Node Solvers}
\label{sec.nodesolvers}
Once the local predictor procedure has been carried out, at each vertex $k$ different velocity vectors $\mathbf{V}_{k,j}^n$ are defined, depending on the number of elements belonging to the Voronoi neighborhood $\mathcal{V}_k$ of  node $k$. In order to obtain a \textit{continuous} mesh configuration at the new time level $t^{n+1}$, one has to fix a \textit{unique} time-averaged node velocity $\overline{\mathbf{V}}_k^n$ through a so called \textit{node solver}. This is 
a common feature in all cell-centered Lagrangian schemes. 
We compare three different methods to compute the node velocity with each other and briefly describe each of them in the following. Once a unique high order accurate time-averaged vertex velocity $\overline{\mathbf{V}}_k^n$ is  known, the new vertex coordinates are simply given by  
\begin{equation} 
	\mathbf{X}^{n+1}_{k}	= \mathbf{X}^{n}_{k}	+ \Delta t \, \overline{\mathbf{V}}_k^n,  
	\label{eqn.vertex.update}
\end{equation}
and we are able to update all the other geometric quantities needed for the computation, e.g. normal vectors, volumes, side lengths, \textit{etc.} Note that \eqref{eqn.vertex.update} is the weak integral form of the ODE that governs the vertex position, hence it is not just simply a first order Euler method,  
but it is high order accurate if the time-averaged node velocity $\overline{\mathbf{V}}_k^n$ is computed with high order of accuracy in time. 

\subsubsection{The node solver of Cheng and Shu $\mathcal{NS}_{cs}$.}
In \cite{chengshu1,chengshu3,chengshu4} Cheng and Shu introduced a very simple and general formulation for obtaining the final node velocity, which is chosen to be the arithmetic average velocity amongst all the 
contributions coming from the neighbor elements. Since the mesh might be locally highly deformed, we propose to define the node solver $\mathcal{NS}_{cs}$ using the idea of Cheng and Shu, as done in  \cite{BoscheriDumbserLag,DumbserBoscheriLagNC}, but taking a \textit{mass weighted} average velocity among the neighborhood $\mathcal{V}_k$ of node $k$, i.e.
\begin{equation}
\overline{\mathbf{V}}_k^n = \frac{1}{\mu_k}\sum \limits_{T_j^n \in \mathcal{V}_k}{\mu_{k,j}\overline{\mathbf{V}}_{k,j}}, 
\label{eqnNScs}
\end{equation}
with
\begin{equation}
\mu_k = \sum \limits_{T_j^n \in \mathcal{V}_k}{\mu_{k,j}}, \qquad \mu_{k,j}=\rho^n_j |T_j^n|.
\label{eqn.NScs.weights}
\end{equation}
The local weights $\mu_{k,j}$, which are the masses of the elements $T_j^n$, are defined multiplying the cell averaged value of density $\rho^n_j$ with the cell area $|T_j^n|$, while the local velocity contributions $\overline{\mathbf{V}}_{k,j}$ are computed integrating in time the high order vertex-extrapolated velocity at node $k$ as
\begin{equation}
\overline{\mathbf{V}}_{k,j} = \left( \int \limits_{0}^{1} \theta_l(\xi^e_{m(k)}, \eta^e_{m(k)}, \tau) d \tau \right) \widehat{\mathbf{V}}_{l,j}, 
\label{NodesVel}
\end{equation} 
where $m(k)$ is a mapping from the global node number $k$ to the local node number in element $T_j^n$. Recall that the $\xi^e_{m}$ and $\eta^e_{m}$ denote the coordinates of the vertices of the reference triangle in space. 

\subsubsection{The node solver of Maire $\mathcal{NS}_{m}$.}
\begin{figure}[htbp]
	\centering
		\includegraphics[width=0.70\textwidth]{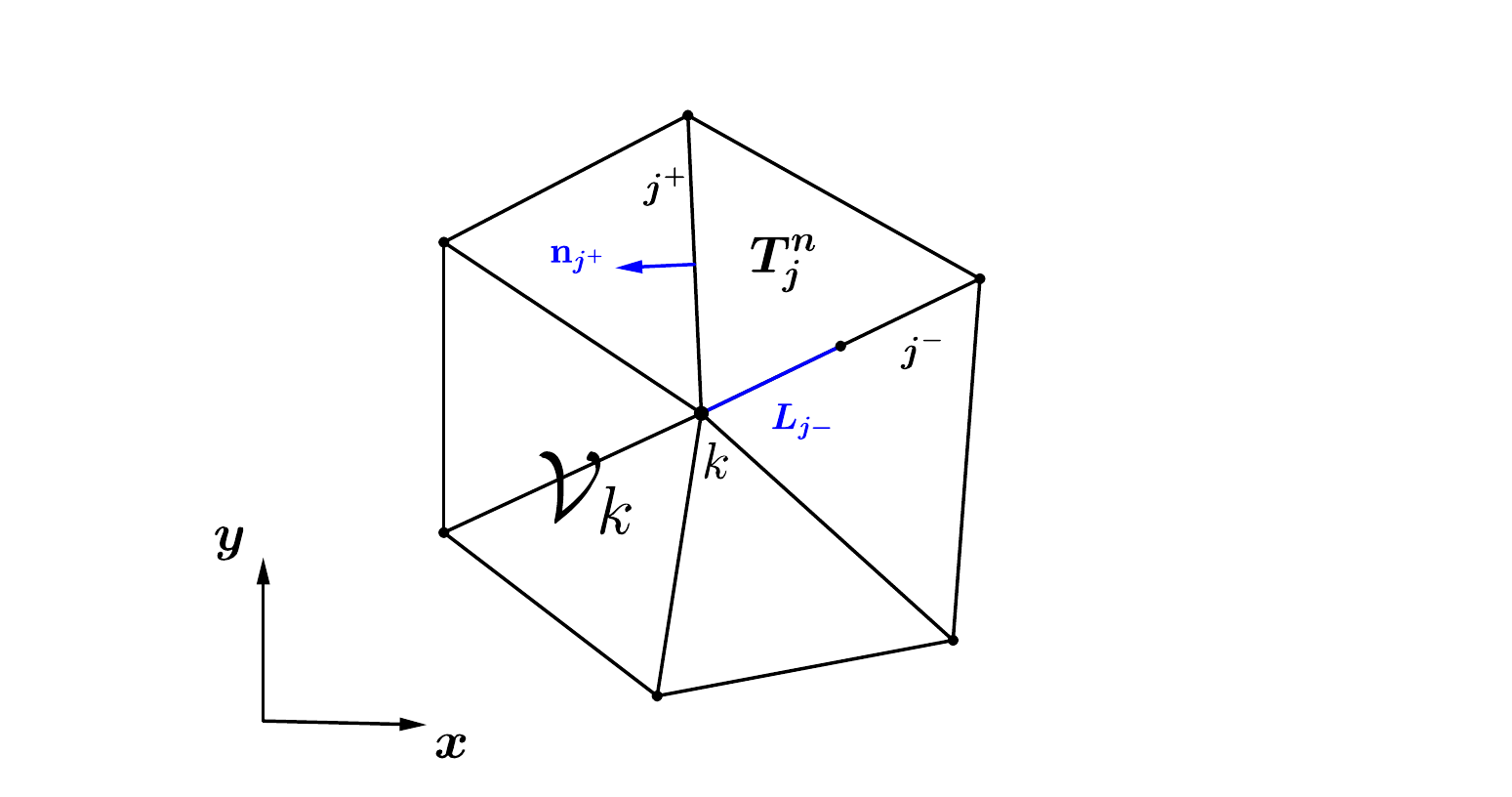}
	\caption{Geometrical notation for the node solver $\mathcal{NS}_{m}$: $k$ is the local node, $T_j^n$ denotes one element of the neighborhood $\mathcal{V}_k$ and $(j^-,j^+)$ are the counterclockwise ordered sides of $T_j^n$ that  share vertex $k$. $\mathbf{n}_{j^+}$ is the outward pointing unit normal vector and $L_{j^+}$ is the half length of side $j^+$.}
	\label{NS_Maire}
\end{figure}

The node solver $\mathcal{NS}_{m}$ used in this paper has been developed by Maire for hydrodynamics and the details of its derivation can be found in \cite{Maire2009,Maire2010,Maire2011}. A similar approach has been 
proposed by  Despr\'es et al. in \cite{Despres2009}. Here, we present only a brief overview of this node solver and refer to the previous references for further details. For the sake of clarity we first summarize the 
notation, according to Figure \ref{NS_Maire}: 
\begin{itemize}
  \item $k$ is the node index;
	\item $T_j^n$ represents the neighbor element $j$ of node $k$ and the subscripts $(j^-,j^+)$  denote the two sides of $T_j^n$ which share node $k$, ordered adopting a counterclockwise convention;
	\item $L_{j^+}$ and $\mathbf{n}_{j^+}$ denote, respectively, the half length and the outward oriented unit normal vector of side ${j^+}$;
	\item finally $c_j$ is the speed of sound for hydrodynamics, or the fastest magnetosonic speed for magneto--hydrodynamics, whose expressions will be explicitly given in Section \ref{sec.validation}. 
\end{itemize}

The formulation of the node solver $\mathcal{NS}_{m}$ is based on the conservation of total energy in the equations for compressible hydrodynamics. 
Consider a generic node $k$ and let $\mathbf{F}_{k,j}$ be the sub--cell force that the neighbor element $T_j^n$ exerts onto node $k$. 
According to \cite{Maire2010}, the sub--cell force is computed solving approximately two half Riemann problems on the sides $(j^+,{j^-})$ with the 
use of the acoustic Riemann solver \cite{DukowiczRS}, hence obtaining the following expression: 
\begin{equation}
\mathbf{F}_{k,j} = L_{k,j}p_{k,j}\mathbf{n}_{k,j} - \mathbf{M}_{k,j}\left(\mathbf{V}_{k} - \mathbf{V}_{k,j}\right),
\label{eqnCellForce}
\end{equation}
where the corner vector related to node $k$ is $L_{k,j}\mathbf{n}_{k,j}=L_{j^-}\mathbf{n}_{j^-}+L_{j^+}\mathbf{n}_{j^+}$ and the matrix $\mathbf{M}_{k,j}$ are computed as 
\begin{equation}
\mathbf{M}_{k,j} = z_{j^-}L_{j^-}\left(\mathbf{n}_{j^-}\otimes\mathbf{n}_{j^-}\right) + z_{j^+}L_{j^+}\left(\mathbf{n}_{j^+}\otimes\mathbf{n}_{j^+}\right),
\label{eqnMkj}
\end{equation}
with the acoustic impedance $z_{j}=\rho_j c_j$. $\mathbf{V}_{k,j}$ is the known vertex-extrapolated velocity of cell $j$, while $\mathbf{V}_{k}$ represents the unknown velocity of node $k$. 

The energy conservation at the generic node $k$ is guaranteed only if the forces acting on node $k$ sum up to zero, i.e.
\begin{equation}
\sum \limits_{T_j^n \in \mathcal{V}_k}{\mathbf{F}_{k,j}} = 0.
\label{eqnEnergycons}
\end{equation}

Inserting Eqn. \eqref{eqnCellForce} into the equation for the total energy conservation \eqref{eqnEnergycons}, yields the following linear algebraic system for the unknown node velocity ${\mathbf{V}}_k$: 
\begin{equation}
\mathbf{M}_k {\mathbf{V}}_k = \sum \limits_{T_j^n \in \mathcal{V}_k}{\left(L_{k,j}p_{k,j}\mathbf{n}_{k,j} + \mathbf{M}_{k,j}\mathbf{V}_{k,j}\right)}, \quad \mathbf{M}_k = \sum \limits_{T_j^n \in \mathcal{V}_k}{\mathbf{M}_{k,j}}.
\label{eq:NS_mLinSyst}
\end{equation}

Matrix $\mathbf{M}_k$ is always invertible since it is symmetric positive definite by construction, hence the system \eqref{eq:NS_mLinSyst} admits a unique solution and the node velocity can always be computed. 
As suggested in \cite{Maire2011}, the acoustic impedance is chosen as originally proposed by Dukowicz in \cite{DukowiczRS}: 
\begin{equation}
z_{j^+} = \rho_j\left[c_j+\Gamma_j |\left(\mathbf{V}_k-\mathbf{V}_{k,j}\right)\cdot \mathbf{n}_{j^+}|\right],
\label{eqnZnonlin}
\end{equation}
where $\Gamma_j=\frac{\gamma+1}{2}$ is a material dependent parameter which is a function of the ratio of specific heats $\gamma$. 
Eqn. \eqref{eqnZnonlin} depends on the unknown node velocity, hence the system \eqref{eq:NS_mLinSyst} becomes nonlinear and has to be resolved with a suitable iterative algorithm. 
For magneto--hydrodynamics we adopt the same procedure as for hydrodynamics, where we add the magnetic pressure in the sub--cell force computation \eqref{eqnCellForce} and we use the fastest magnetospeed 
for the acoustic impedance evaluation \eqref{eqnZnonlin}. The final time-averaged node velocity $\overline{\mathbf{V}}_k^n$ is obtained using Gaussian quadrature in time, 
where the node solver is invoked at each Gaussian point in time with the corresponding vertex-extrapolated states from the cells surrounding node $k$. 

\subsubsection{The node solver of Balsara $\mathcal{NS}_{b}$.}
\begin{figure}[htbp]
	\centering
		\includegraphics[width=0.50\textwidth]{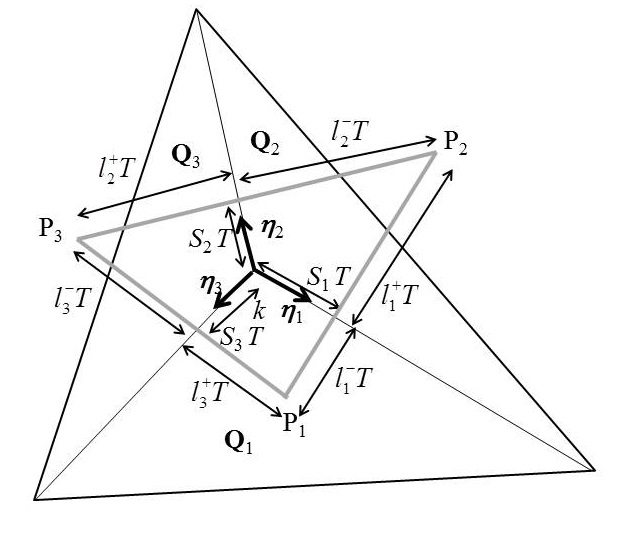}
	\caption{Multidimensional Riemann problem at vertex $k$, where three different states $(\Q_1,\Q_2,\Q_3)$ come together. The control volume generated by the propagation of the wavespeeds $(\S_1,\S_2,\S_3)$ within a time step $\Delta t$ is highlighted by the grey lines.}
	\label{NSb1}
\end{figure}

\begin{figure}[htbp]
	\centering
		\includegraphics[width=0.50\textwidth]{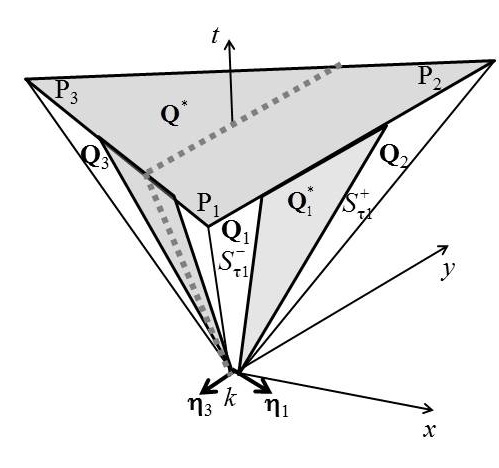}
	\caption{Inverted prism in space and time where the strongly interacting state $\Q^*$ is on the top surface. Along the side panels are depicted the one--dimensional Riemann problems.}
	\label{NSb2}
\end{figure}

In a recent series of papers \cite{BalsaraMultiDRS,BalsaraHLLE,BalsaraHLLC} Balsara et al. have proposed a genuinely multidimensional formulation of HLL and HLLC Riemann solvers for nonlinear hyperbolic conservation laws 
on Cartesian grids and general unstructured meshes in two space dimensions. There, a family of node--based HLL Riemann solvers is developed that considers a genuinely multidimensional flow structure developing at each grid vertex, 
in contrast to the classical edge-based Riemann solvers used in traditional Godunov--type finite volume schemes. At a grid vertex $k$ one can indeed take into account more physical information because multiple elements $T_j^n$ 
come together from all possible directions. In this paper, we rely on the fact that the genuinely multidimensional HLL Riemann solver can be used as one of the essential building blocks in the cell-centered ALE framework, namely 
as alternative node solver to the two previously mentioned ones. Hence, the multi-dimensional HLL Riemann solver also allows the node to be assigned a unique node velocity vector $\overline{\mathbf{V}}_k^n$ after the element-local  space-time predictor stage. Once the multidimensional HLL state $\Q^*$ is computed, the velocity components can be extracted from this so-called strongly interacting state and can be integrated in time in order to move the node 
to its location at the new time level $t^{n+1}$. 
For our purposes we always use the HLL version of the multidimensional Riemann solver proposed in \cite{BalsaraMultiDRS}. Figure \ref{NSb1}, which is taken from \cite{BalsaraMultiDRS}, shows the neighborhood $\mathcal{V}_k$ of  vertex $k$, where three different states $(\Q_1,\Q_2,\Q_3)$ come together at a node. The method is designed to handle an arbitrary number of states coming together at a node, hence we will use the generic states $\Q_j$ being the  vertex-extrapolated states from element $T_j^n$ at node $k$. 
The edge-aligned unit vector $\mathbf{\eta}_j$ separates the states $\Q_j$ and $\Q_{j+1}$, which have to be ordered in a counterclockwise fashion. Associated with vectors $\mathbf{\eta}_j$, we define $\mathbf{\tau}_j$ in such a way that $\mathbf{\eta}_j \cdot \mathbf{\tau}_j = 0$. The fastest waves propagate along the $\mathbf{\eta}_j$ direction with speeds $\mathbf{S}_j$ and within the time interval $\mathbf{T}=\Delta t = t^{n+1}-t^n$ they are contained in  the polygon bounded by vertexes $P_j$, defined as the intersection between the lines orthogonal to $\mathbf{\eta}_j$ and located at a distance $d_j=\mathbf{S}_j \mathbf{T}$ from vertex $k$ along direction $\mathbf{\eta}_j$. These  wavefronts define a polygonal area $\Omega_{HLL}$ which circumscribes the strongly interacting state and which evolves in time. In the space--time coordinate system it forms an inverted prism, as depicted in Figure \ref{NSb2}. 

The multidimensional state $\Q^*$ can be computed following three main steps:
\begin{enumerate}
	\item first we solve the one--dimensional Riemann problems perpendicular to $\mathbf{\eta}_j$, hence along the $\mathbf{\tau}_j$ directions. For this purpose we adopt a rotated reference system to solve the one-dimensional 
	Riemann problems arising at each side $j$ using a classical one-dimensional HLL solver. The dark shaded areas on the side panels of Figure \ref{NSb2} represent the resolved one--dimensional states; 
	\item the interacting state $\Q^*$ should then fully contain all the wave speeds starting from vertex $k$, originated from all the one--dimensional Riemann problems resolved during step 1. Thus, we use the wave speeds to obtain the \textit{multidimensional wave model} as shown in Figure \ref{NSb2} and the extremal wavefronts move with speed $\mathbf{S}_j$; 
	\item finally, the two--dimensional conservation law \eqref{PDE} is integrated over the three-dimensional prism in space--time (an inverted triangular pyramid in Figure \ref{NSb2}) in order to calculate the strongly 
	interacting multidimensional HLL state $\Q^*$. 
\end{enumerate}

The details of the computation of the above reported steps can be found in \cite{BalsaraMultiDRS} and \cite{BalsaraHLLE,BalsaraHLLC}. The final value of the velocity vector $\overline{\mathbf{V}}_k^n$ for node $k$ is then easily  extracted from the multidimensional state $\Q^*$. Also in this case, the final time-averaged node velocity $\overline{\mathbf{V}}_k^n$ is obtained by Gaussian quadrature in time. 

\subsection{Rezoning}
\label{sec.rezoning}
The aim of any Lagrangian scheme consists in computing the flow variables by moving together with the fluid. This implies that the mesh has to follow the flow motion as close as possible, hence the mesh velocity should be computed as accurately as possible, using one of the node solver algorithms described in the previous Section \ref{sec.nodesolvers}. When high deformations occur, namely high compression, high stretching or even twisting, the computational grid may become very badly shaped, where some elements tend to be very small or distorted and where mesh tangling or element crossing may occur. This usually leads to a failure of the computation. In order to avoid this problem,  
a suitable rezoning algorithm may be used, which consists in improving the geometric quality of the grid during the computation.

Let us consider the local node $k$ with its coordinate vector $\mathbf{x}_k^{n+1,Lag}$ computed at the end of the predictor step with \eqref{eqn.vertex.update} and its neighborhood $\mathcal{V}_k$. Let $T_j^{n+1}$ be one element of the neighborhood, which for sake of simplicity will be indicated with $j$, and let $\mathbf{x}_{j,l}=(x_{j,l},y_{j,l},z_{j,l})$ be the three nodes $l=1,2,3$ associated with $T_j^{n+1}$, which are counterclockwise ordered and in  such a way that node $k$ corresponds to $l=1$. Then the Jacobian matrix $\mathbf{J}_{j}$ of the mapping from the reference triangle to the physical element $j$ reads
\begin{equation}
\mathbf{J}_{j} = \left( \begin{array}{cc} x_{j,2}-x_k & y_{j,2}-y_k \\ x_{j,3}-x_k & y_{j,3}-y_k \end{array} \right),
\label{eq:locJac}
\end{equation}
and $\kappa_j = \left\| \mathbf{J}_{j}^{-1} \right\| \left\| \mathbf{J}_{j} \right\|$ represents its condition number.
According to \cite{KnuppRezoning}, we define the goal function $\mathcal{K}_k$ 
\begin{equation}
\mathcal{K}_k = \sum\limits_{T_j^{n+1} \in \mathcal{V}_k}{ \kappa_{j} },
\end{equation}
whose minimization leads to a \textit{locally} optimal position of the free vertex $k$. As done by Galera et al. \cite{MaireRezoning}, we use the first step of a Newton algorithm to compute the optimized rezoned coordinates  $\mathbf{x}_k^{Rez}$ 
for  node $k$, i.e. 
\begin{equation}
 \mathbf{x}_k^{Rez} = \mathbf{x}_k^{n+1,Lag} - \mathbf{H}_k^{-1}\left(\mathcal{K}_k\right) \cdot \nabla \mathcal{K}_k,
\label{eqn.vertex.rez}
\end{equation}
where $\mathbf{H}_k$ and $\nabla \mathcal{K}_k$ represent the Hessian and the gradient of the goal function $\mathcal{K}_k$, respectively: 
\begin{equation}
\mathbf{H}_k = \sum\limits_{T_j^{n+1} \in \mathcal{V}_k}{\left( \begin{array}{cc} \frac{\partial^2 \kappa_{j}}{\partial x^2} & \frac{\partial^2 \kappa_{j}}{\partial x \partial y} \\ \frac{\partial^2 \kappa_{j}}{\partial y \partial x} & \frac{\partial^2 \kappa_{j}}{\partial y^2} \end{array} \right)}, \quad \nabla \mathcal{K}_k = \sum\limits_{T_j^{n+1} \in \mathcal{V}_k}{\left( \frac{\partial \kappa_{j}}{\partial x}, \frac{\partial \kappa_{j}}{\partial y}\right)}.
\label{eqn.HessGrad}
\end{equation}

The final node location $\mathbf{x}_k^{n+1}$ is computed as a convex combination between its Lagrangian position and its rezoned position, that is
\begin{equation}
\mathbf{x}_k^{n+1} = \mathbf{x}_k^{n+1,Lag} + \omega_k \left( \mathbf{x}_k^{Rez} - \mathbf{x}_k^{n+1,Lag} \right),
\label{eqn.relaxation}
\end{equation}
where $\omega_k$ is a coefficient that lies within the interval $[0,1]$. It is related to the deformation of the Lagrangian grid over a time step $\Delta t$ and for rigid body motion, namely rigid translation and rigid rotation, $\omega_k$ results in $\omega_k=0$ so that the fully Lagrangian motion of the mesh is guaranteed. All the details of the computation of $\omega_k$ can be found in \cite{MaireRezoning}. 

Since our scheme is supposed to be as Lagrangian as possible, the rezoning step is applied only if it is strictly necessary to carry on the computation. 
Thus, for those test problems presented in Section \ref{sec.validation} where it is necessary, we will write it explicitly, otherwise, no rezoning has been used.  
We stress that in our one-step predictor-corrector formulation of the high order finite volume algorithm, the rezoning is done directly after the predictor stage and still \textit{before} the final evaluation 
of the fluxes in the finite volume scheme (corrector stage). This means that the ALE fluxes of the final finite volume scheme already take into account the rezoned geometry at the new time level. In this 
sense, our scheme can be classified as a \textit{direct} ALE method, which does not need a separate remapping step.   

\subsection{Finite Volume Scheme}
\label{sec.SolAlg}

\begin{figure}[htbp]
	\centering
		\includegraphics[width=0.98\textwidth]{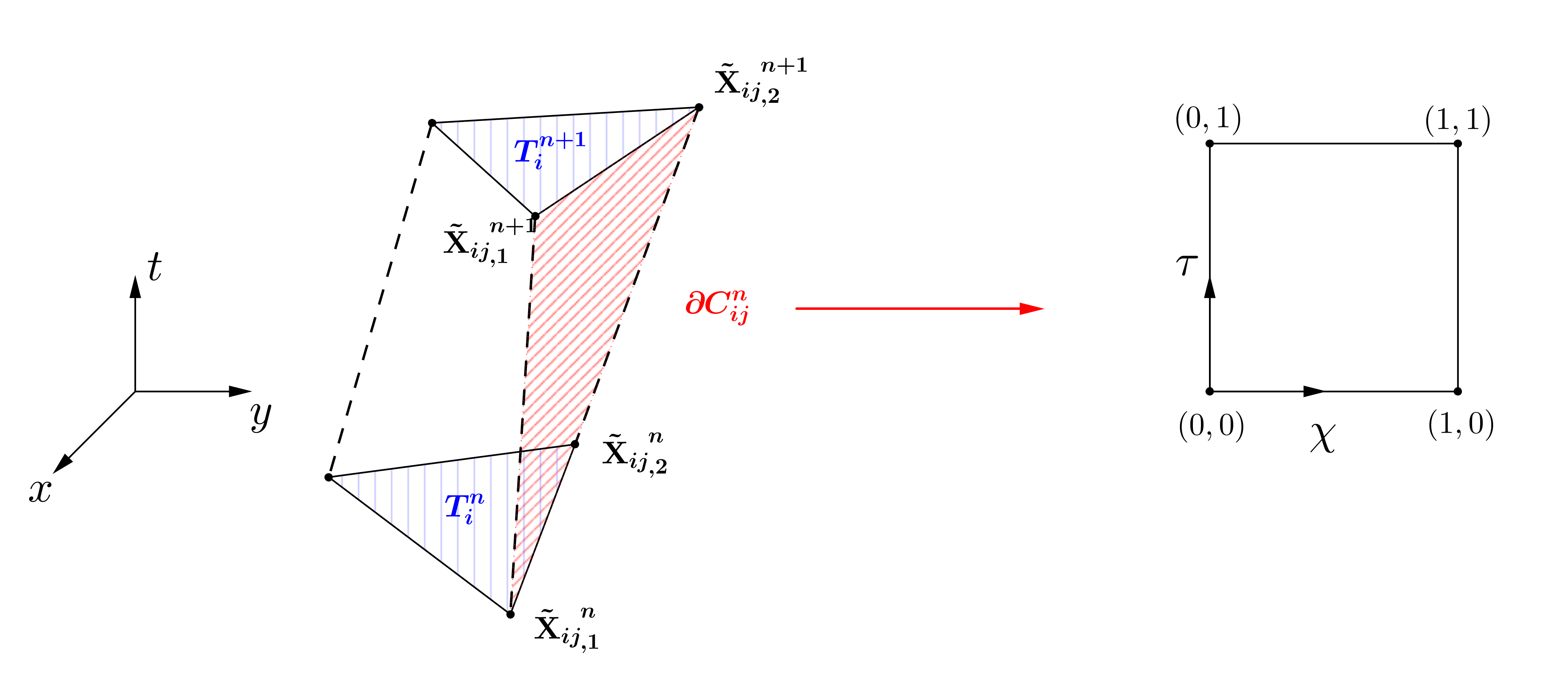}
	\caption{Physical space--time volume $C^n_i$ and reference system $(\chi,\tau)$ adopted for the bilinear parametrization of the lateral sub--surfaces $\partial C^n_{ij}$.}
	\label{Fvsurf}
\end{figure}

Once the new vertex coordinates have been fixed through the node solver algorithm and eventually the rezoning procedure, the physical space-time control volume $C^n_i = T_i(t) \times \left[t^{n}; t^{n+1}\right]$ is then obtained by connecting each vertex of triangle $T_i^n$ at time $t^n$ via \textit{straight} line segments with the corresponding vertex of triangle $T_i^{n+1}$ at time $t^{n+1}$, building a prism shaped control volume in space-time, 
see Figure \ref{Fvsurf}. The space-time control volume $C^n_i$ contains overall five space--time sub--surfaces, namely the triangle at the current time level $T_i^{n}$ as bottom, the triangle at the new time level 
$T_i^{n+1}$ as top and a total number of $\mathcal{N}_i=3$ lateral sub--surfaces $\partial C^n_{ij} = \partial T_{ij}(t) \times [t^n;t^{n+1}]$. 

In order to develop a finite volume scheme, we need to compute the space--time normal vectors for all the sub--surfaces. Let $\mathbf{X}_{ik}^n$ be the old vertex coordinates of element $T_i$ and let $\mathbf{X}_{ik}^{n+1}$ be the new ones. Let furthermore $\mathbf{\tilde{X}}_{ij,k}^n$ represent the physical space--time coordinate vectors for the four vertices that define the lateral space--time sub--surface  $\partial C_{ij}^n$, according to Figure \ref{Fvsurf}. Next, the lateral space--time sub--surfaces are parametrized using a set of bilinear basis functions as
\begin{equation}
\partial C_{ij}^n = \mathbf{\tilde{x}} \left( \chi,\tau \right) = 
 \sum\limits_{k=1}^{4}{\beta_k(\chi,\tau) \, \mathbf{\tilde{X}}_{ij,k}^n },	
 \qquad 0 \leq \chi \leq 1,  \quad	0 \leq \tau \leq 1, 										 
\label{SurfPar}
\end{equation}
where $(\chi,\tau)$ represents a side-aligned local reference system and the $\beta_k(\chi,\tau)$ functions read
\begin{eqnarray}
 \beta_1(\chi,\tau) &=& (1-\chi)(1-\tau), \nonumber\\
 \beta_2(\chi,\tau) &=& \chi(1-\tau), \nonumber\\
 \beta_3(\chi,\tau) &=& \chi\tau, \nonumber\\
 \beta_4(\chi,\tau) &=& (1-\chi)\tau.
 \label{BetaBaseFunc}
\end{eqnarray}
The mapping in time is again given by the transformation \eqref{timeTransf}, hence the Jacobian matrix $J_{\partial C_{ij}^n}$ of the parametrization is 
\begin{equation}
J_{\partial C_{ij}^n} = \left( \begin{array}{ccc} \vec{e}_x & \vec{e}_y & \vec{e}_t \\ 
\frac{\partial x}{\partial \chi} & \frac{\partial y}{\partial \chi} & \frac{\partial t}{\partial \chi} \\ 
\frac{\partial x}{\partial \tau } & \frac{\partial y}{\partial \tau } & \frac{\partial t}{\partial \tau } 
\end{array} \right) = \left( \begin{array}{c} \mathbf{\tilde{e}} \\ \frac{\partial \mathbf{\tilde{x}}}{\partial \chi} \\ \frac{\partial \mathbf{\tilde{x}}}{\partial \tau} \end{array} \right), 
\label{JacSTsurf}
\end{equation}
and the space--time unit normal vector $\mathbf{\tilde n}_{ij}$ can be evaluated computing the normalized cross product between the transformation vectors of the mapping \eqref{SurfPar}, i.e.
\begin{equation}
| \partial C_{ij}^n| = \left| \frac{\partial \mathbf{\tilde{x}}}{\partial \chi} \times \frac{\partial \mathbf{\tilde{x}}}{\partial \tau} \right|, 
\quad 
\mathbf{\tilde n}_{ij} = \left( \frac{\partial \mathbf{\tilde{x}}}{\partial \chi} \times \frac{\partial \mathbf{\tilde{x}}}{\partial \tau}\right) / | \partial C_{ij}^n|,
\label{n_lateral}
\end{equation}
where $| \partial C_{ij}^n|$ is the determinant of the Jacobian matrix $J_{\partial C_{ij}^n}$.

The parametrization of the upper space--time sub--surface $T_i^{n+1}$ and the lower space--time sub--surface $T_i^{n}$ is more simple, since these faces are orthogonal to the time coordinate. Therefore we use the mapping to the reference triangle given by \eqref{xietaTransf} and the space--time unit normal vectors read $\mathbf{\tilde n} = (0,0,1)$ for $T_i^{n+1}$ and $\mathbf{\tilde n} = (0,0,-1)$ for $T_i^{n}$.

We are now in the position to design a suitable conservative direct ALE finite volume scheme for the conservation law \eqref{PDE} on moving meshes. As proposed in \cite{BoscheriDumbserLag}, the governing PDE \eqref{PDE} is first  reformulated in a space--time divergence form as
\begin{equation}
\tilde \nabla \cdot \tilde{\F} = \mathbf{S}(\Q), 
\label{PDEdiv3D}
\end{equation} 
with
\begin{equation}
\tilde \nabla  = \left( \frac{\partial}{\partial x}, \, \frac{\partial}{\partial y}, \, \frac{\partial}{\partial t} \right)^T,  \qquad 
\tilde{\F}  = \left( \mathbf{F}, \, \Q \right) = \left( \mathbf{f}, \, \mathbf{g}, \, \Q \right),
\end{equation}
and it is then integrated in time over the space--time control volume $C^n_i$
\begin{equation}
\int\limits_{t^{n}}^{t^{n+1}} \int \limits_{T_i(t)} \tilde \nabla \cdot \tilde{\F} \, d\mathbf{x} dt = \int\limits_{t^{n}}^{t^{n+1}} \int \limits_{T_i(t)} \S \, d\mathbf{x} dt.   
\label{STPDE}
\end{equation} 
The volume integral on the left-hand side can be rewritten using the Gauss theorem as
\begin{equation}
\int \limits_{\partial C^n_i} \tilde{\F} \cdot \ \mathbf{\tilde n} \, \, dS = 
\int\limits_{t^{n}}^{t^{n+1}} \int \limits_{T_i(t)} \S \, d\mathbf{x} dt,   
\label{I1}
\end{equation}    
where $\mathbf{\tilde n} = (\tilde n_x,\tilde n_y,\tilde n_t)$ is the previously defined \eqref{n_lateral} outward pointing space--time unit normal vector on the space--time surface $\partial C^n_i$, which is given by the union of the five space--time sub--surfaces, i.e.
\begin{equation}
\partial C^n_i = \left( \bigcup \limits_{T_j(t) \in \mathcal{N}_i} \partial C^n_{ij} \right) 
\,\, \cup \,\, T_i^{n} \,\, \cup \,\, T_i^{n+1}.  
\label{dCi}
\end{equation}

The final high order ALE one--step finite volume scheme is obtained from Eqn. \eqref{I1} as
\begin{equation}
|T_i^{n+1}| \, \Q_i^{n+1} = |T_i^n| \, \Q_i^n - \sum \limits_{T_j \in \mathcal{N}_i} \,\, {\int \limits_0^1 \int \limits_0^1 
| \partial C_{ij}^n| \tilde{\F}_{ij} \cdot \mathbf{\tilde n}_{ij} \, d\chi d\tau}
+ \int\limits_{t^{n}}^{t^{n+1}} \int \limits_{T_i(t)} \S(\mathbf{q}_h) \, d\mathbf{x} dt, 
\label{PDEfinal}
\end{equation}
where the discontinuity of the predictor solution $\mathbf{q}_h$ at the space--time sub--face $\partial C_{ij}^n$ is resolved by an ALE numerical flux function $\tilde{\F}_{ij} \cdot \mathbf{\tilde n}_{ij}$. As done also for the local predictor step, the integrals in \eqref{PDEfinal} are approximated using multidimensional Gaussian quadrature rules, see \cite{stroud} for details. In \eqref{PDEfinal} the time step $\Delta t$ is given by 
\begin{equation}
\Delta t = \textnormal{CFL} \, \min \limits_{T_i^n} \frac{d_i}{|\lambda_{\max,i}|}, \qquad \forall T_i^n \in \Omega^n, 
\label{eq:timestep}
\end{equation}
where $\textnormal{CFL}$ is the Courant-Friedrichs-Levy number, $d_i$ represents the incircle diameter of element $T_i^n$ and $|\lambda_{\max,i}|$ is the maximum absolute value of the eigenvalues computed from the solution $\Q_i^n$ in $T_i^n$. As stated in \cite{toro-book}, for linear stability in two space dimensions the Courant number must satisfy $\textnormal{CFL} \leq 0.5$.

For the test problems shown in Section \ref{sec.validation} we use either a simple Rusanov--type ALE flux or a more complex Osher--type ALE flux. Let $\q_h^-$ be the numerical solution inside element $T_i(t)$ and $\q_h^+$ be the numerical solution inside the neighbor element $T_j(t)$. The expression for the Rusanov flux is given by
\begin{equation}
  \tilde{\F}_{ij} \cdot \mathbf{\tilde n}_{ij} =  
  \frac{1}{2} \left( \tilde{\F}(\q_h^+) + \tilde{\F}(\q_h^-)  \right) \cdot \mathbf{\tilde n}_{ij}  - 
  \frac{1}{2} s_{\max} \left( \q_h^+ - \q_h^- \right),  
  \label{eqn.rusanov} 
\end{equation} 
where $s_{\max}$ is the maximum eigenvalue of the ALE Jacobian matrix w.r.t. the normal direction in space, i.e. 
\begin{equation} 
\mathbf{A}^{\!\! \mathbf{V}}_{\mathbf{n}}(\Q)=\left(\sqrt{\tilde n_x^2 + \tilde n_y^2}\right)\left[\frac{\partial \mathbf{F}}{\partial \Q} \cdot \mathbf{n}  - 
(\mathbf{V} \cdot \mathbf{n}) \,  \mathbf{I}\right], \qquad    
\mathbf{n} = \frac{(\tilde n_x, \tilde n_y)^T}{\sqrt{\tilde n_x^2 + \tilde n_y^2}},  
\end{equation} 
with $\mathbf{I}$ representing the identity matrix and $\mathbf{V} \cdot \mathbf{n}$ denoting the local normal mesh velocity. 

The Osher--type flux formulation has been proposed in the Eulerian framework in \cite{OsherUniversal} and has been subsequently extended to moving meshes in one and two space dimensions in \cite{Dumbser2012} and \cite{BoscheriDumbserLag}, respectively. It reads 
\begin{equation}
  \tilde{\F}_{ij} \cdot \mathbf{\tilde n}_{ij} =  
  \frac{1}{2} \left( \tilde{\F}(\q_h^+) + \tilde{\F}(\q_h^-)  \right) \cdot \mathbf{\tilde n}_{ij}  - 
  \frac{1}{2} \left( \int \limits_0^1 \left| \mathbf{A}^{\!\! \mathbf{V}}_{\mathbf{n}}(\boldsymbol{\Psi}(s)) \right| ds \right) \left( \q_h^+ - \q_h^- \right),  
  \label{eqn.osher} 
\end{equation} 
where we choose to connect the left and the right state across the discontinuity using a simple straight--line segment path  
\begin{equation}
\boldsymbol{\Psi}(s) = \q_h^- + s \left( \q_h^+ - \q_h^- \right), \qquad 0 \leq s \leq 1.  
\label{eqn.path} 
\end{equation} 
According to \cite{OsherUniversal} the integral in \eqref{eqn.osher} is evaluated numerically using Gaussian quadrature. 
The absolute value of the dissipation matrix in \eqref{eqn.osher} is evaluated as usual as 
\begin{equation}
 |\mathbf{A}| = \mathbf{R} |\boldsymbol{\Lambda}| \mathbf{R}^{-1},  \qquad |\boldsymbol{\Lambda}| = \textnormal{diag}\left( |\lambda_1|, |\lambda_2|, ..., |\lambda_\nu| \right),  
\end{equation}
where $\mathbf{R}$ and $\mathbf{R}^{-1}$ denote the right eigenvector matrix and its inverse, respectively.

We underline that the integration over a closed space--time control volume, as done above, automatically satisfies the so-called geometric conservation law (GCL),  
since from the Gauss theorem follows 
\begin{equation}
 \int_{\partial \mathcal{C}_i^n} \mathbf{\tilde n} \, dS = 0. 
 \label{eqn.gcl} 
\end{equation} 
For all the numerical test problems shown later in this paper it has been explicitly verified that property \eqref{eqn.gcl} holds for all elements and for all time steps up to machine precision.  


\section{Test problems}
\label{sec.validation} 
\vspace{-2pt}

In this Section we solve some numerical test problems in order to validate the high order cell-centered direct ALE finite volume scheme presented above. The test cases will be carried out using the different node solvers  presented in Section \ref{sec.nodesolvers} and the two different numerical fluxes, namely the Osher--type flux \eqref{eqn.osher} and the Rusanov--type flux \eqref{eqn.rusanov}. We consider three hyperbolic systems of 
conservation laws, namely the Euler equations of compressible gasdynamics, the classical equations of magnetohydrodynamics (MHD) and the relativistic MHD equations (RMHD). 

\subsection{The Euler equations of compressible gasdynamics} 
Let $\rho$ denote the fluid density and $\mathbf{v}=(u,v)$ the fluid velocity vector and let $\rho E$ be the total energy density and $p$ the fluid pressure. Let furthermore $\gamma$ represent the ratio of specific heats of the gas and $c=\sqrt{\frac{\gamma p}{\rho}}$ the speed of sound. The Euler equations of compressible gas dynamics can be cast into form \eqref{PDE}, with 
\begin{equation}
\label{eulerTerms}
\Q = \left( \begin{array}{c} \rho \\ \rho u \\ \rho v \\ \rho E \end{array} \right), \quad \f = \left( \begin{array}{c} \rho u \\ \rho u^2 + p \\ \rho uv \\ u(\rho E + p) \end{array} \right), \quad \g = \left( \begin{array}{c} \rho v \\ \rho uv \\ \rho v^2 + p  \\ v(\rho E + p) \end{array} \right).  
\end{equation}
The system is closed by the following equation of state for an ideal gas:
\begin{equation}
\label{eqn.eos} 
p = (\gamma-1)\left(\rho E - \frac{1}{2} \rho (u^2+v^2) \right).  
\end{equation}

\subsubsection{Numerical convergence results.} 
\label{sec.conv.Rates-Eul}

In order to carry out the numerical convergence studies for the two--dimensional Lagrangian finite volume scheme, we consider the classical isentropic vortex test problem, see e.g. \cite{HuShuVortex1999}. It consists 
in a smooth isentropic vortex that is furthermore convected with velocity $\v_c=(1,1)$. The initial condition is given as a linear superposition of a homogeneous background field and some perturbations $\delta$: 
\begin{equation}
\label{ShuVortIC}
(\rho, u, v, p) = (1+\delta \rho, 1+\delta u, 1+\delta v, 1+\delta p).
\end{equation} 
We assume the entropy perturbation to be zero, i.e. $S=\frac{p}{\rho^\gamma}=0$, while the perturbations of velocity $\mathbf{v}=(u,v)$ and temperature $T$ are expressed as
\begin{equation}
\label{ShuVortDelta}
\left(\begin{array}{c} \delta u \\ \delta v \end{array}\right) = \frac{\epsilon}{2\pi}e^{\frac{1-r^2}{2}} \left(\begin{array}{c} -(y-5) \\ \phantom{-}(x-5) \end{array}\right), \qquad \delta T = -\frac{(\gamma-1)\epsilon^2}{8\gamma\pi^2}e^{1-r^2},
\end{equation} 
with the vortex radius $r^2=(x-5)^2+(y-5)^2$, the vortex strength $\epsilon=5$ and the ratio of specific heats $\gamma=1.4$. The perturbations for density and pressure are given by 
\begin{equation}
\label{rhopressDelta}
\delta \rho = (1+\delta T)^{\frac{1}{\gamma-1}}-1, \quad \delta p = (1+\delta T)^{\frac{\gamma}{\gamma-1}}-1. 
\end{equation}

\begin{figure}[!htbp]
	\centering
		\includegraphics[width=0.85\textwidth]{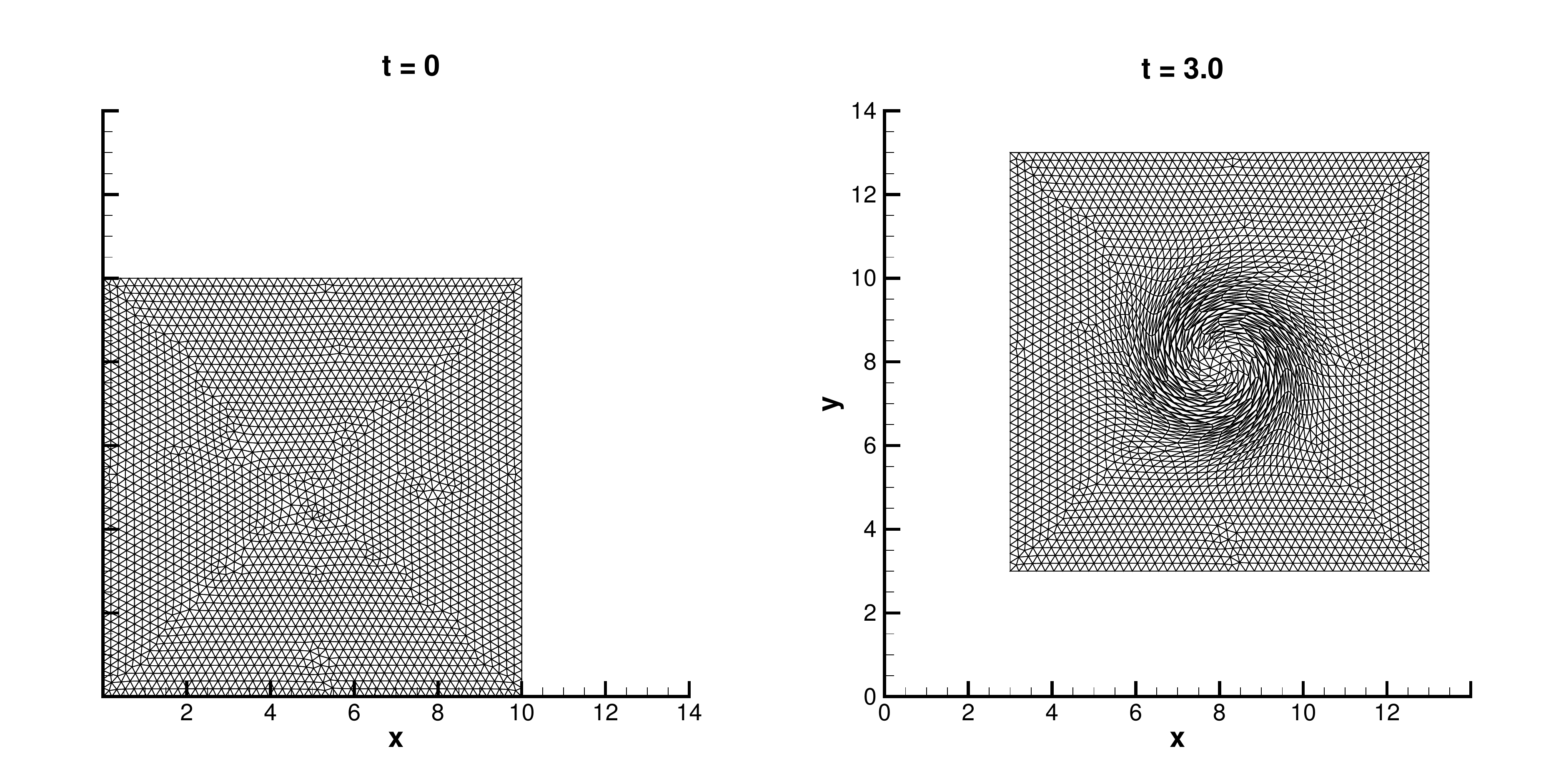}
	\caption{Mesh configuration for the isentropic vortex at time $t=0$ and $t=3.0$ with $\mathcal{NS}_{cs}$.}
	\label{fig:ShVgrid}
\end{figure}

The initially square--shaped computational domain is $\Omega(0)=[0;10]\times[0;10]$ and we assign periodic boundary conditions to each side. Due to the use of a Lagrangian scheme, the mesh motion follows the fluid flow and quickly  the grid becomes highly twisted, as depicted in Figure \ref{fig:ShVgrid}, so that it is not possible to run this test problem for large times unless the rezoning stage is performed. Here, we want to analyze the numerical  convergence of the purely Lagrangian algorithm, hence no rezoning is carried out and the simulation is run until the final time $t_f=1.0$ with the exact solution $\Q_e$ simply given by the  
time--shifted initial condition, e.g. $\Q_e(\x,t_f)=\Q(\x-\v_c t_f,0)$, where the convective mean velocity has been previously defined. We run this test case on a series of successively refined meshes and the corresponding 
error is expressed in the continuous $L_2$ norm as  
\begin{equation}
  \epsilon_{L_2} = \sqrt{ \int \limits_{\Omega(t_f)} \left( \Q_e(x,y,t_f) - \w_h(x,y,t_f) \right)^2 dxdy },  
	\label{eqnL2error}
\end{equation}
where $\w_h(x,y,t_f)$ represents the high order reconstructed solution at the final time. The mesh size $h(\Omega(t_f))$ is taken to be the maximum diameter of the circumcircles of the triangles in the final domain $\Omega(t_f)$. Table \ref{tab.convEul} shows the resulting numerical convergence rates from first up to fifth order of accuracy for each type of node solver. We can notice that the order of accuracy is well preserved by each scheme and essentially there are no differences due to the node solver choice.

\begin{table}  
\caption{Numerical convergence results for the compressible Euler equations. The first up to fifth order version of the two--dimensional Lagrangian one--step WENO finite volume 
scheme has been used for each node solver type. The error norms refer to the variable $\rho$ (density) at time $t=1.0$.} 
\begin{center} 
\begin{small}
\renewcommand{\arraystretch}{1.0}
\begin{tabular}{ccccccccc} 
\hline
                   &  $\mathcal{NS}_{cs}$  &		&   & $\mathcal{NS}_{m}$  &   &   &  $\mathcal{NS}_{b}$  &  \\  
  $h(\Omega(t_f))$ & $\epsilon_{L_2}$ & $\mathcal{O}(L_2)$ & $h(\Omega,t_f)$ & $\epsilon_{L_2}$ & $\mathcal{O}(L_2)$ & $h(\Omega,t_f)$ & $\epsilon_{L_2}$ & $\mathcal{O}(L_2)$ \\ 
\hline
\hline
   \multicolumn{3}{c} {} &        \multicolumn{3}{c}{$\mathcal{O}1$} & \multicolumn{3}{c}{} \\  
3.73E-01 & 6.9161E-02 & -    & 3.71E-01  & 6.9535E-02 & -    &  3.45E-01 & 7.0045E-02 & -     \\ 
2.62E-01 & 5.0030E-02 & 0.92 & 2.52E-01  & 5.0715E-02 & 0.82 &  2.52E-01 & 5.1427E-02 & 0.99  \\ 
1.76E-01 & 3.4607E-02 & 0.92 & 1.69E-01  & 3.5310E-03 & 0.90 &  1.71E-01 & 3.5513E-02 & 0.95 \\ 
1.36E-01 & 2.5724E-02 & 1.14 & 1.29E-01  & 2.6151E-03 & 1.11 &  1.29E-01 & 2.6277E-02 & 1.07  \\ 
\hline 
  \multicolumn{3}{c} {} &        \multicolumn{3}{c}{$\mathcal{O}2$} & \multicolumn{3}{c}{} \\  
3.49E-01 & 4.0451E-02 & -    & 3.40E-01  & 3.9160E-02 & -    &  3.39E-01 & 3.9585E-02 & -     \\ 
2.49E-01 & 2.6261E-02 & 1.28 & 2.49E-01  & 2.5787E-02 & 1.34 &  2.51E-01 & 2.5758E-02 & 1.44  \\ 
1.69E-01 & 1.5884E-02 & 1.29 & 1.67E-01  & 1.5721E-03 & 1.24 &  1.68E-01 & 1.5625E-02 & 1.24 \\ 
1.28E-01 & 1.0355E-02 & 1.55 & 1.28E-01  & 1.0300E-03 & 1.58 &  1.28E-01 & 1.0258E-02 & 1.54  \\
\hline 
  \multicolumn{3}{c} {} &        \multicolumn{3}{c}{$\mathcal{O}3$} & \multicolumn{3}{c}{} \\  
3.28E-01 & 1.6140E-02 & -    & 3.30E-01  & 1.6172E-02 & -    &  3.29E-01 & 1.6206E-02 & -     \\ 
2.51E-01 & 6.9455E-03 & 3.16 & 2.51E-01  & 6.9570E-03 & 3.10 &  2.51E-01 & 6.9638E-03 & 3.15  \\ 
1.68E-01 & 2.2904E-03 & 2.75 & 1.68E-01  & 2.2913E-03 & 2.74 &  1.68E-01 & 2.2925E-03 & 2.75 \\ 
1.28E-01 & 9.2805E-04 & 3.33 & 1.28E-01  & 9.2763E-04 & 3.34 &  1.28E-01 & 9.2806E-04 & 3.33  \\
\hline 
  \multicolumn{3}{c} {} &        \multicolumn{3}{c}{$\mathcal{O}4$} & \multicolumn{3}{c}{} \\  
3.29E-01 & 4.4613E-03 & -    & 3.29E-01  & 4.4636E-03 & -    &  3.29E-01 & 4.4627E-03 & -     \\ 
2.51E-01 & 1.7186E-03 & 3.54 & 2.51E-01  & 1.7189E-03 & 3.53 &  2.51E-01 & 1.7188E-03 & 3.55  \\ 
1.68E-01 & 4.2840E-04 & 3.43 & 1.68E-01  & 4.2834E-04 & 3.44 &  1.68E-01 & 4.2845E-03 & 3.43 \\ 
1.28E-01 & 1.3480E-04 & 4.27 & 1.28E-01  & 1.3477E-04 & 4.27 &  1.28E-01 & 1.3483E-04 & 4.27  \\ 
\hline 
  \multicolumn{3}{c} {} &        \multicolumn{3}{c}{$\mathcal{O}5$} & \multicolumn{3}{c}{} \\  
3.29E-01 & 4.4850E-03 & -    & 3.30E-01  & 4.4811E-03 & -    &  3.29E-01 & 4.4823E-03 & -     \\ 
2.51E-01 & 1.2711E-03 & 4.65 & 2.51E-01  & 1.2705E-03 & 4.63 &  2.51E-01 & 1.2705E-03 & 4.66  \\ 
1.68E-01 & 2.2531E-04 & 4.28 & 1.68E-01  & 2.2516E-04 & 4.28 &  1.68E-01 & 2.2520E-03 & 4.27 \\ 
1.28E-01 & 5.7869E-05 & 5.02 & 1.28E-01  & 5.7821E-05 & 5.02 &  1.28E-01 & 5.7838E-04 & 5.02  \\ 

\hline 
\end{tabular}
\end{small}
\end{center}
\label{tab.convEul}
\end{table}

\subsubsection{The Kidder problem.} 
\label{sec.Kidder}

The Kidder test problem is an isentropic compression of a shell filled with perfect gas. It constitutes a classical benchmark problem for Lagrangian schemes because it allows to verify whether the scheme produces spurious entropy  during the isentropic compression, or not. Moreover a self-similar analytical solution has been provided by Kidder in \cite{Kidder1976}. We consider the portion of a shell delimited by $r_i(t) \leq r \leq r_e(t)$, where $r_i(t),r_e(t)$ denote the time--dependent internal and external radius, respectively, and $r$ represents the general radial coordinate. At the initial time $t=0$ we set $r_i(0)=r_{i,0}=0.9$ and $r_e(0)=r_{e,0}=1.0$. Let $\rho_{i,0}=1$ and $\rho_{e,0}=2$ be the initial values of density defined at the internal and at the external frontier of the shell, respectively, and let the ratio of specific heats be $\gamma=2$. Let furthermore $c_l=\sqrt{\gamma\frac{p_l}{\rho_l}}$ be the sound speed defined at the general space location $l$. The initial density distribution is then given by 
\begin{equation}
\rho_0 = \rho(r,0) = \left(\frac{r_{e,0}^2-r^2}{r_{e,0}^2-r_{i,0}^2}\rho_{i,0}^{\gamma-1}+\frac{r^2-r_{i,0}^2}{r_{e,0}^2-r_{e,0}^2}\rho_{e,0}^{\gamma-1}\right)^{\frac{1}{\gamma-1}},
\end{equation}
and the initial entropy is assumed to be uniform, i.e. $s_0= \frac{p_0}{\rho_0^\gamma} = 1$, so that the initial pressure distribution can be expressed as
\begin{equation}
p_0(r) = s_0\rho_0(r)^\gamma. 
\end{equation}
Initially the velocity field is $u=v=0$. We impose sliding wall boundary conditions on the lateral faces, while on the internal and on the external frontier a space--time dependent state is assigned according to the exact analytical solution $R(r,t)$ (see \cite{Kidder1976} for details), which for a fluid particle initially located at radius $r$ is expressed as a function of the radius and the homothety rate $h(t)$, i.e.
\begin{equation}
  R(r,t) = h(t)r, \qquad h(t) = \sqrt{1-\frac{t^2}{\tau^2}},
\label{eqKidderEx}
\end{equation} 
with $\tau$ denoting the focalisation time 
\begin{equation}
\tau = \sqrt{\frac{\gamma-1}{2}\frac{(r_{e,0}^2-r_{i,0}^2)}{c_{e,0}^2-c_{i,0}^2}}.
\end{equation}  
The final time is taken to be $t_f=\frac{\sqrt{3}}{2}\tau$, according to \cite{Despres2009,Maire2009}, therefore the compression rate is $h(t_f)=0.5$ and the exact location of the shell is bounded with $0.45 \leq R \leq 0.5$.

\begin{figure}[!htbp]
	\centering
		\includegraphics[width=0.85\textwidth]{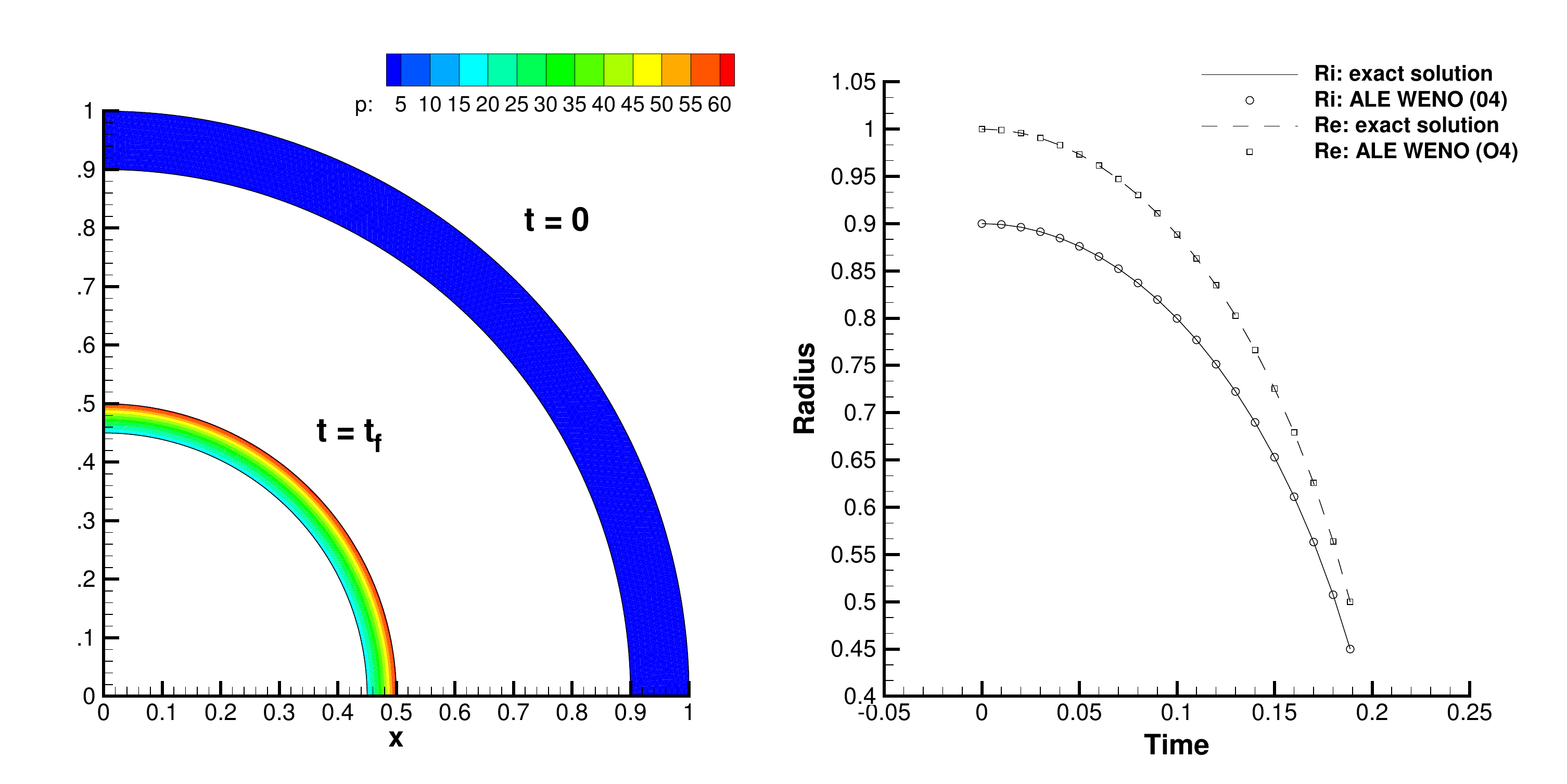}
	\caption{Left: pressure distribution at times $t=0$ and at $t=t_f$. Right: Evolution of the internal and external radius of the shell and comparison between analytical and numerical solution.}
	\label{fig:Kidder}
\end{figure}

We use a fourth order version of the ALE ADER-WENO scheme together with the Osher--type numerical flux \eqref{eqn.osher} with a characteristic mesh size $h=1/100$ and the numerical results are depicted in Figure \ref{fig:Kidder}. The Kidder problem has been run with each node solver type and we computed the absolute error $|err|$, reported in Table \ref{tab:KidErr} and defined as the difference between the analytical and the numerical location of the internal and external radius at the final time.

\begin{table}[!htbp]
  \caption{Absolute error for the internal and external radius location between exact and numerical solution for the three different node solvers. The numerical value has been evaluated as an average of the position of all the nodes lying on the internal and on the external frontier.}
	\begin{center}
		\begin{tabular}{|c|c|c|c|}
		\hline
		    		  			    & $\mathcal{NS}_{cs}$ 		& $\mathcal{NS}_{m}$  & $\mathcal{NS}_{b}$     \\
		\hline
		\textit{$|err_{int}|$}	& 7.72443E-06 	& 7.72460E-06  & 7.73181E-06 \\
		\hline
		\textit{$|err_{ext}|$}	& 1.01812E-05 	& 1.01811E-05  & 1.01867E-05 \\
		\hline
		\end{tabular}
	\end{center}
	
	\label{tab:KidErr}
\end{table}

\subsubsection{The Saltzman problem.} 
\label{sec.Saltzman}

We consider now a strong shock wave caused by the one--dimensional motion of a piston which compresses the gas contained in a rectangular channel. This test case was initially proposed by Dukowicz et al. in \cite{SaltzmanOrg} for a Cartesian grid that has been skewed, so that the mesh is no more aligned with the fluid flow. The Lagrangian algorithm has to be very robust to run such a difficult test problem, which is usually proposed in literature \cite{Maire2009,chengshu2}. The initial two--dimensional domain is $\Omega(0)=[0;1]\times[0;0.1]$ and the computational mesh is composed of $200 \times 20$ triangular elements, obtained as follows:
\begin{itemize} 
	\item first we build a Cartesian mesh with $100 \times 10$ square elements, as done in \cite{Maire2009,chengshu2};
	\item each square element is then split into two right triangles;
	
	\item finally the uniform grid, defined by the coordinate vector $\mathbf{x}=(x,y)$, is skewed with the mapping
	  \begin{eqnarray}
    	x' &=& x + \left( 0.1 - y \right) \sin(\pi x) \nonumber, \\
    	y' &=& y,
	   \label{eqSaltzSkew}
	  \end{eqnarray}
\end{itemize}
where $x'$ and $y'$ denote the deformed coordinates, respectively. 

The ideal gas is highly compressed by the piston, which travels from the left to the right with constant velocity $\mathbf{v}_p = (1,0)$. According to \cite{chengshu2}, the ratio of specific heats is taken to be $\gamma = \frac{5}{3}$ and the initial condition $\Q_{0}$ is given by the state
\begin{equation}
 \Q_{0} = \left( \rho_0, u_0, v_0, \rho E \right) = \left( 1, 0, 0, 10^{-4} \right).
\label{eqSaltz_ini}
\end{equation}

We set a moving slip wall boundary condition for the piston on the left side of the domain and sliding wall boundary conditions on the remaining boundaries. The piston starts moving very fast towards the fluid, which is initially at rest, therefore one has to be aware that the geometric $\textnormal{CFL}$ condition is observed and elements crossing over does not occur. For this reason we set an initial Courant number of $\textnormal{CFL}=0.01$ which is increased to its usual value of $\textnormal{CFL}=0.5$ at time $t=0.01$, as done by Cheng and Shu \cite{chengshu2}.

The numerical results are compared with the exact solution $\Q_{ex}$, 
\begin{equation}
  \Q_{ex}(\x,t_f) = \left\{ \begin{array}{ccc} \left( 4, 1, 0, 2.5     \right) & \textnormal{ if } & x \leq x_f, \\
                                               \left( 1, 0, 0, 10^{-4} \right) & \textnormal{ if } & x > x_f,        
                      \end{array}  \right. 
\end{equation}
where $x_f=0.8$ is the shock location at time $t = 0.6$. The details of the algorithm that computes the exact solution of the Saltzman problem can be found in \cite{BoscheriDumbserLag,ToroBook}. 

\begin{figure}[!htbp]
\begin{center}
\begin{tabular}{cc} 
\multicolumn{2}{c}{
\includegraphics[width=0.85\textwidth]{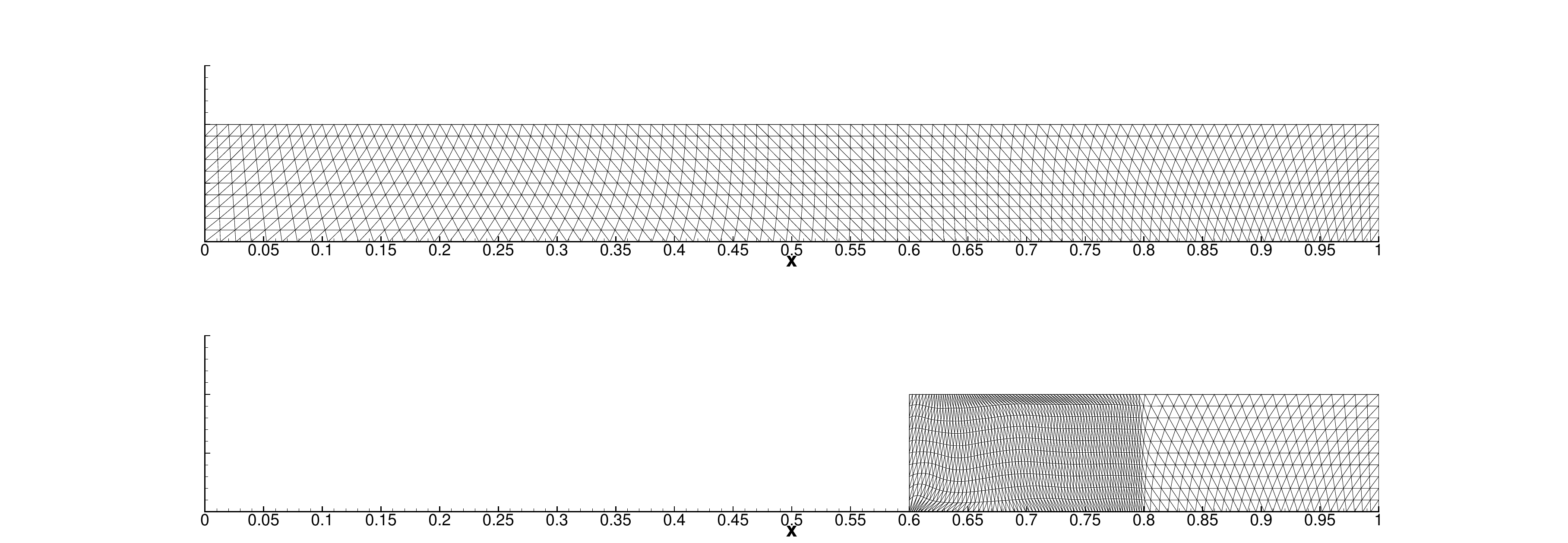}    
} \\
\includegraphics[width=0.47\textwidth]{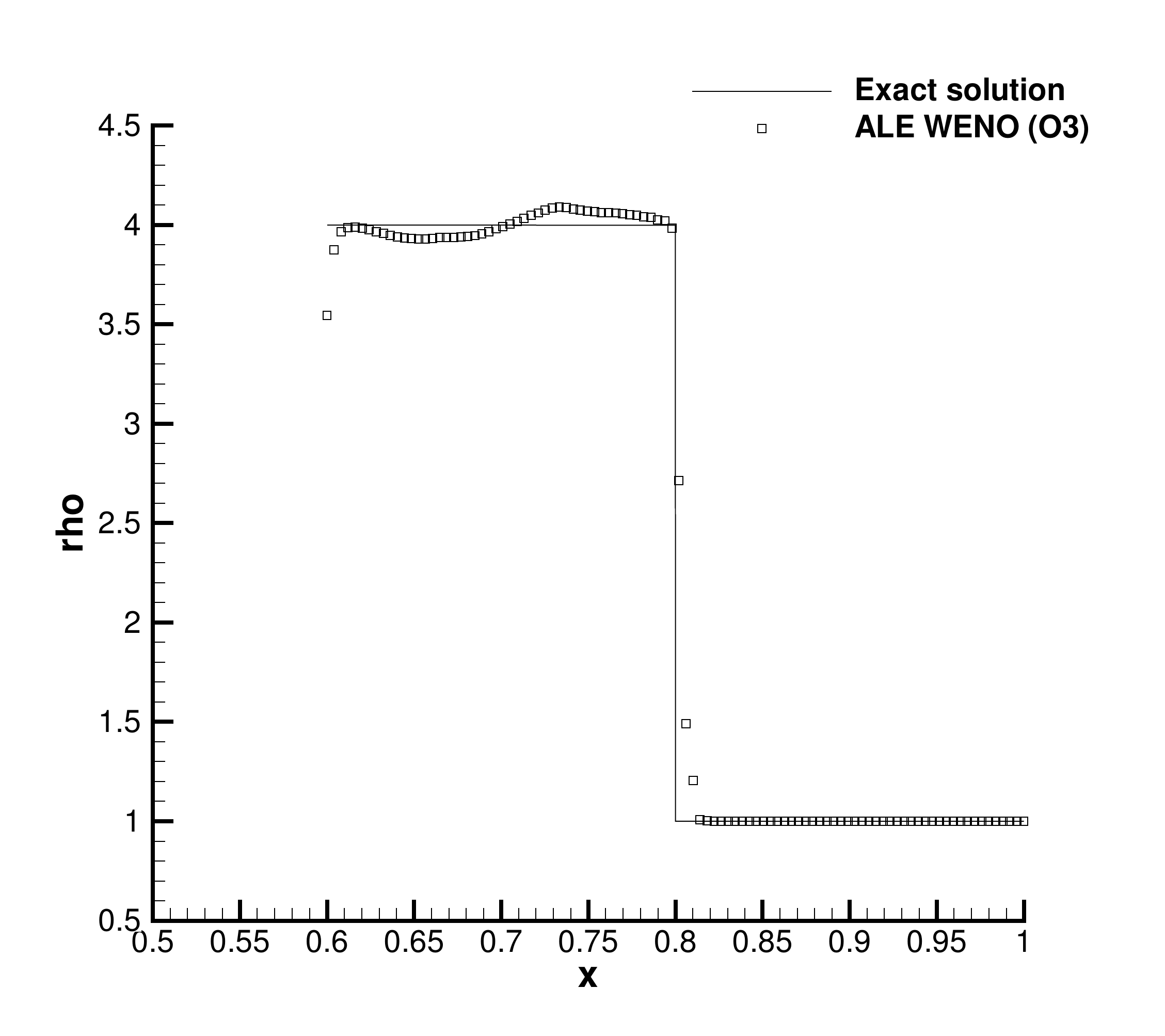}  &           
\includegraphics[width=0.47\textwidth]{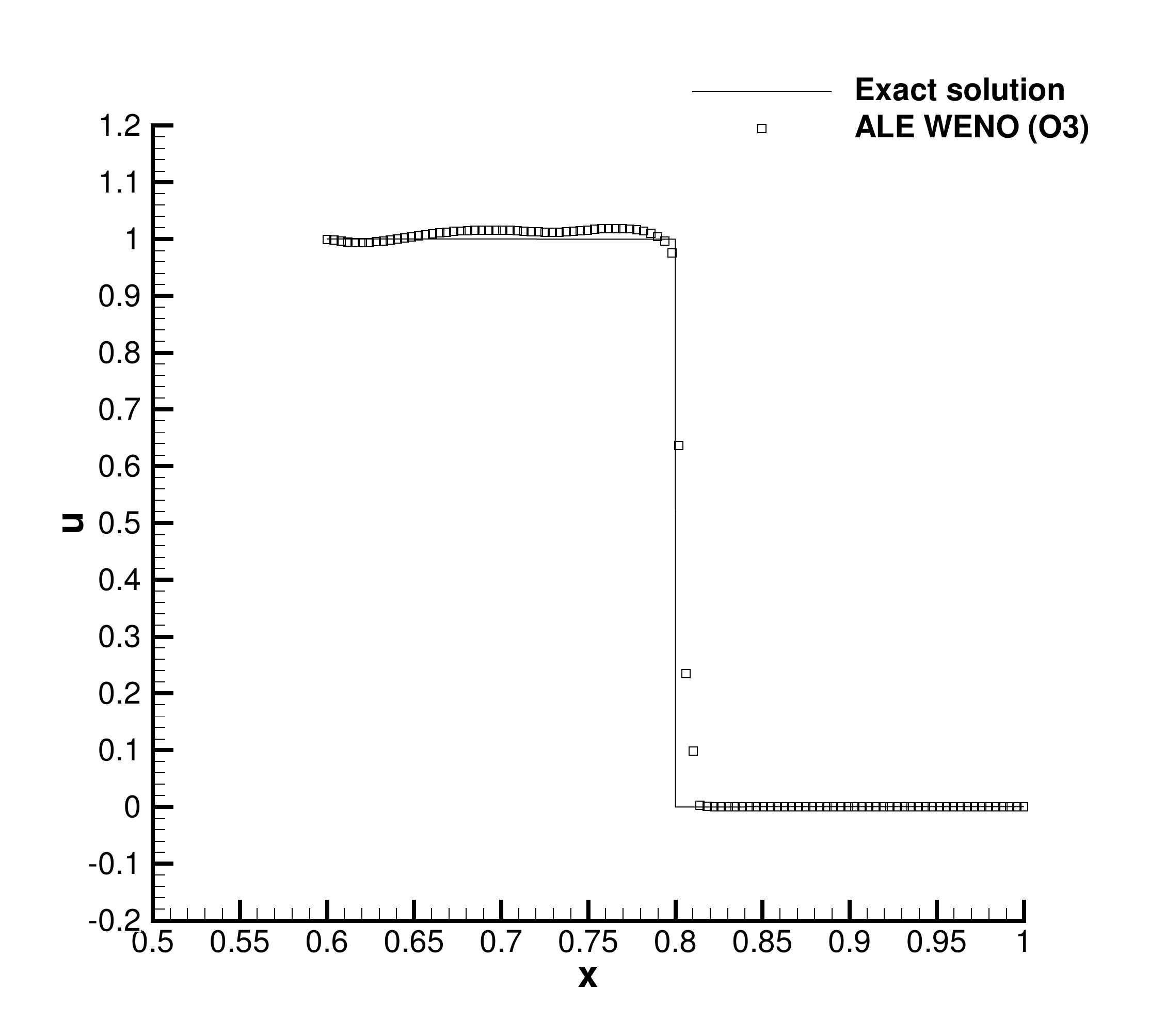} \\   
\end{tabular} 
\caption{Top: initial($t=0$) and final ($t=0.6$) mesh configuration for the Saltzman problem. Bottom: comparison between numerical and analytical solution at time $t=0.6$ for the variables $\rho$ (density) and $u$ (horizontal velocity).} 
\label{fig.Saltz2D}
\end{center}
\end{figure}

\begin{figure}[!htbp]
\begin{center}
\begin{tabular}{cc} 
\includegraphics[width=0.47\textwidth]{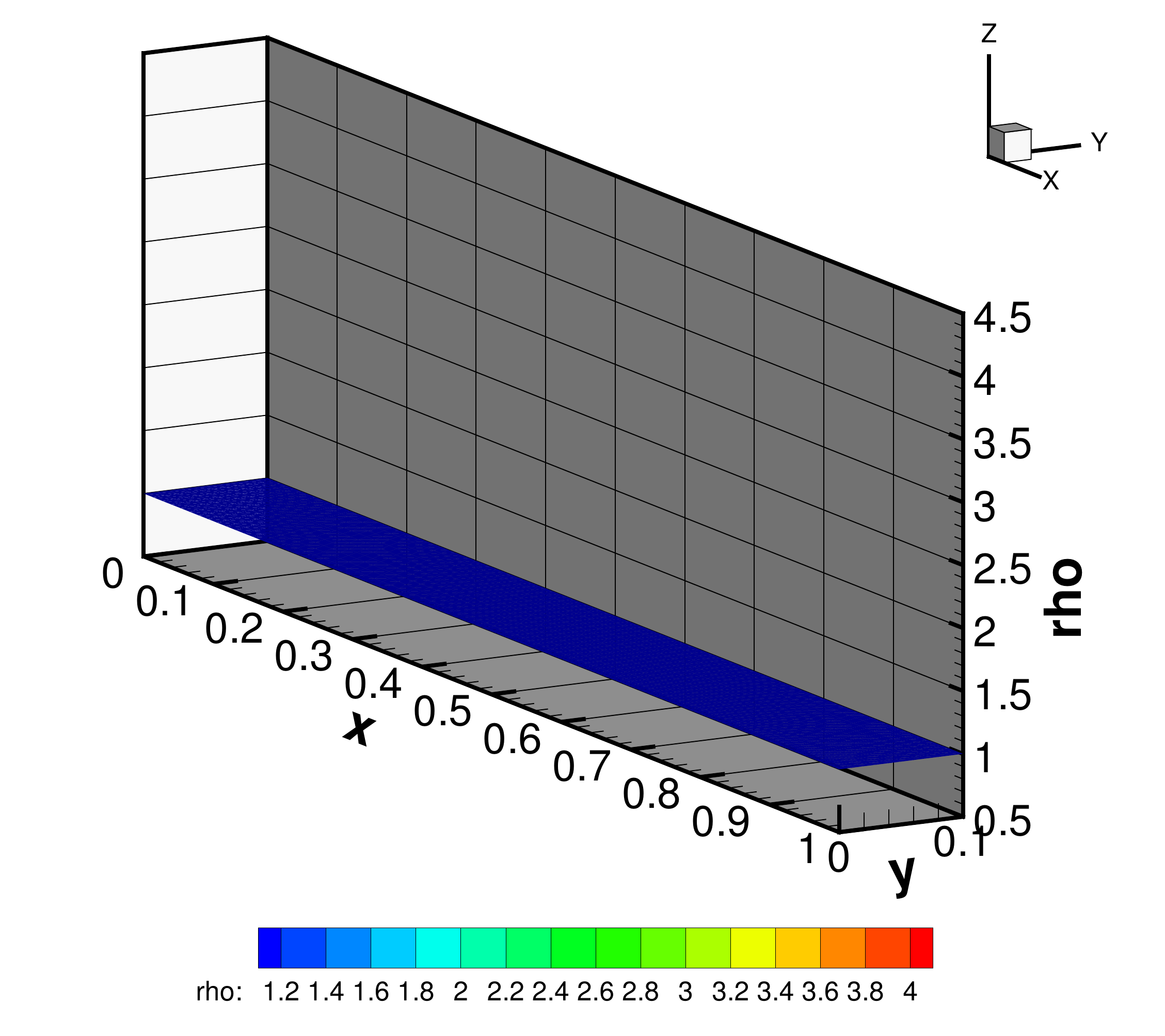}  &           
\includegraphics[width=0.47\textwidth]{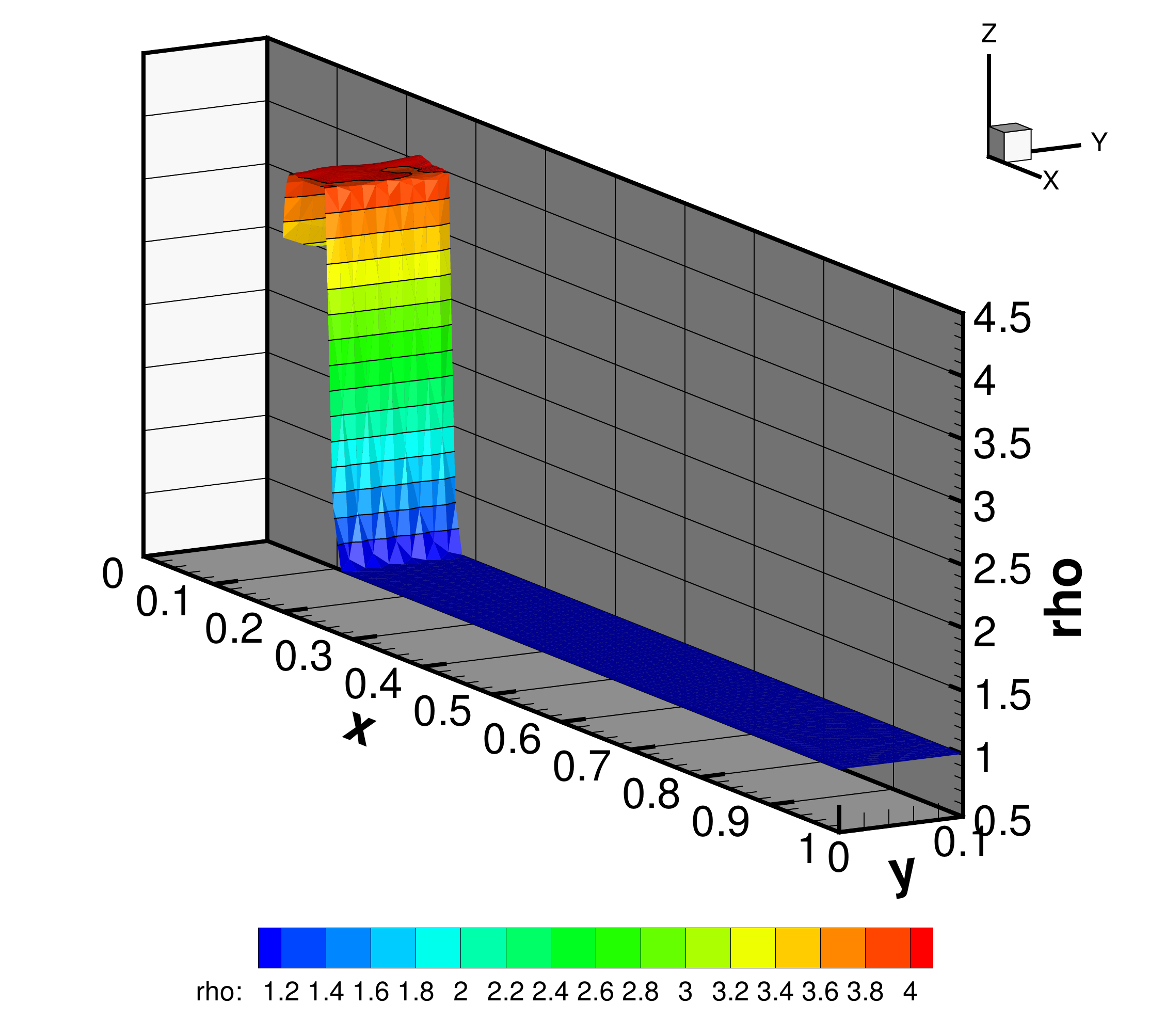} \\
\includegraphics[width=0.47\textwidth]{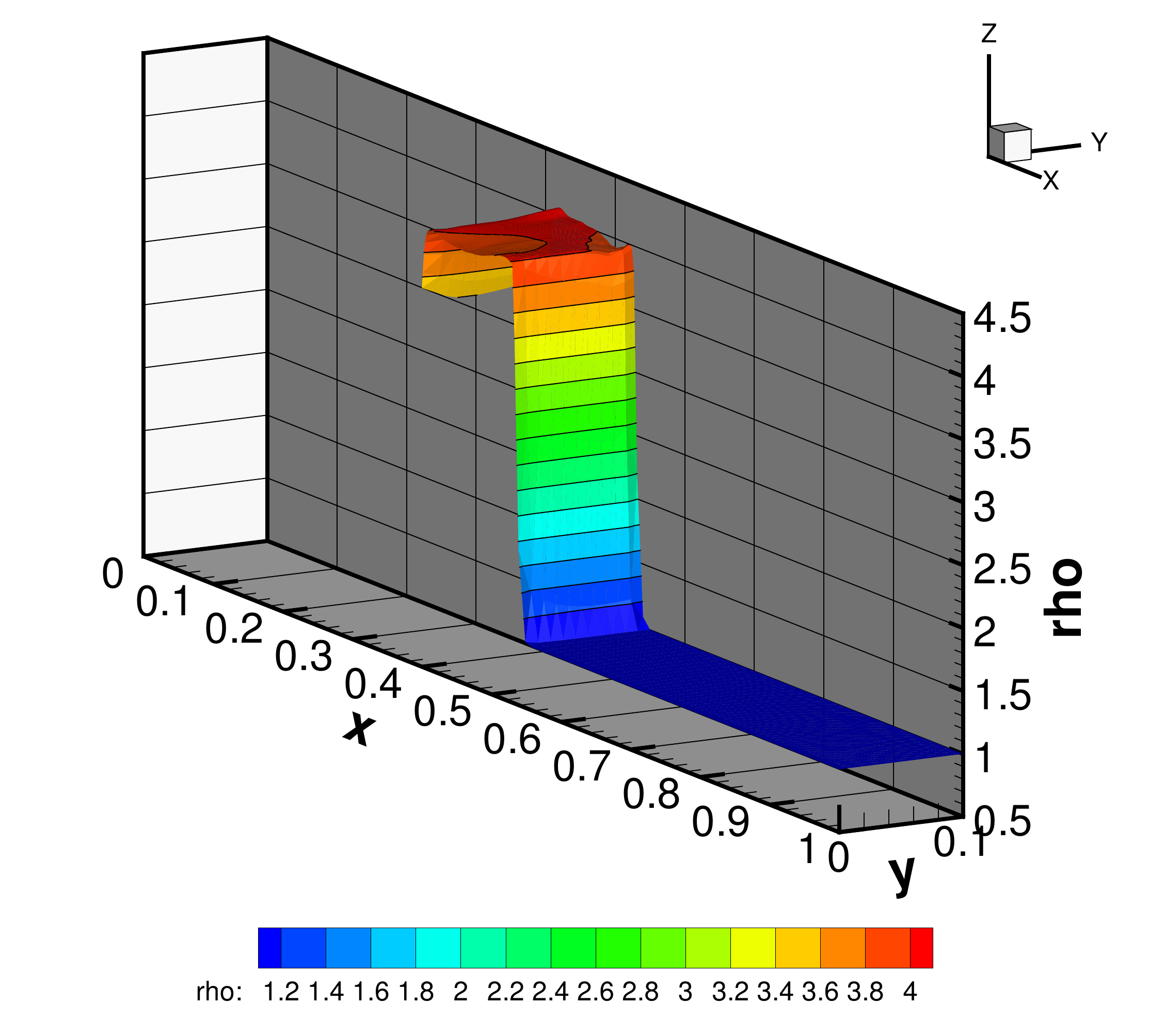}  &           
\includegraphics[width=0.47\textwidth]{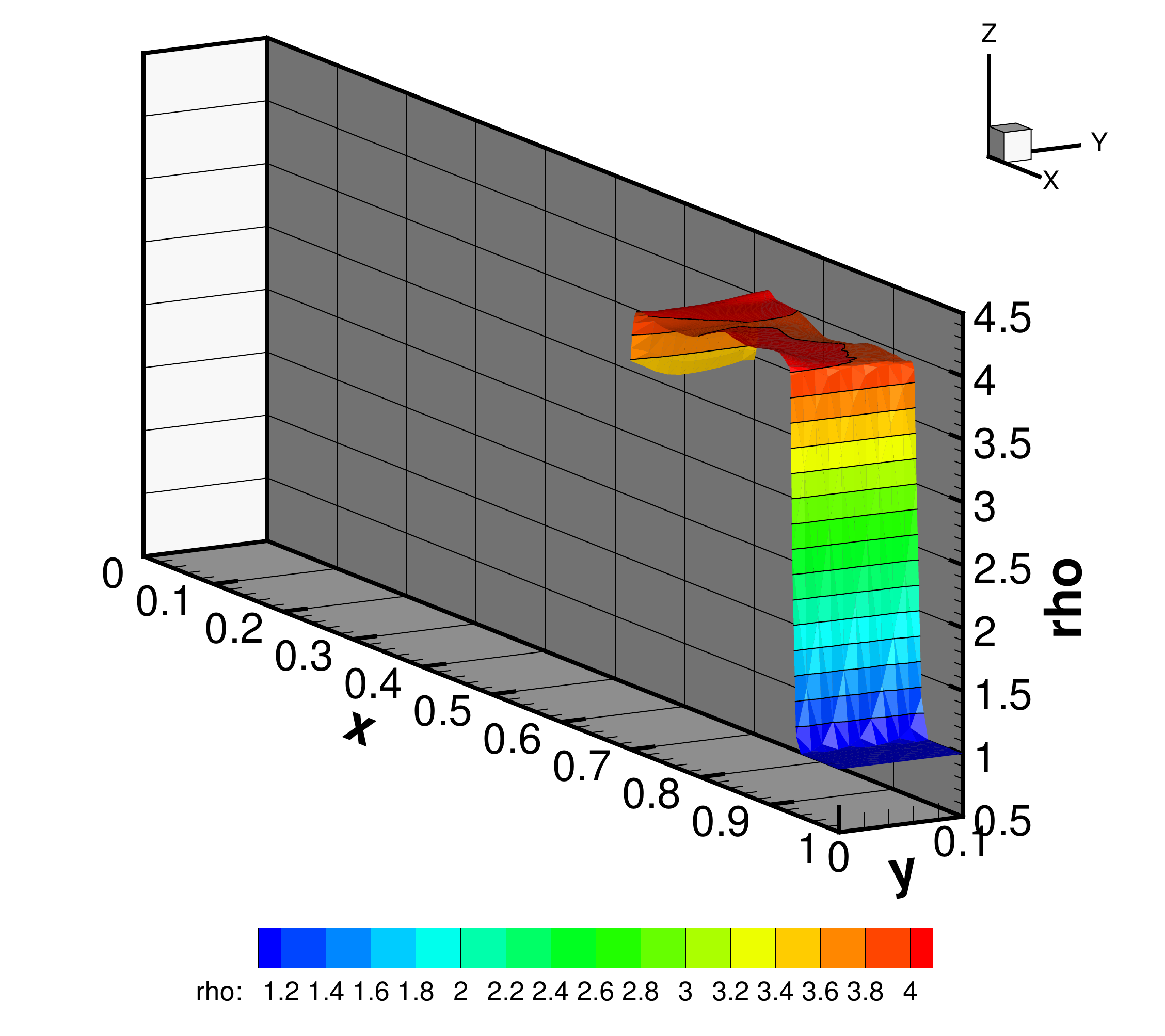} \\    
\end{tabular} 
\caption{Evolution of the density solution for the Saltzman problem at output times $t=0$, $t=0.2$, $t=0.4$ and $t=0.7$ (top left to bottom right).} 
\label{fig.SaltzView3d}
\end{center}
\end{figure}

The numerical results obtained with the third order ALE ADER-WENO method are shown at time $t=0.6$ in Figure \ref{fig.Saltz2D}, together with the initial and the final mesh configuration. We use a third order scheme with a robust Rusanov--type numerical flux \eqref{eqn.rusanov} and the node solver $\mathcal{NS}_b$. A good agreement of the density distribution with the exact solution can be observed and the decrease of the density which occurs near the  piston is due to the well known \textit{wall--heating problem}, see \cite{toro.anomalies.2002}. The evolution of density is depicted in Figure \ref{fig.SaltzView3d}. Using node solver $\mathcal{NS}_b$ we were able to run the simulation with a time step size that was 10\% larger with respect to the other node solvers. Furthermore, with the node solver $\mathcal{NS}_b$ it was possible to run the simulation until a final time of $t=0.74$,  which was not possible  with the other node solvers $\mathcal{NS}_m$ and $\mathcal{NS}_{cs}$, which required smaller time steps and reached only $t_f=0.69$. 

\subsubsection{The Sedov problem.} 
\label{sec.Sedov}

\begin{figure}[!htbp]
\begin{center}
\begin{tabular}{cc} 
\includegraphics[width=0.47\textwidth]{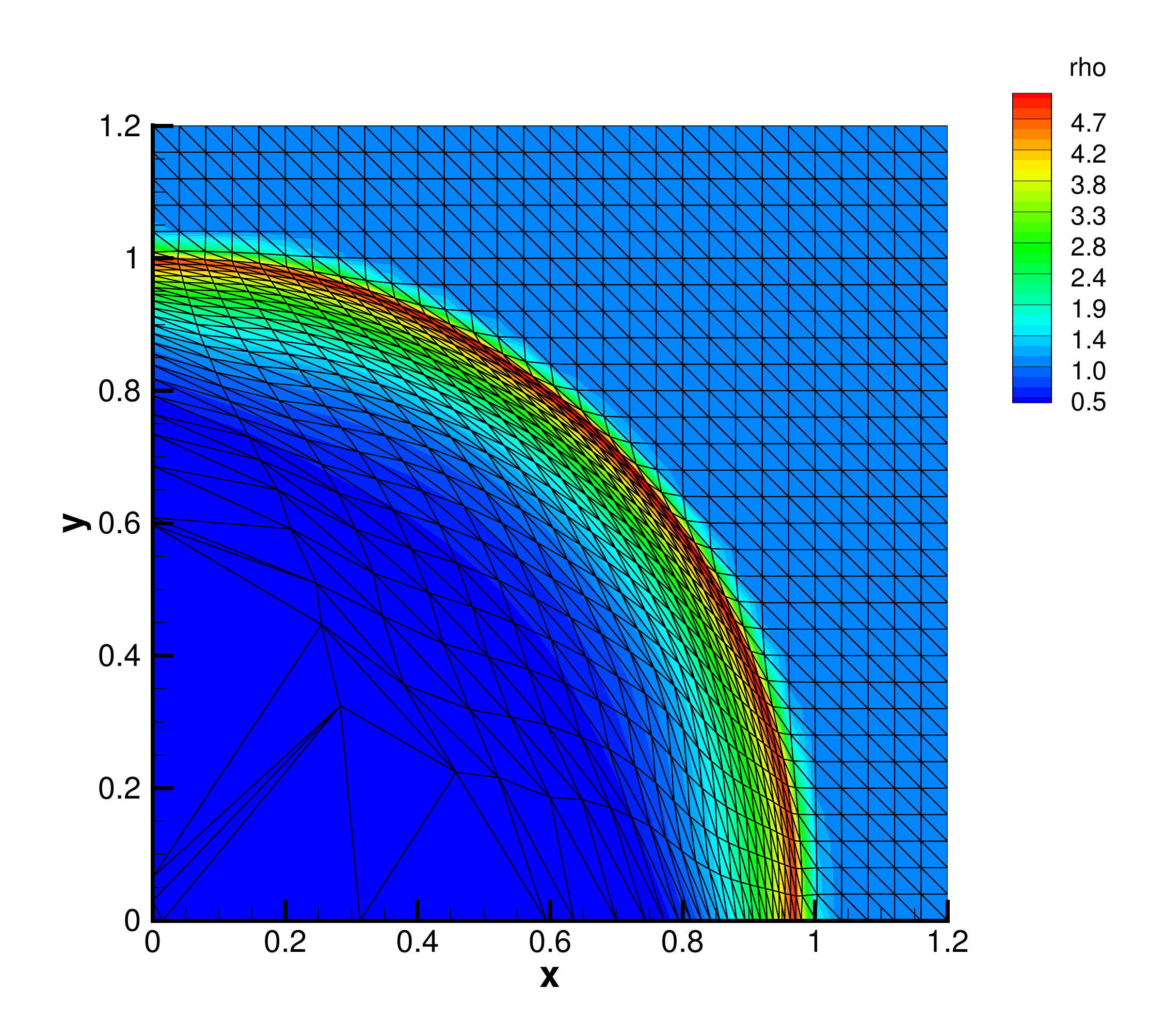}  &           
\includegraphics[width=0.47\textwidth]{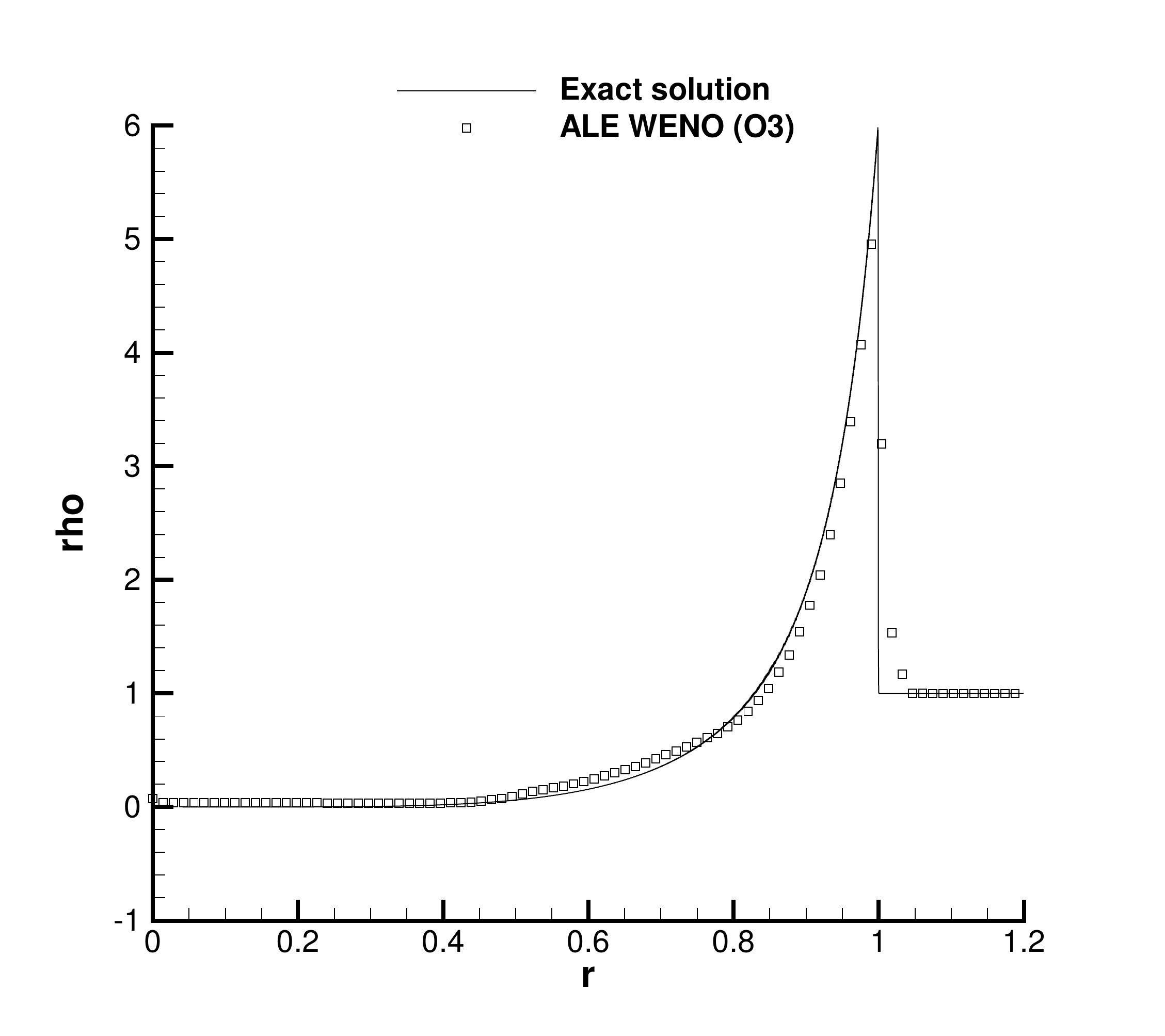} \\    
\end{tabular} 
\caption{Left: density distribution at the final time of the simulation $t_f=1$. Right: comparison between analytical and numerical solution of density along the radial direction at $t_f=1$.} 
\label{fig.Sedov2D}
\end{center}
\end{figure}

The Sedov problem consists in the evolution of a blast wave with cylindrical symmetry. It is a classical test case for Lagrangian schemes \cite{Maire2009,Maire2009b} and Kamm et al. \cite{SedovExact} proposed an exact solution which is based on self--similarity arguments. The initial computational domain is a square $\Omega(0)=[0;1.2]\times[0;1.2]$ and the initial mesh is composed by $(30\times30)$ square elements, each of those has been split into two triangles, hence the total number of elements is $N_E=1800$. The gas is initially at rest and has a uniform unity density distribution $\rho_0=1$. The initial pressure is $p_0=10^{-6}$ everywhere, except in the cell $c_{or}$ containing the origin of the domain, i.e. $O=(0,0)$. There, an energy source is collocated and the initial pressure $p_{or}$ is computed as
\begin{equation}
p_{or} = (\gamma-1)\rho_0 \frac{\epsilon_0}{V_{or}},
\label{eqn.p0.sedov}
\end{equation} 
where the ratio of specific heats is taken to be $\gamma=1.4$, $V_{or}$ is the volume of the cell $c_{or}$, which is composed by two triangles, and $\epsilon_0=0.244816$ is the total amount of released energy. The parameters have been set according to \cite{SedovExact}, hence the solution is given by a radially traveling shock wave that is located at radius $r=\sqrt{x^2+y^2}=1$ at the final time of the simulation $t_f=1$. Sliding wall boundary conditions have been imposed at each side of the domain. We use the third order version of the ALE ADER-WENO finite volume schemes and the node solver $\mathcal{NS}_{m}$ together with a Rusanov--type numerical flux \eqref{eqn.rusanov} and the
rezoning strategy of Section \ref{sec.rezoning}. Figure \ref{fig.Sedov2D} shows the density distribution at the final time and a comparison between the analytical and the numerical solution of density along the radial direction, where the one--dimensional cylindrical solution is reasonably well preserved by the scheme. 

\subsubsection{The Noh problem.} 
\label{sec.Noh}

\begin{figure}[!htbp]
\begin{center}
\begin{tabular}{cc} 
\includegraphics[width=0.47\textwidth]{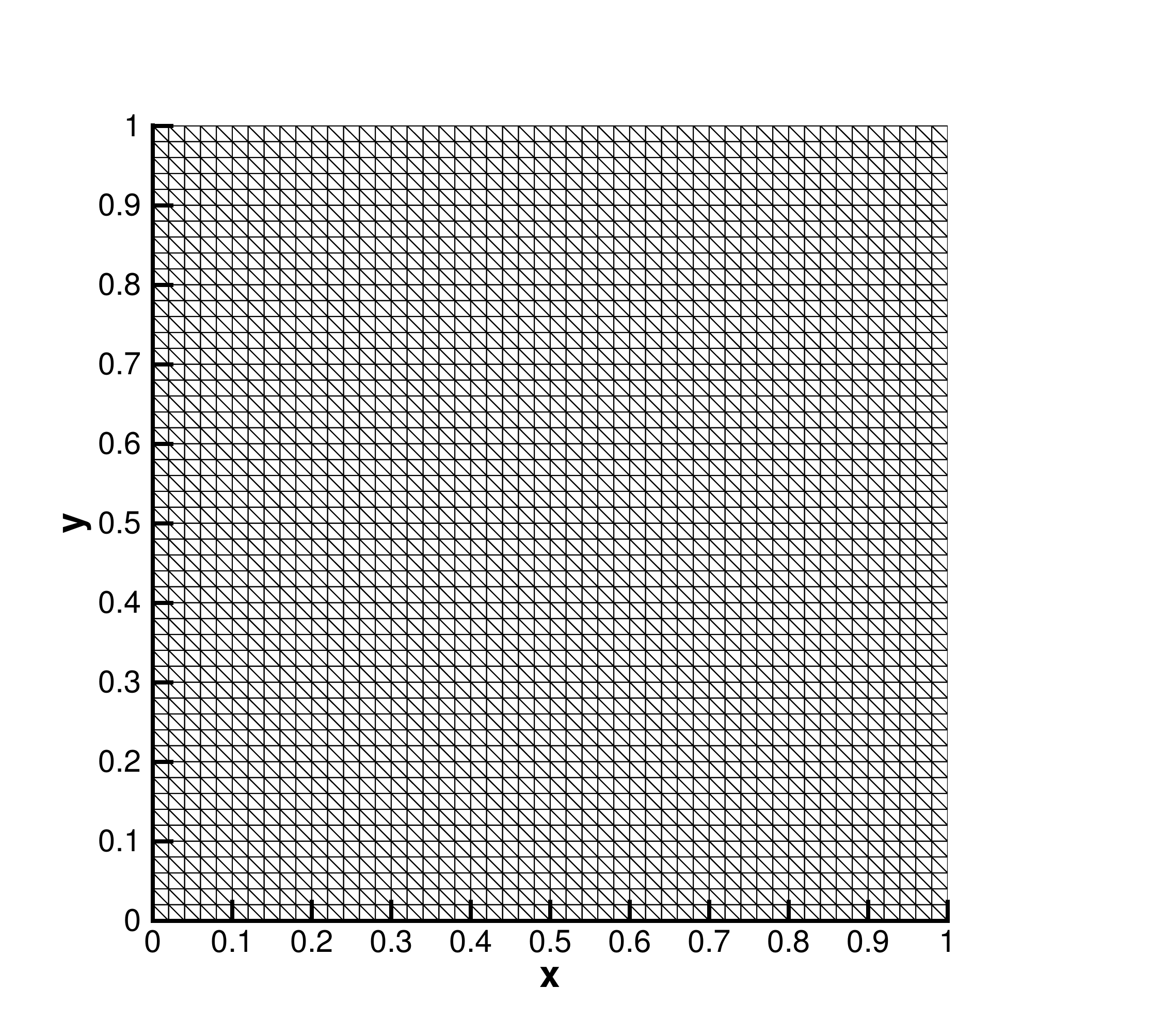}  &           
\includegraphics[width=0.47\textwidth]{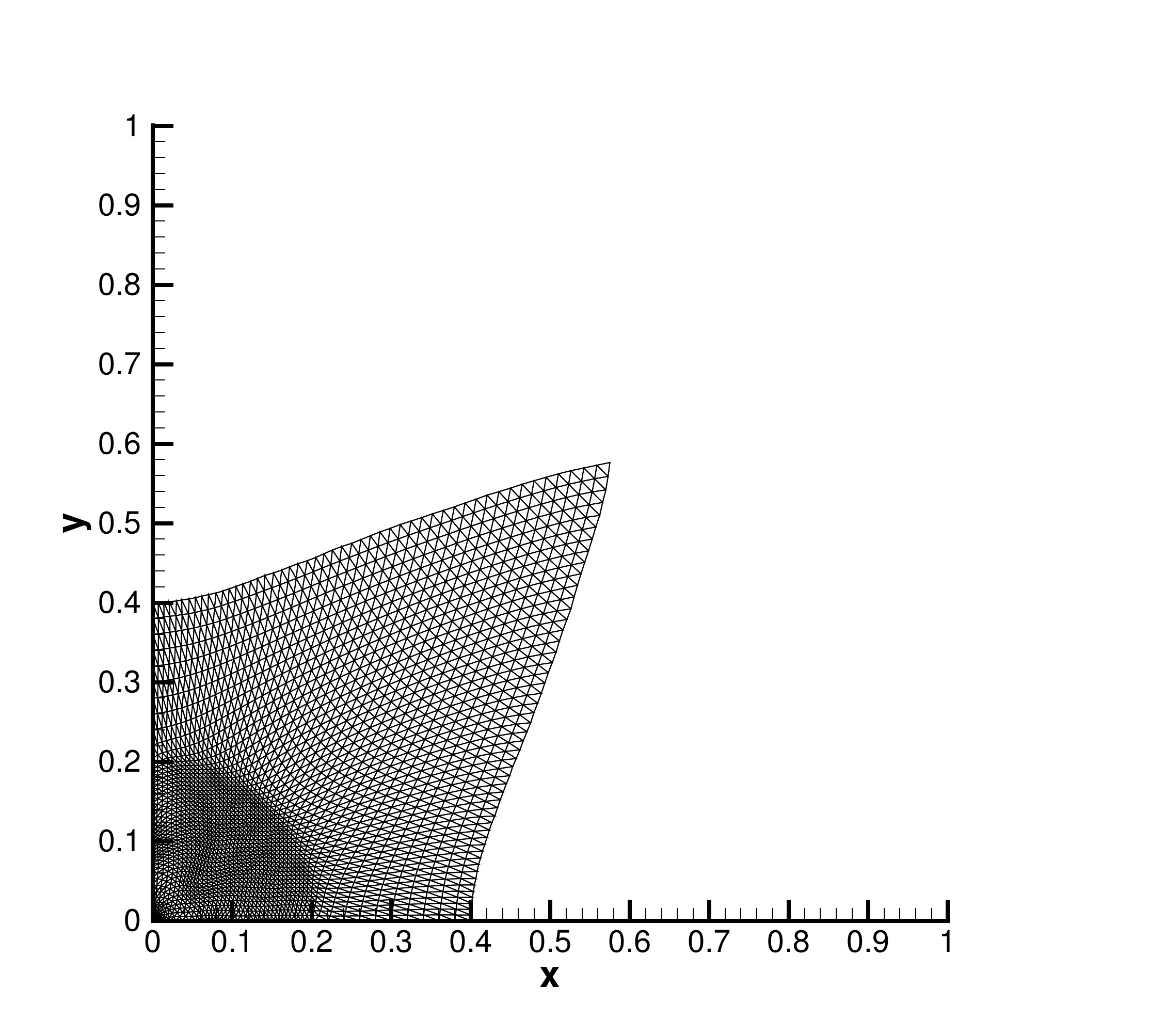} \\ 
\includegraphics[width=0.47\textwidth]{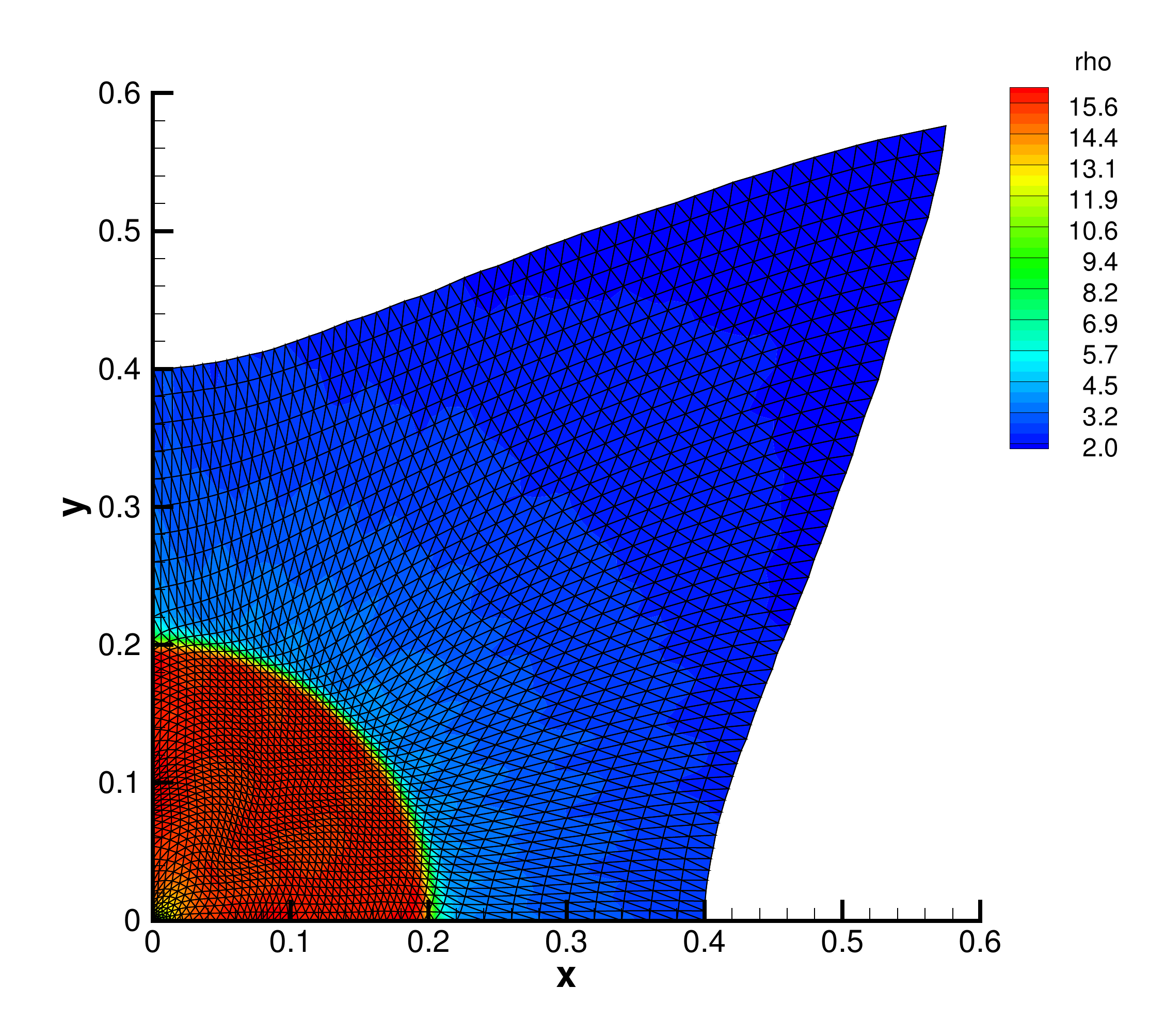}  &           
\includegraphics[width=0.47\textwidth]{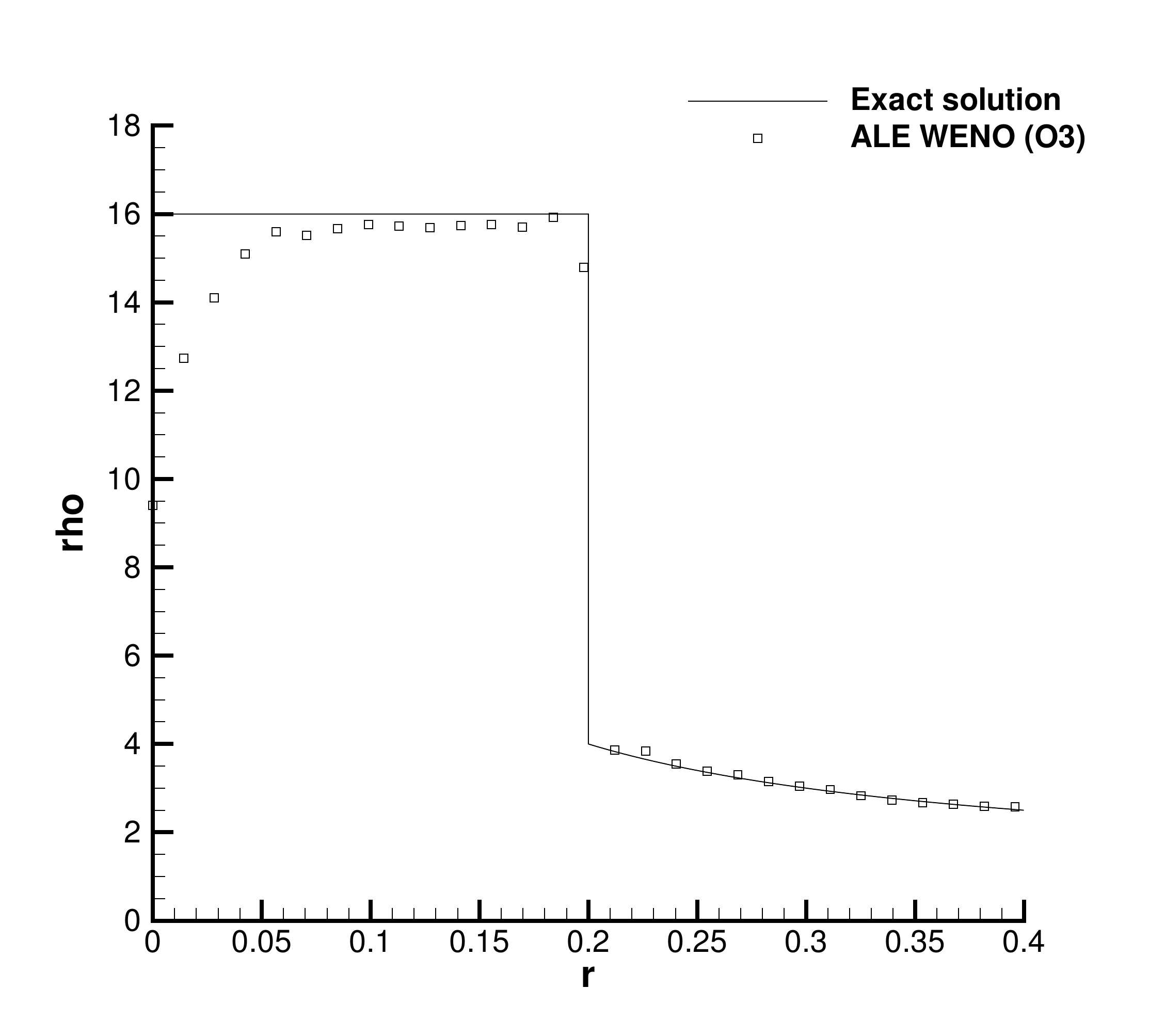} \\    
\end{tabular} 
\caption{Top: initial and final mesh configuration for the Noh problem. Bottom: density contour colors (left) and comparison with the analytical solution along the radial direction at the final time $t_f=0.6$ (right).} 
\label{fig.Noh2D}
\end{center}
\end{figure}

In \cite{Noh} Noh proposed this test case in order to validate Lagrangian schemes in the regime of strong shock waves. The initial computational domain is in our case given by $\Omega(0)=[0;1.2]\times[0;1.2]$ and the initial 
mesh is composed of $50\times50$ square elements which have been split into triangles, obtaining a total number of elements of $N_E=5000$. A gas with $\gamma=\frac{5}{3}$ is initially assigned with a unity density $\rho_0=1$ 
and a unity radial velocity which is moving the gas towards the origin of the domain $O=(0,0)$. Hence the horizontal $u$ and vertical $v$ velocity components are initialized with
\begin{equation}
u = -\frac{x}{r}, \qquad v = -\frac{y}{r}, 
\end{equation} 
and the initial pressure is $p=10^{-6}$ everywhere. As time advances, an outward moving cylindrical shock wave is generated which travels with velocity $v_{sh}=\frac{1}{3}$ in radial direction. According to  \cite{Noh,Maire2009,Maire2009b}, the final time is chosen to be $t_f=0.6$, therefore the shock wave is located at radius $R=0.2$ and the maximum density value is $\rho_f=16$, which occurs on the plateau behind the shock wave. We  impose no--slip wall boundary conditions on the left and on the bottom of the domain, while moving boundaries have been used on the remaining sides. Figure \ref{fig.Noh2D} shows the initial and the final mesh configuration, together with the final density distribution compared with the exact solution. The node solver $\mathcal{NS}_{m}$ and the Rusanov--type numerical flux \eqref{eqn.rusanov} allow the third order ALE ADER-WENO scheme to obtain a good result. The rezoning algorithm presented in Section \ref{sec.rezoning} has been used in order to limit the strong mesh deformation caused by the shock wave.

\subsection{The MHD equations}

We now apply the ALE ADER-WENO finite volume method to a more complicated hyperbolic system, namely to the well known equations of ideal classical magnetohydrodynamics (MHD), where an additional difficulty is given by the constraint  related to the divergence of the magnetic field $\mathbf{B}=(B_x,B_y,B_z)$ which must remain zero for all times, i.e.
\begin{equation}
\nabla \cdot \mathbf{B} = 0.
\label{eq:divB}
\end{equation} 
On the continuous level, Eqn. \eqref{eq:divB} is always satisfied if the initial data of $\mathbf{B}$ are ensured to be divergence--free, but on the discrete level this might not be necessarily guaranteed. Therefore we adopt the  hyperbolic version of the generalized Lagrangian multiplier (GLM) divergence cleaning approach proposed by Dedner et al. \cite{Dedneretal}. The main idea is that one additional variable $\Psi$ as well as one linear scalar PDE are added to the MHD system in order to transport divergence errors outside the computational domain with an artificial divergence cleaning speed $c_h$. Therefore the state vector, as well as the flux tensor for the augmented  GLM-MHD system read  
\begin{equation}
\Q^{T} = \left( \rho , \rho u , \rho v , \rho E , B_x , B_y, \Psi \right), \nonumber
\end{equation}
\begin{equation}
\label{MHDTerms}
\f = \left( \begin{array}{c} \rho u \\ \rho u^2 + \left(p+\frac{1}{8\pi}\mathbf{B}^2 \right) - \frac{B_x B_x}{4\pi} \\ \rho uv - \frac{B_y B_x}{4\pi} \\ u\left(\rho E + p+\frac{1}{8\pi}\mathbf{B}^2\right) - \frac{B_x\left(\mathbf{v} \cdot \mathbf{B} \right)}{4\pi} \\ B_x u - u B_x + \Psi \\ B_y u - v B_x \\ c_{h}^{2} B_x \end{array} \right), \qquad \g = \left( \begin{array}{c} \rho v \\ \rho uv - \frac{B_x B_y}{4\pi} \\ \rho v^2 + \left(p+\frac{1}{8\pi}\mathbf{B}^2 \right) - \frac{B_y B_y}{4\pi} \\ v\left(\rho E + p+\frac{1}{8\pi}\mathbf{B}^2\right) - \frac{B_y\left(\mathbf{v} \cdot \mathbf{B} \right)}{4\pi} \\ B_x v - u B_y \\ B_y v - v B_y + \Psi \\ c_{h}^{2} B_y \end{array} \right),  
\end{equation}
which is closed by the following equation of state
\begin{equation}
p = (\gamma-1)\left(\rho E - \frac{1}{2} (u^2+v^2) - \frac{\mathbf{B}^2}{8\pi} \right).  
\label{eqn.MHDeos}
\end{equation}
Let furthermore the fastest magnetosonic speed be defined as 
\begin{equation}
c = \sqrt{ \frac{1}{2} \left( \frac{\gamma p}{\rho} + (B_x+B_y) + \sqrt{ \left(\frac{\gamma p}{\rho} + (B_x+B_y)\right)^2 - 4 \frac{\gamma p}{\rho} \frac{B_x^2}{4\pi\rho}} \right) },
\label{eq:cMHD}
\end{equation}
which will be used for the node solver $\mathcal{NS}_md$.

\subsubsection{Numerical convergence results.} 
\label{sec.conv.Rates-MHD}

The numerical convergence studies for the ideal MHD equations are carried out using a smooth vortex problem similar to the isentropic vortex test case used previously for the Euler equations of compressible gas dynamics. It was  first proposed by Balsara in \cite{Balsara2004vortex}, where one can find the details for setting up this test case. The initial computational domain is the square $\Omega(0)=[0;10]\times[0;10]$ and it is periodic in both  directions. According to \cite{Balsara2004vortex}, the ratio of specific heats is taken to be $\gamma=\frac{5}{3}$ and we need to define the parameters $\epsilon=1$ and $\mu=\sqrt{4\pi}$ in order to assign the initial condition. 
As done for the hydrodynamic vortex described in Section \ref{sec.conv.Rates-Eul}, the initial condition is given again as a superposition of a constant flow plus some fluctuations. In terms of primitive variables it reads 
\begin{equation}
\label{MHDVortIC}
(\rho, u, v, p, B_x, B_y, \Psi) = (1+\delta \rho, 1+\delta u, 1+\delta v, 1+\delta p, 1+\delta B_x, 1+\delta B_y, 0),
\end{equation} 
with the perturbations defined as follows:
\begin{eqnarray}
\left(\begin{array}{c} \delta u \\ \delta v \\ \delta p \\ \delta B_x \\ \delta B_y \end{array}\right) &=& \left(\begin{array}{c} \frac{\epsilon}{2\pi}e^{\frac{1}{2}(1-r^2)}(5-y) \\ \frac{\epsilon}{2\pi}e^{\frac{1}{2}(1-r^2)}(x-5) \\    \frac{1}{8\pi} \left(\frac{\mu}{2\pi}\right)^2(1-r^2) e^{(1-r^2)}-\frac{1}{2}\left(\frac{\epsilon}{2\pi}\right)^2 e^{(1-r^2)} \\ \frac{\mu}{2\pi}e^{\frac{1}{2}(1-r^2)}(5-y)
\\ \frac{\mu}{2\pi}e^{\frac{1}{2}(1-r^2)}(x-5)
\end{array}\right).  
\end{eqnarray}
The speed for the divergence cleaning is set to $c_h=2$ and the final time is $t_f=1.0$. The vortex is furthermore convected with velocity $\v_c=(1,1)$ and the exact solution is given by the initial condition shifted in space by  $\mathbf{s}=\mathbf{v}_c\cdot t_f$. The error norms for density are reported in Table \ref{tab.convMHD} and have been computed in $L_2$ norm using \eqref{eqnL2error}. We show from first up to fifth order accurate numerical results for each of the three different node solvers, obtained with an Osher--type numerical flux.
 
\begin{table}  
\caption{Numerical convergence results for the ideal MHD equations. The first up to fifth order version of the two--dimensional Lagrangian one--step WENO finite volume 
scheme has been used for each node solver type. The error norms refer to the variable $\rho$ (density) at time $t=1.0$.} 
\begin{center} 
\begin{small}
\renewcommand{\arraystretch}{1.0}
\begin{tabular}{ccccccccc} 
\hline
                   &  $\mathcal{NS}_{cs}$  &		&   & $\mathcal{NS}_{m}$  &   &   &  $\mathcal{NS}_{b}$  &  \\  
  $h(\Omega(t_f))$ & $\epsilon_{L_2}$ & $\mathcal{O}(L_2)$ & $h(\Omega,t_f)$ & $\epsilon_{L_2}$ & $\mathcal{O}(L_2)$ & $h(\Omega,t_f)$ & $\epsilon_{L_2}$ & $\mathcal{O}(L_2)$ \\ 
\hline
\hline
   \multicolumn{3}{c} {} &        \multicolumn{3}{c}{$\mathcal{O}1$} & \multicolumn{3}{c}{} \\  
3.26E-01 & 2.7330E-03 & -    & 3.25E-01  & 2.7059E-03 & -    &  3.26E-01 & 2.7381E-03 & -     \\ 
2.37E-01 & 2.0111E-03 & 0.96 & 2.35E-01  & 2.0173E-03 & 0.90 &  2.35E-01 & 2.0173E-03 & 0.93  \\ 
1.64E-01 & 1.3081E-03 & 1.17 & 1.64E-01  & 1.3055E-03 & 1.20 &  1.64E-01 & 1.3113E-03 & 1.20 \\ 
1.28E-01 & 9.5497E-04 & 1.26 & 1.28E-01  & 9.5150E-04 & 1.30 &  1.28E-01 & 9.5617E-04 & 1.28  \\ 
\hline 
  \multicolumn{3}{c} {} &        \multicolumn{3}{c}{$\mathcal{O}2$} & \multicolumn{3}{c}{} \\  
3.26E-01 & 4.8091E-03 & -    & 3.27E-01  & 4.7707E-03 & -    &  3.26E-01 & 5.5971E-03 & -     \\ 
2.35E-01 & 2.8382E-03 & 1.61 & 2.37E-01  & 2.8571E-03 & 1.58 &  2.35E-01 & 2.7874E-03 & 2.13  \\ 
1.64E-01 & 1.4212E-03 & 1.91 & 1.63E-01  & 1.4239E-03 & 1.88 &  1.63E-01 & 1.3789E-03 & 1.94 \\ 
1.28E-01 & 6.4686E-04 & 3.24 & 1.28E-01  & 6.4610E-04 & 3.26 &  1.28E-01 & 7.2141E-04 & 2.67  \\
\hline 
  \multicolumn{3}{c} {} &        \multicolumn{3}{c}{$\mathcal{O}3$} & \multicolumn{3}{c}{} \\  
3.25E-01 & 1.1417E-03 & -    & 3.25E-01  & 1.1376E-03 & -    &  3.26E-01 & 1.1265E-03 & -     \\ 
2.36E-01 & 1.8935E-04 & 5.57 & 2.36E-01  & 1.8930E-04 & 5.56 &  2.36E-01 & 1.8632E-04 & 5.56  \\ 
1.63E-01 & 7.1734E-05 & 2.65 & 1.63E-01  & 7.1740E-05 & 2.65 &  1.63E-01 & 7.1912E-05 & 2.60 \\ 
1.28E-01 & 3.1651E-05 & 3.38 & 1.28E-01  & 3.1653E-05 & 3.38 &  1.28E-01 & 3.1738E-05 & 3.38  \\
\hline 
  \multicolumn{3}{c} {} &        \multicolumn{3}{c}{$\mathcal{O}4$} & \multicolumn{3}{c}{} \\  
3.26E-01 & 2.4858E-04 & -    & 3.26E-01  & 2.4864E-04 & -    &  3.26E-01 & 2.4472E-04 & -     \\ 
2.35E-01 & 7.9871E-05 & 3.50 & 2.35E-01  & 7.9875E-05 & 3.50 &  2.35E-01 & 7.9884E-05 & 3.45  \\ 
1.63E-01 & 2.1790E-05 & 3.55 & 1.63E-01  & 2.1791E-05 & 3.55 &  1.63E-01 & 2.1795E-05 & 3.55 \\ 
1.28E-01 & 8.2013E-06 & 4.03 & 1.28E-01  & 8.2014E-06 & 4.03 &  1.28E-01 & 8.1998E-06 & 4.03  \\ 
\hline 
  \multicolumn{3}{c} {} &        \multicolumn{3}{c}{$\mathcal{O}5$} & \multicolumn{3}{c}{} \\  
3.26E-01 & 1.2010E-04 & -    & 3.26E-01  & 1.2010E-04 & -    &  3.26E-01 & 1.1992E-04 & -     \\ 
2.35E-01 & 2.7365E-05 & 4.56 & 2.35E-01  & 2.7359E-05 & 4.56 &  2.35E-01 & 2.7327E-05 & 4.56  \\ 
1.63E-01 & 4.8779E-06 & 4.71 & 1.63E-01  & 4.8778E-06 & 4.71 &  1.63E-01 & 4.8898E-06 & 4.70 \\ 
1.28E-01 & 1.3947E-06 & 5.17 & 1.28E-01  & 1.3947E-06 & 5.17 &  1.28E-01 & 1.3935E-06 & 5.18  \\ 

\hline 
\end{tabular}
\end{small}
\end{center}
\label{tab.convMHD}
\end{table}

\subsubsection{The MHD rotor problem.} 
\label{sec.MHDRotor}

\begin{figure}[!htbp]
\begin{center}
\begin{tabular}{cc} 
\includegraphics[width=0.47\textwidth]{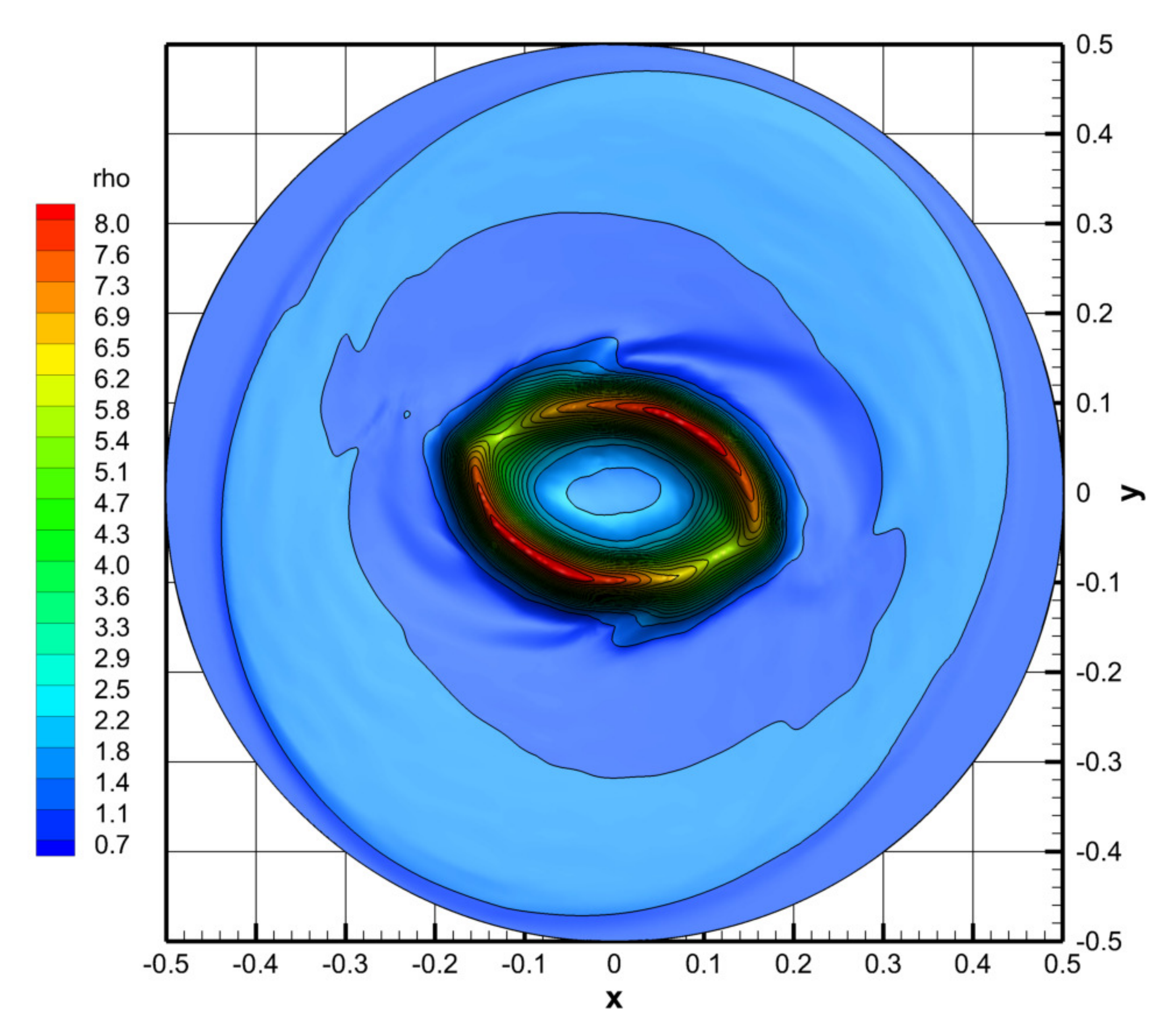}  &           
\includegraphics[width=0.47\textwidth]{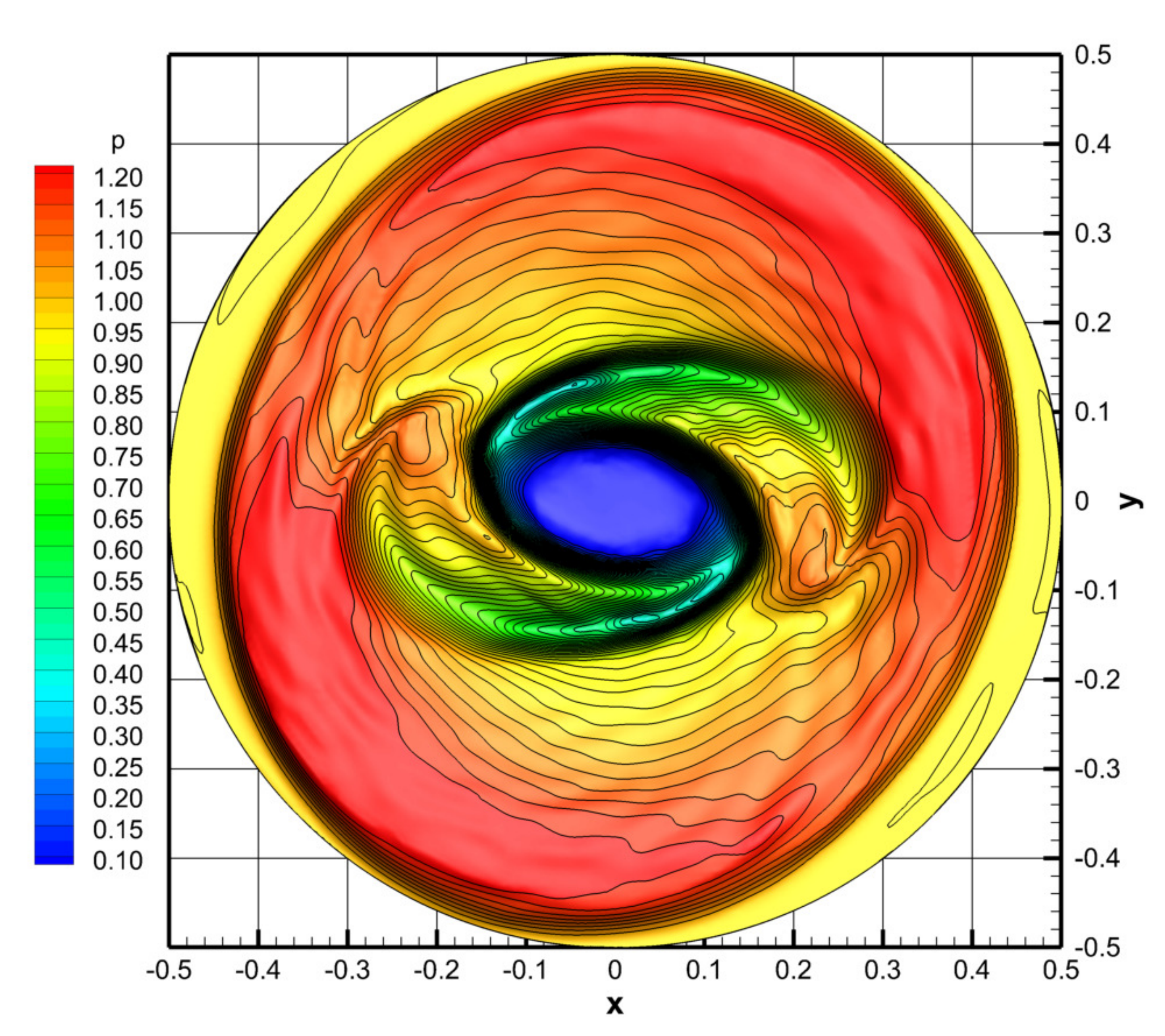} \\
\includegraphics[width=0.47\textwidth]{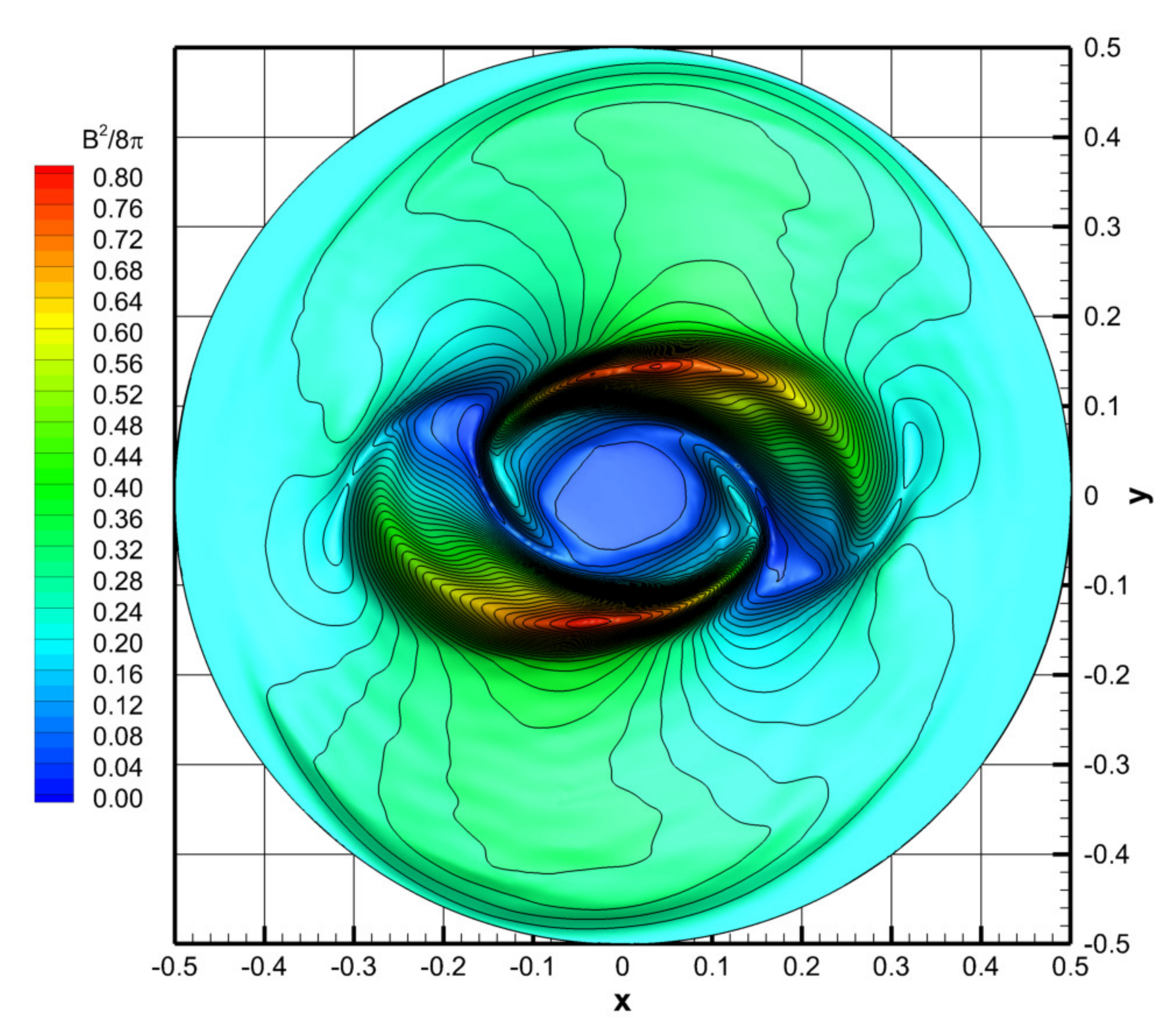} & 
\includegraphics[width=0.47\textwidth]{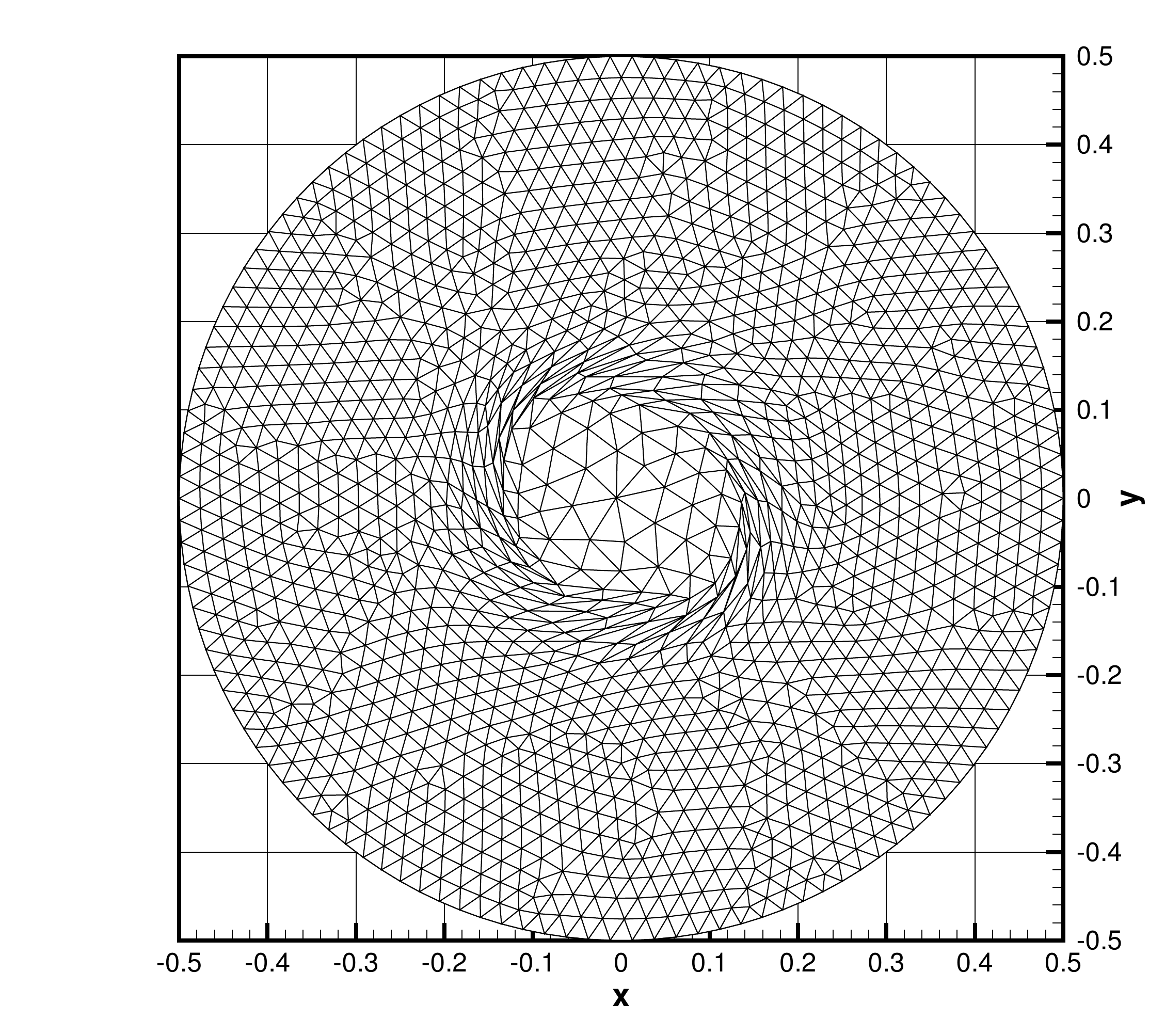}  \\  
\end{tabular} 
\caption{Numerical results for the ideal MHD rotor problem: density, pressure, magnetic pressure and a coarse mesh configuration at time $t=0.25$. A $4^{th}$ order Lagrangian WENO scheme has been used with the rezoning stage and the multidimensional node solver $\mathcal{NS}_{b}$. } 
\label{fig.MHDRotor}
\end{center}
\end{figure}

A classical test case for the ideal MHD equations is the MHD rotor problem \cite{BalsaraSpicer1999}. The initial computational domain $\Omega(0)$ is a circle of radius $R_0=0.5$ which is split into an \textit{internal} and an \textit{external} region by the internal frontier located at radius $R=0.1$. Let $r=\sqrt{x^2+y^2}$ be the generic radial position in the computational domain. Initially in the inner state a high density fluid is rotating while the outer state is filled with a low density fluid at rest. The angular velocity $\omega$ of the rotor is constant, so that at $r=R$ the toroidal velocity is $v_t=\omega R =1$. A constant magnetic field $\mathbf{B}=(2.5,0,0)^T$ is applied to the whole domain as well as a constant pressure value of $p=1$ and the initial density distribution is $\rho=10$ for $0 \leq r \leq R$ and $\rho=1$ elsewhere. As the simulation goes on, the angular momentum of the rotor is diminishing, because of the Alfv\'en waves produced by the rotor. The ratio of specific heats is taken to be $\gamma=1.4$, while the divergence cleaning velocity is set to $c_h=2$ and the final time is $t_f=0.25$. According to \cite{BalsaraSpicer1999}, in order to smear the initial discontinuity between the internal and the external region, we use a linear taper bounded by $0.1 < r \leq 0.13$ for the velocity and the density field in such a way that at radius $r=0.13$ density and velocity match exactly the values of the outer region. We use a computational grid with a characteristic mesh size of $h=1/200$ and we set transmissive boundary conditions at the external boundary. Numerical results obtained with a fourth order ALE ADER-WENO scheme with the the Rusanov--type flux \eqref{eqn.rusanov} and the node solver $\mathcal{NS}_b$ are depicted in Figure \ref{fig.MHDRotor}. We can notice a good agreement with the solution presented in \cite{BalsaraSpicer1999}, although the mesh used for the simulation is coarser than the one adopted by Balsara and Spicer.

\subsubsection{The MHD blast wave problem.} 
\label{sec.MHDBlast}

The MHD blast wave problem involves a strong circular fast magnetosonic shock wave which is propagating from the center to the boundaries of the initial circular domain $\Omega(0)$ of radius $R_0=1.0$. It is a well known difficult test case proposed in \cite{BalsaraSpicer1999b} and the initial condition reads
\begin{equation}
  \Q(\x,0) = \left\{ \begin{array}{ccc} \Q_i & \textnormal{ if } & r \leq R, \\ 
                                        \Q_o & \textnormal{ if } & r > R,        
                      \end{array}  \right. 
\end{equation}
with $r=\sqrt{x^2+y^2}$. The \textit{inner state} is defined in the central circle of radius $R=0.1$, while the \textit{outer state} $\Q_o$ is defined outside. We assume $\gamma = 1.4$ and the final time of the simulation is chosen to be $t_f=10^{-3}$. We use the initial condition reported in Table \ref{tab:Blast_IC} and a grid with a characteristic mesh size of $h=1/200$. Transmissive boundary conditions are imposed.
 
\begin{table}[!htbp]
	\caption{Initial condition for the MHD blast wave problem.}
	\centering
		\begin{tabular}{cccccccc}
		\hline
		                     & $\rho$ & $u$ & $v$ & $p$   & $B_x$ & $B_y$ & $\Psi$ \\  
		\hline
		Inner state ($\Q_i$) & 1.0    & 0.0 & 0.0 &  1000 & 70    & 0.0 &  0.0   \\ 
		Outer state ($\Q_o$) & 1.0    & 0.0 & 0.0 &  0.1  & 70    & 0.0 &  0.0   \\
		\hline
		\end{tabular}
	\label{tab:Blast_IC}
\end{table}

\begin{figure}[!htbp]
\begin{center}
\begin{tabular}{cc} 
\includegraphics[width=0.47\textwidth]{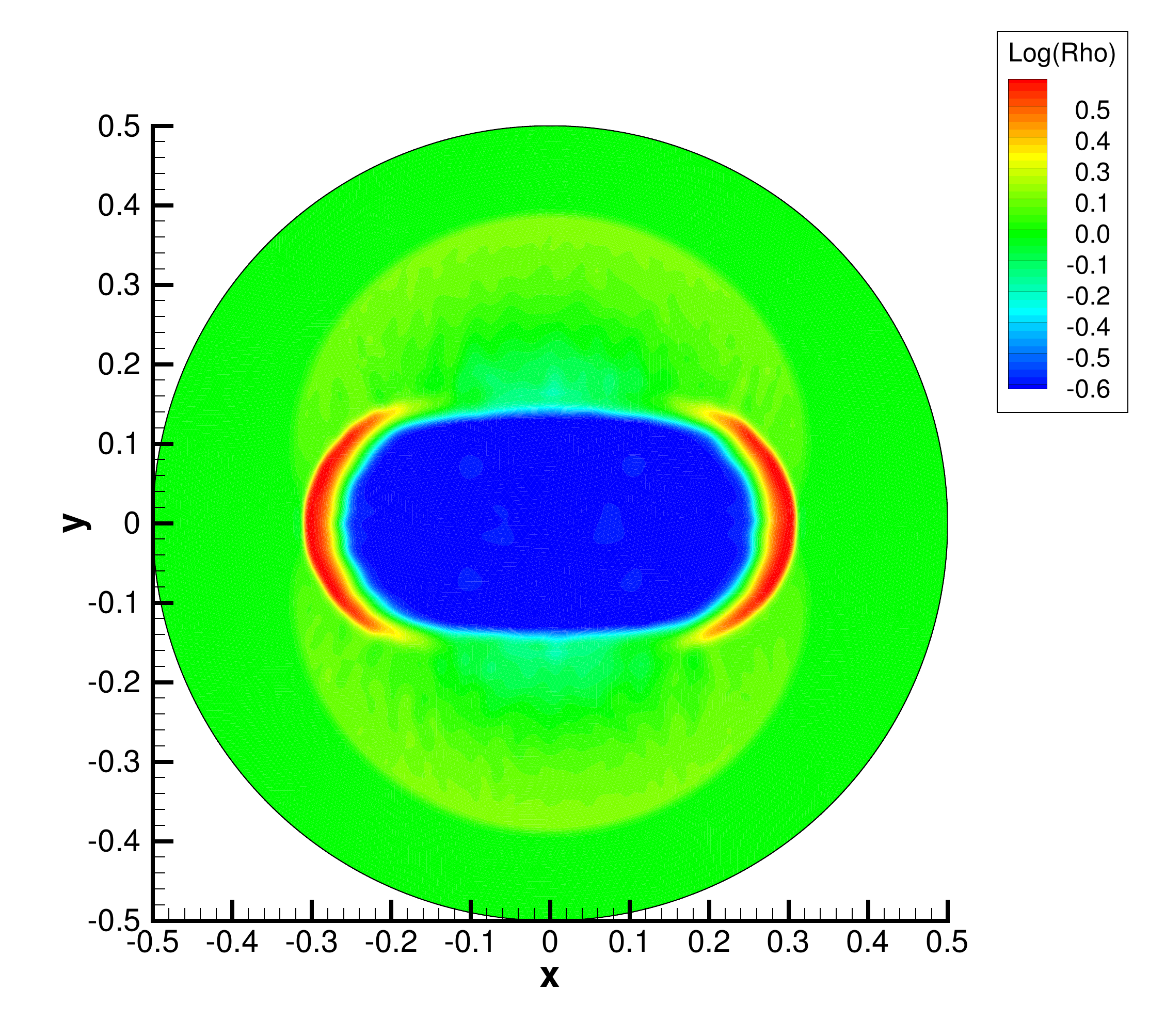}  &           
\includegraphics[width=0.47\textwidth]{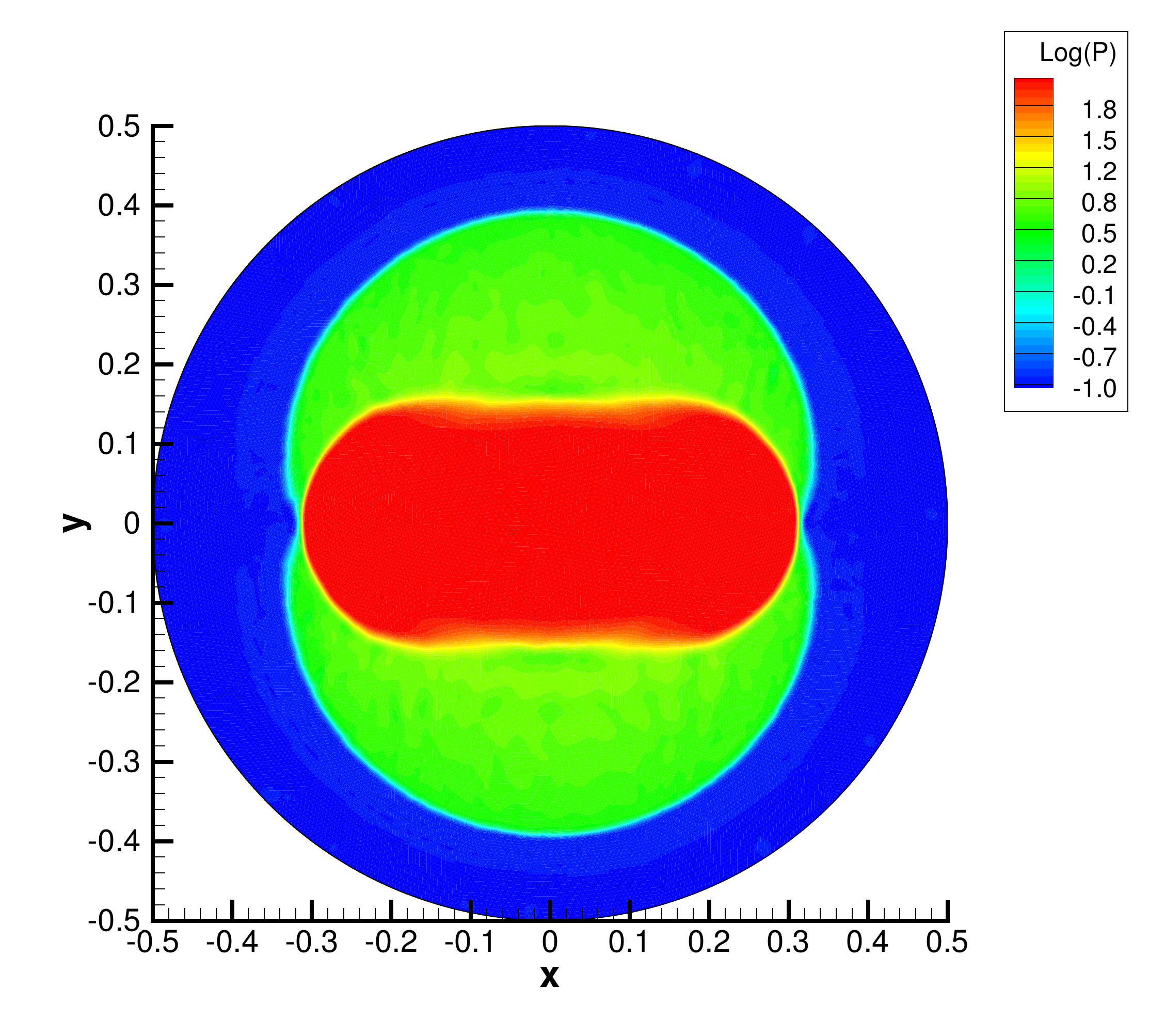} \\ 
\end{tabular} 
\caption{Numerical results for the Blast problem at time $t=0.01$ with $\mathcal{NS}_{m}$. Left: logarithm (base 10) of the density. Right: logarithm (base 10) of the pressure.}
\label{fig.Blast}
\end{center}
\end{figure}

We use a third order accurate version of the ALE WENO finite volume scheme with the Rusanov--type flux \eqref{eqn.rusanov} and the node solver $\mathcal{NS}_m$ to obtain the numerical results depicted in Figure 
\ref{fig.Blast}. The logarithm of density and pressure are reported, and a good agreement with the solution given in \cite{Balsara2004} can be noticed.

\begin{figure}[!htbp]
\begin{center}
\begin{tabular}{cc} 
\includegraphics[width=0.47\textwidth]{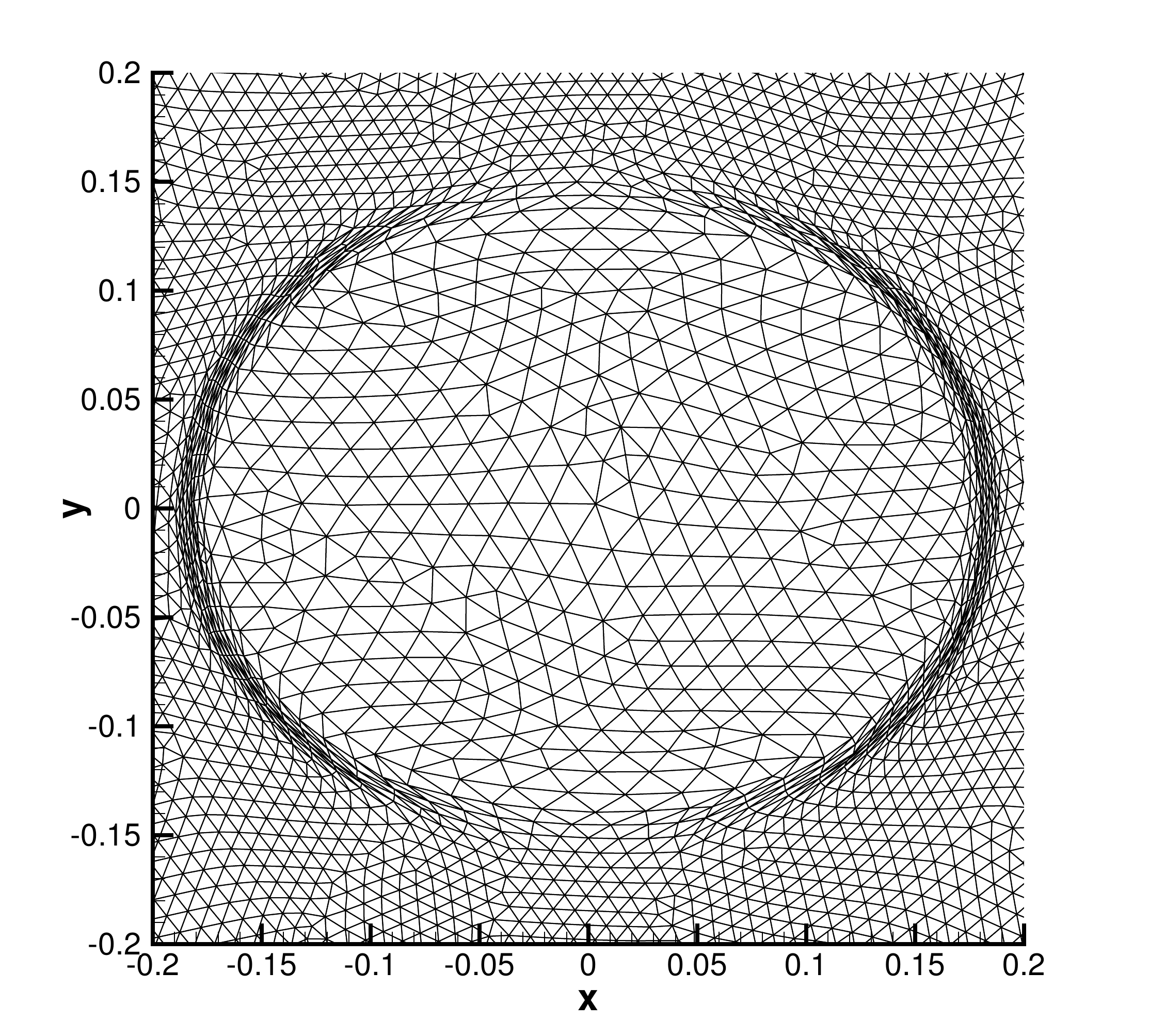}  &           
\includegraphics[width=0.47\textwidth]{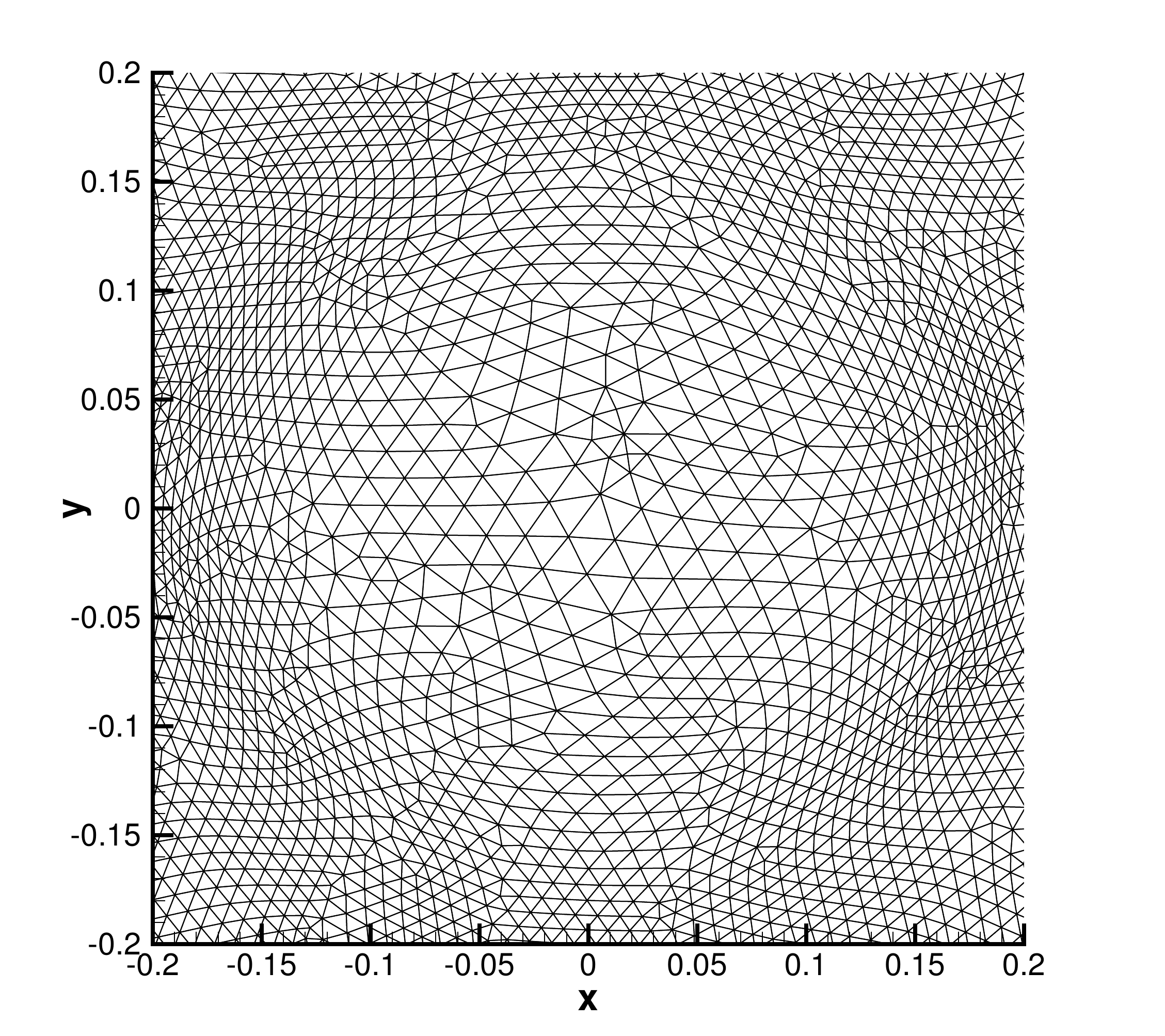} \\   
\end{tabular} 
\caption{Mesh configurations for the MHD blast wave problem at time $t=0.004$. Left: fully Lagrangian mesh motion. Right: Lagrangian mesh motion with the rezoning stage.} 
\label{fig.BlastGrid}
\end{center}
\end{figure}

Due to the very strong shock wave, the velocity of the flow is quite high and the fluid is pushed towards the left and the right part of the computational domain. Therefore we used the rezoning algorithm presented in Section \ref{sec.rezoning}, which allows the mesh elements to recover a more regular shape in order to carry on the simulation until the final time $t_f$. Figure \ref{fig.BlastGrid} shows a comparison between the fully Lagrangian mesh configuration and the rezoned mesh configuration at time $t=0.004$.

\subsection{The relativistic MHD equations (RMHD)} 
\label{sec.RMHD} 
The last test cases concern an even more complicated hyperbolic system, namely the relativistic MHD equations (RMHD). All the details regarding this physical model can be found in  \cite{BalsaraRMHD,RMHD,GiacomazzoRezzolla,RezzollaZanotti}. Let $\rho$ be the density and $\mathbf{v}=(u,v,w)$ be the velocity vector, then $p$ is the hydrodynamic pressure while $p_{tot}$ is the total pressure, obtained adding to $p$ also the contribution of the magnetic pressure. Furthermore let $e$ represent the internal energy, $E$ the total energy and let denote the magnetic field with $\mathbf{B}=(B_x,B_y,B_z)$ and the Lorenz factor with $\gamma$, while for the ratio of specific heats in this Section we use the symbol $\Gamma$. Again we take care of the divergence constraint for the magnetic field using the hyperbolic divergence--cleaning approach of Dedner et al. \cite{Dedneretal}, as done for the ideal MHD equations presented in Section \ref{sec.validation}. The vector of conserved variables of the RMHD system reads
\begin{equation}
\Q = \left( \begin{array}{c} D \\ \mathbf{q} \\ E \\ \mathbf{B} \\ \Psi \end{array} \right) = \left( \begin{array}{c} \gamma \rho \\ \gamma w_{tot} \mathbf{v} - b^0 \mathbf{b} \\ \gamma^2 w_{tot} - b^0b^0 - p_{tot} \\ \mathbf{B} \\ \Psi \end{array} \right),
\label{eqn.Q_RMHD}
\end{equation}
and the flux tensor $\F(\Q)$ is given by
\begin{equation}
\F = \left( \begin{array}{c} \gamma \rho \mathbf{v}^T \\ \gamma^2  w_{tot} \mathbf{v} \mathbf{v}  - \mathbf{b} \mathbf{b} + p_{tot} \mathbf{I} \\ \gamma^2  w_{tot} \mathbf{v}^T  - b^0 \mathbf{b}^T \\ \mathbf{v} \mathbf{B} - \mathbf{B} \mathbf{v} + \Psi \mathbf{I}  \\ c_h^2 \mathbf{B}^T \end{array} \right).  
\label{RMHDTerms}
\end{equation}
Here, $\mathbf{I}$ is the identity matrix, the enthalpy $w_{tot}$ and the total pressure $p_{tot}$ are defined as
\begin{equation}
w_{tot} = e + p + |b|^2, \qquad p_{tot} = p + \frac{1}{2}|b|^2,
\label{eqn.wtot.ptot}
\end{equation}
where the internal energy is given by the following equation of state
\begin{equation}
e = \rho + \frac{p}{\Gamma - 1}.
\label{eqn.EOS.RMHD}
\end{equation}
The Lorenz factor is
\begin{equation}
\gamma = \frac{1}{\sqrt{1-\mathbf{v}^2}},
\label{eqn.Lorenz}
\end{equation}
and the other quantities appearing in \eqref{RMHDTerms} are
\begin{equation}
b^0 = \gamma \left( \mathbf{v} \cdot  \mathbf{B}\right), \quad \mathbf{b} = \frac{\mathbf{B}}{\gamma} + \gamma \mathbf{v} \left(\mathbf{v} \cdot \mathbf{B}\right), \quad |b^2|=\frac{\mathbf{B}^2}{\gamma}+\left(\mathbf{v} \cdot \mathbf{B}\right)^2.
\end{equation}
We assume a speed of light normalized to unity. The computation of the primitive variables $\mathbf{W}=(\rho,\mathbf{v},p,\mathbf{B})$ from the conserved quantities $\mathbf{Q}$ has to be done \textit{numerically}, 
by using an iterative Newton or bisection method, as explained in \cite{RMHD,Dumbser20088209}. 

\subsubsection{Large Amplitude Alfv\'en wave.} 
\label{sec.RMHD.conv} 
The relativistic MHD equations are an extremely challenging and highly nonlinear hyperbolic system, for which the development of accurate and robust 
numerical methods is very difficult. To assess the accuracy of our high order cell-centered one-step Lagrangian finite volume schemes we perform 
a numerical convergence study of the third, fourth and fifth order version of our scheme on a very nice time-dependent test case proposed originally by Del Zanna et al. in  
\cite{ZannaZanotti} and which has subsequently also been used for the assessment of other high order schemes in \cite{Dumbser20088209,DumbserZanotti,Palenzuela2009,Dumbser2012}. 
It consists in a space-time periodic Alfv\'en wave with large amplitude. The initial condition for the primitive variables is chosen as the exact solution of the problem
at time $t=0$. In particular, one has 
$\rho=p=1$, $u=B_x=\Psi=0$, $B_y = \eta B_0 \cos\left(k x - v_A t \right)$, $ B_z = \eta B_0 \sin \left(kx - v_A t \right)$ and 
$v = -v_A B_y/B_0$, $w = -v_A  B_z/B_0$. 
We use the wavenumber $k=2 \pi$, the 2D computational domain is $\Omega = [0;1] \times [-0.1;+0.1]$ with four periodic boundary conditions 
and $\Gamma=\frac{5}{3}$. With these parameters and $B_0=\eta=1$, the speed of the Alfv\'en wave in positive $x$-direction is $v_A = 0.433892047069424$, 
see \cite{ZannaZanotti} for a closed analytical expression for $v_A$. 
The final computation time is set to $t=0.5$ and the mesh velocity is defined as $\mathbf{V}=(\frac{1}{10}\left(1+\cos(\pi x \right)^2,0)$ so that the total 
computational domain $\Omega(t)$ remains constant in time and the periodic boundary conditions can be applied. In the ALE framework proposed in this paper, 
the mesh velocity can indeed be chosen independently of the fluid velocity. Due to the smooth mesh motion imposed here, a rezoning strategy is not needed for 
this test problem. 
Note that in the RMHD system, the minimum and maximum eigenvalues of the ALE Jacobian must remain between $\lambda_{\min}=-1$ and $\lambda_{\max}=+1$, since the relativistic MHD equations are no longer
Galilean invariant as the previous PDE systems based on classical Newtonian mechanics. In all the computations we use a Courant number of $0.5$.
Table \ref{tab.rmhd.conv} shows the errors $\epsilon_{L2}$ and the measured convergence orders $\mathcal{O}_{L2}$ in $L^2$ norm for the flow variable $B_y$. The number $1/h$ denotes the reciprocal 
characteristic mesh spacing along each coordinate direction. We underline that a very high level of accuracy can be achieved on very coarse meshes with the fourth and fifth 
order scheme compared to the third order method even if the latter is run on much finer grids. 

\begin{table}
\caption{Numerical convergence study of third, fourth and fifth order Lagrangian ALE ADER-WENO finite volume schemes for the relativistic MHD equations (RMHD). Errors refer to the variable $B_y$.}
\begin{center}

\begin{tabular}{ccccccccc}
\hline
\multicolumn{3}{c}{$\mathcal{O}3$} & \multicolumn{3}{c}{$\mathcal{O}4$} & \multicolumn{3}{c}{$\mathcal{O}5$} \\
\hline
 $1/h$ &  $\epsilon_{L2}$ & $\mathcal{O}_{L2}$ & $1/h$ &  $\epsilon_{L2}$ & $\mathcal{O}_{L2}$ & $1/h$ &  $\epsilon_{L2}$ & $\mathcal{O}_{L2}$ \\ 
\hline 
  50  & 1.4270E-04 &       &   25  & 6.9640E-05 &       &   25  & 1.0749E-05 &        \\
  100 & 1.7436E-05 & 3.03  &   50  & 3.0158E-06 & 4.53  &   50  & 4.5265E-07 & 4.57   \\
  150 & 5.1826E-06 & 2.99  &   75  & 5.8315E-07 & 4.05  &   75  & 4.1669E-08 & 5.88   \\
  200 & 2.1831E-06 & 3.01  &  100  & 1.7717E-07 & 4.09  &  100  & 9.8553E-09 & 5.52   \\
\hline  
\end{tabular}

\end{center}
\label{tab.rmhd.conv}
\end{table}

\subsubsection{The RMHD rotor problem.} 
\label{sec.RMHD.rotor}
\begin{figure}[!htbp]
\begin{center}
\begin{tabular}{cc} 
\includegraphics[width=0.47\textwidth]{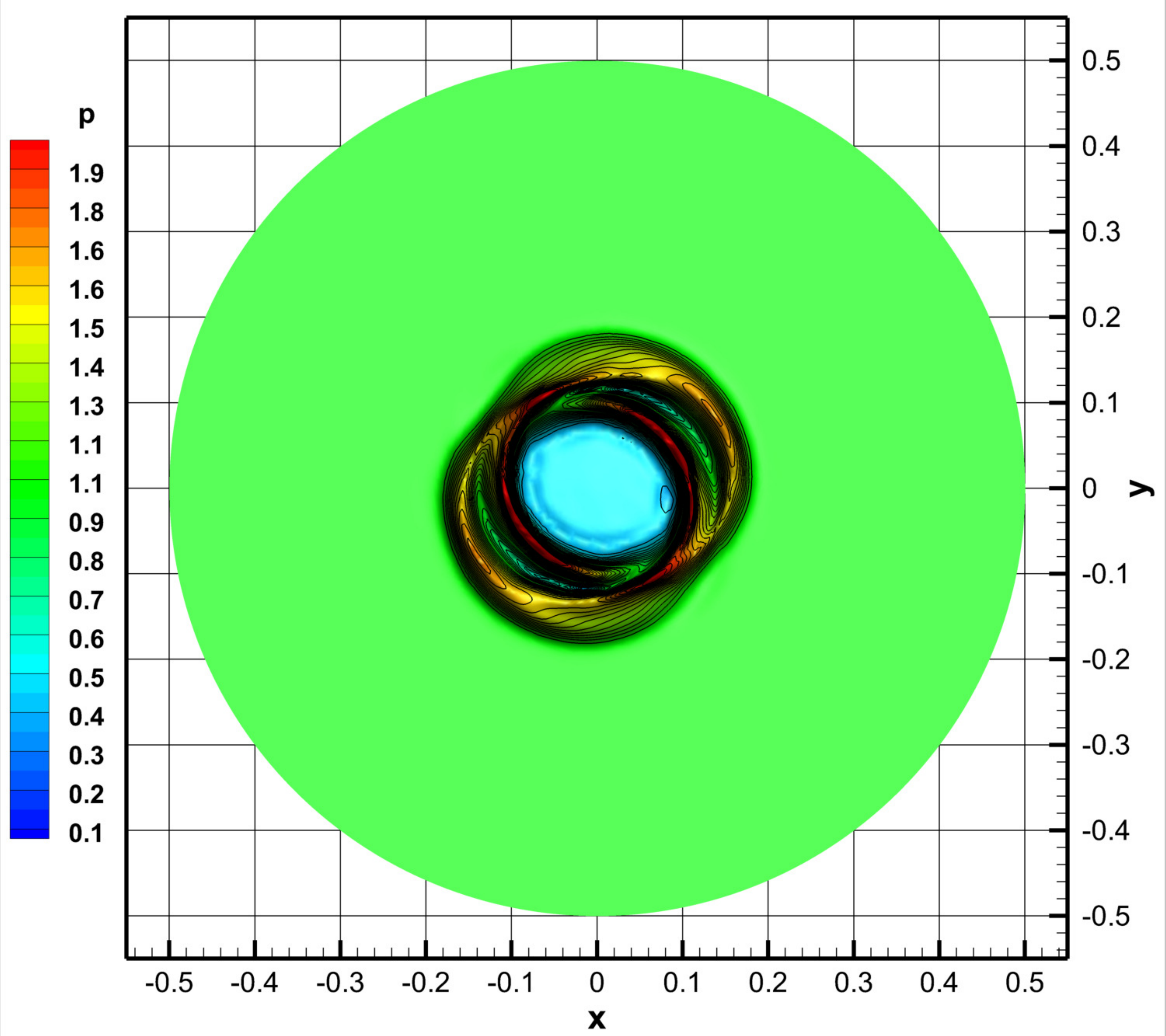}  &           
\includegraphics[width=0.47\textwidth]{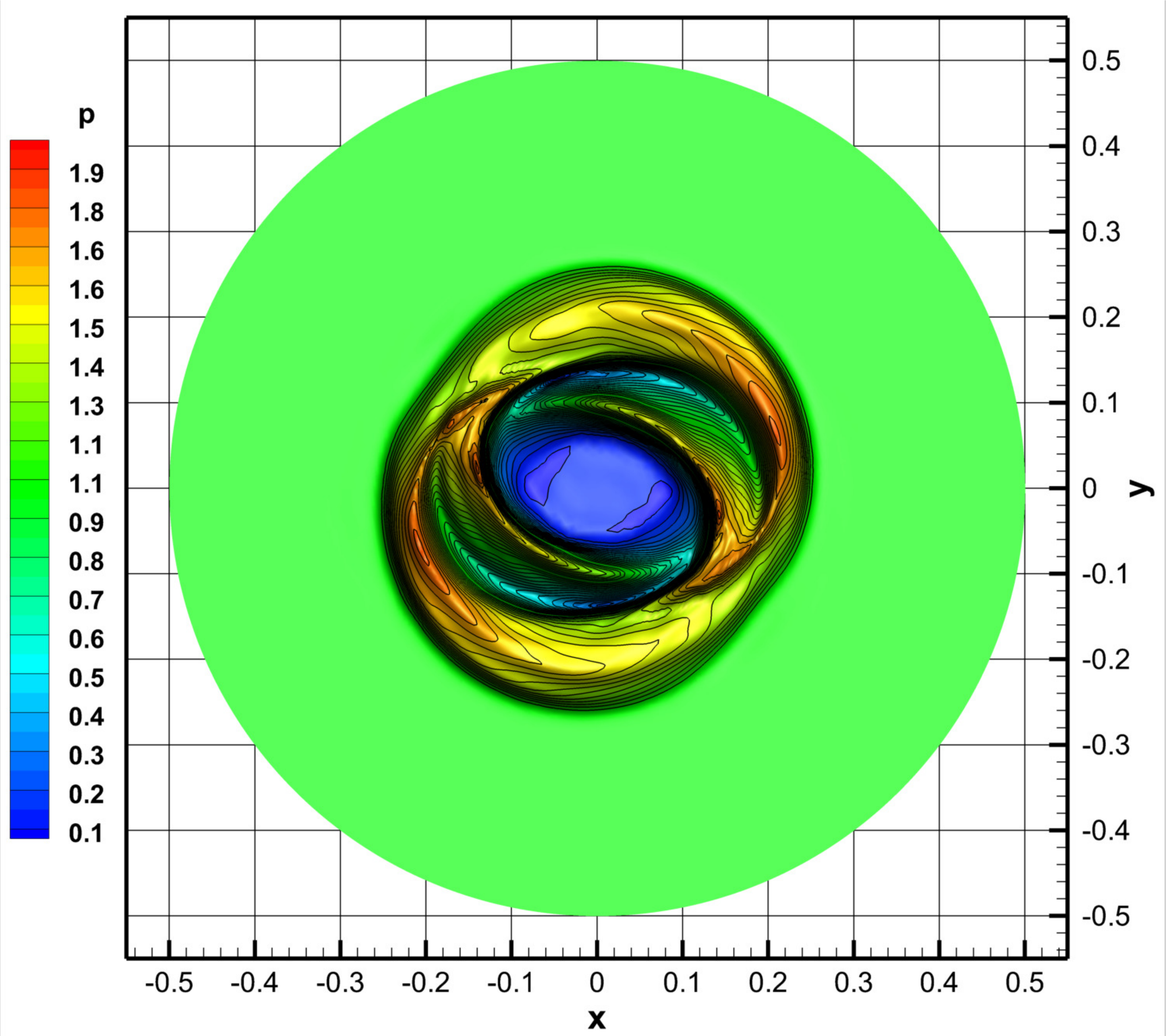} \\
\includegraphics[width=0.47\textwidth]{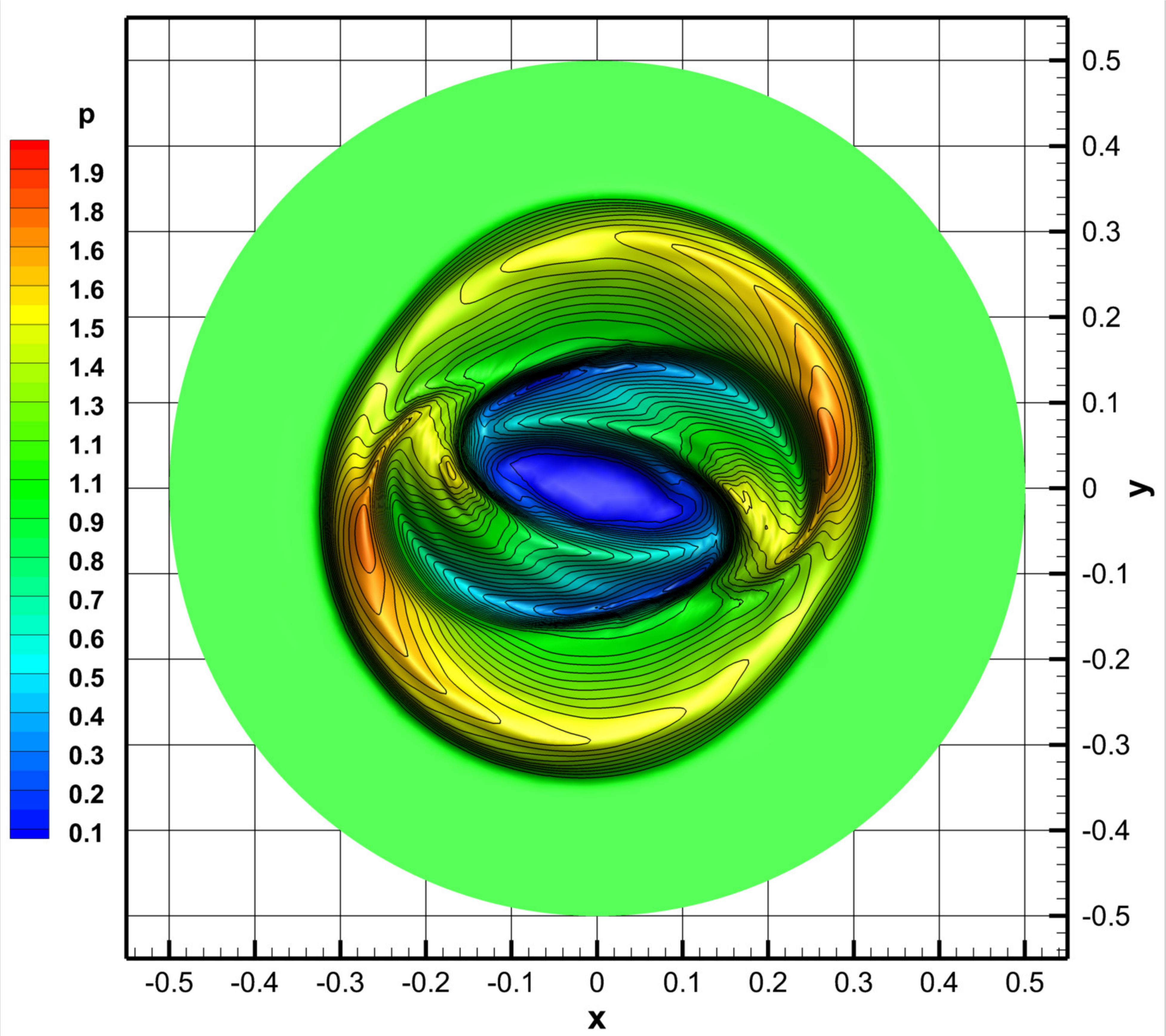}  &           
\includegraphics[width=0.47\textwidth]{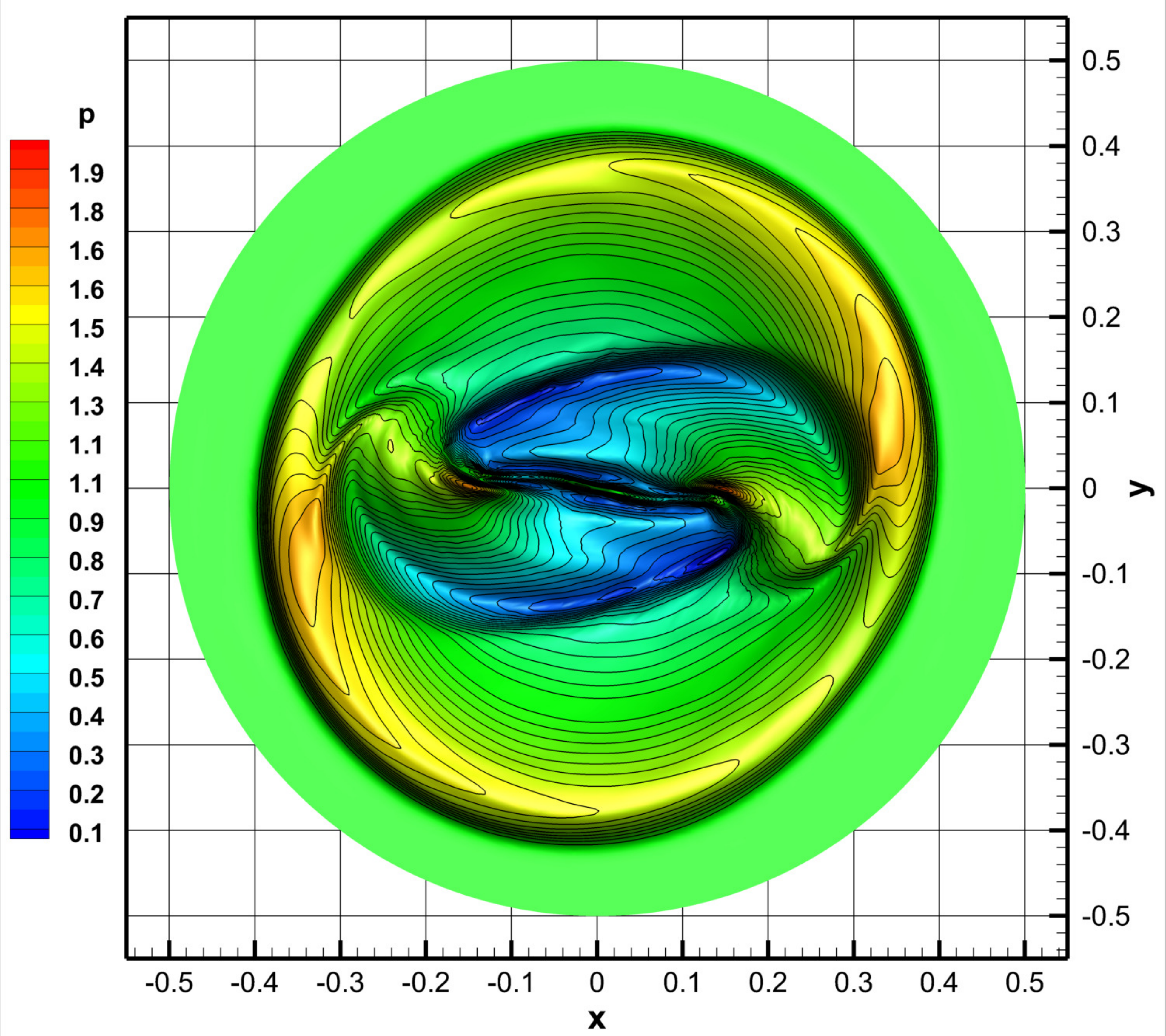} \\   
\end{tabular} 
\caption{Results for the pressure $p$ for the RMHD rotor problem at output times $t=0.10$, $t=0.20$, $t=0.30$ and $t=0.40$.} 
\label{fig.RMHDRotor}
\end{center}
\end{figure}

The initially circular computational domain is of radius $R_0=0.5$ and we use a mesh with a total number of elements $N_E=71046$. As for the ideal MHD rotor problem, radius $R=0.1$ splits again the domain into an internal and an external region. The rotor, which is in the internal region, is here spinning with an angular frequency of $\omega=8$, hence yielding the maximal toroidal velocities of $v_t=(\omega \cdot R)=0.8$. The initial density is $\rho=1$ 
in the external region and $\rho=10$ in the inner state, while the pressure $p=1$ is constant throughout the entire computational domain, as well as the magnetic field $\mathbf{B}=(1,0)$. We use again the taper described in Section \ref{sec.MHDRotor}. The speed of divergence--cleaning is set to $c_h=1$ and the ratio of specific heats is taken to be $\Gamma=\frac{5}{3}$. We use the node solver $\mathcal{NS}_{cs}$ for the calculation of the mesh velocity, due to its simple and very general formulation, which allows this node solver to be applied to any general nonlinear hyperbolic conservation law. This flexibility is not available with the other two node solvers, which have to be designed specifically for each hyperbolic system under consideration. For this test problem, a rezoning is necessary according to Section \ref{sec.rezoning}. 
Figure \ref{fig.RMHDRotor} displays the evolution of the pressure distribution up to the final time $t_f=0.4$ obtained with a third order  ALE WENO scheme and a Rusanov--type \eqref{eqn.rusanov} flux. The results obtained with the high order Lagrangian WENO scheme on a moving unstructured mesh agree qualitatively well with those obtained previously by an Eulerian WENO method on a fixed mesh in \cite{Dumbser20088209}. As far as we know, these are the first results obtained for the RMHD equations with a high order Lagrangian finite volume scheme. 

\subsubsection{The RMHD blast wave problem.} 
\label{sec.RMHD.blast}
This problem is similar to the classical MHD blast wave problem described in Section \ref{sec.MHDBlast}. It was also used in the context of resistive RMHD equations in \cite{DumbserZanotti}. 
The initial computational domain is again a circle of radius $R_0=0.5$ and a mesh with a characteristic mesh size of $h=1/200$ and a total number of $N_E=71046$ elements is used. 
The initial condition reads 
\begin{equation}
  \Q(\x,0) = \left\{ \begin{array}{ccc} \Q_i & \textnormal{ if } & r \leq R, \\ 
                                        \Q_o & \textnormal{ if } & r > R.        
                      \end{array}  \right. 
\end{equation}
The \textit{inner state} is defined in the central circle of radius $R=0.1$, while the \textit{outer state} $\Q_o$ is defined outside. We assume $\gamma = 4/3$ and the final time of the simulation is chosen to be $t_f=0.3$. 
The divergence cleaning speed is $c_h=1$. 
We use the initial condition reported in Table \ref{tab:RMHDBlast_IC} and transmissive boundary conditions are imposed everywhere.
 
\begin{table}[!htbp]
	\caption{Initial condition for the RMHD blast wave problem.}
	\centering
		\begin{tabular}{cccccccccc}
		\hline
		                     & $\rho$ & $u$ & $v$ & $w$ & $p$   & $B_x$ & $B_y$ & $B_z$ & $\Psi$ \\  
		\hline
		Inner state ($\Q_i$) & 1.0    & 0.0 & 0.0 & 0.0 & 1.0        & 0.05  & 0.0   &  0.0  &  0.0  \\ 
		Outer state ($\Q_o$) & 1.0    & 0.0 & 0.0 & 0.0 & $10^{-3}$  & 0.05  & 0.0   &  0.0  &  0.0  \\
		\hline
		\end{tabular}
	\label{tab:RMHDBlast_IC}
\end{table}

\begin{figure}[!htbp]
\begin{center}
\begin{tabular}{cc} 
\includegraphics[width=0.48\textwidth]{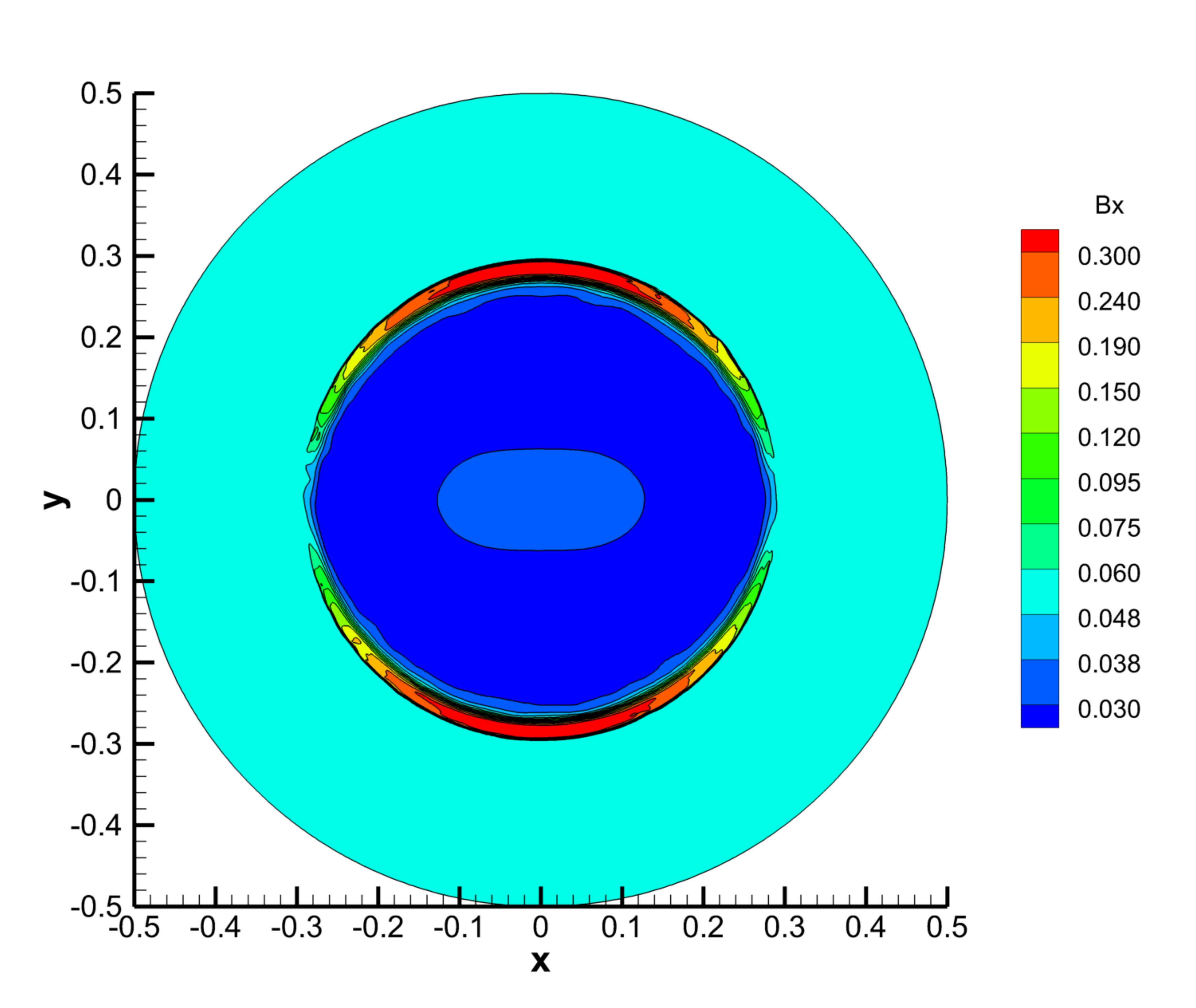}  &           
\includegraphics[width=0.48\textwidth]{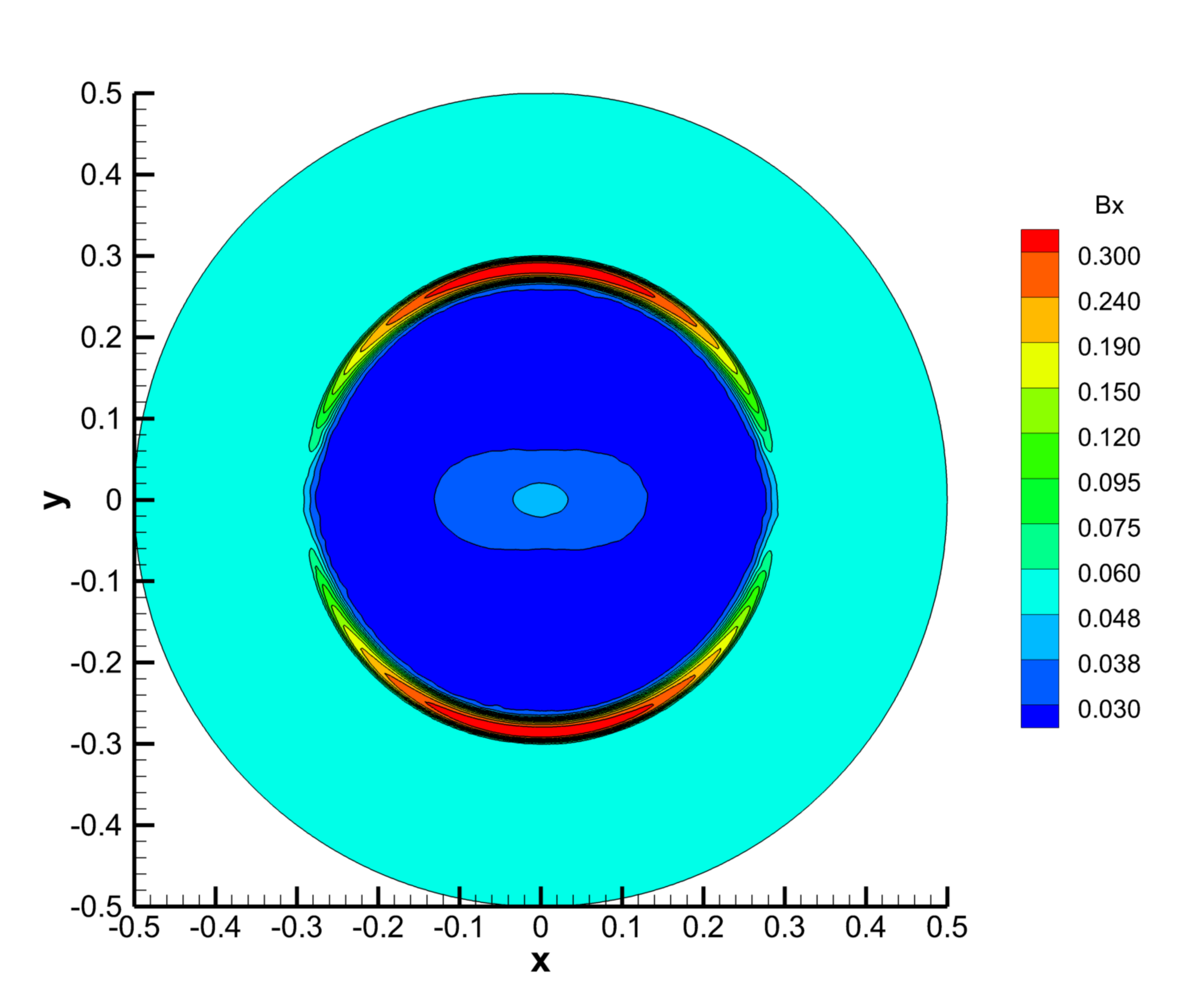}   
\end{tabular} 
\caption{Results for the magnetic field component $B_x$ for the RMHD blast wave problem at the final time $t=0.30$. 11 color contours are exponentially distributed between 0.03 and 0.3. 
Left: Third order Lagrangian ALE ADER-WENO scheme on a grid with $h=1/200$. Right: Eulerian reference solution computed with a third order ADER-WENO scheme on a fine grid ($h=1/400$).} 
\label{fig.RMHDBlast}
\end{center}
\end{figure}

We use the third order accurate version of the ALE WENO finite volume scheme with the simple Rusanov--type flux \eqref{eqn.rusanov} and the simple node solver $\mathcal{NS}_{cs}$, 
since more sophisticated Riemann solvers and node solvers are very difficult to obtain for this very complicated system. For an HLLC-type Riemann solver of the RMHD equations see 
the papers by Mignone and Bodo \cite{HLLCRMHD2} and of Honkkila and Janhunen \cite{HLLCRMHD}. An Osher-Solomon-type flux for RMHD has been recently proposed 
by Dumbser and Toro in the framework of universal Osher-type fluxes for general nonlinear hyperbolic conservation laws in \cite{OsherUniversal}, see eqn. \eqref{eqn.osher}, however, 
for this stringent test problem the more dissipative and more robust Rusanov flux \eqref{eqn.rusanov} was needed. The numerical results obtained with rezoning switched on are depicted in Figure 
\ref{fig.RMHDBlast}. The contour colors of the magnetic field component $B_y$ are reported, together with a fine grid Eulerian reference simulation carried out with a third order 
ADER-WENO scheme on a mesh with 282860 elements and characteristic mesh spacing $h=1/400$. 

\begin{figure}[!htbp]
\begin{center}
\begin{tabular}{cc} 
\includegraphics[width=0.45\textwidth]{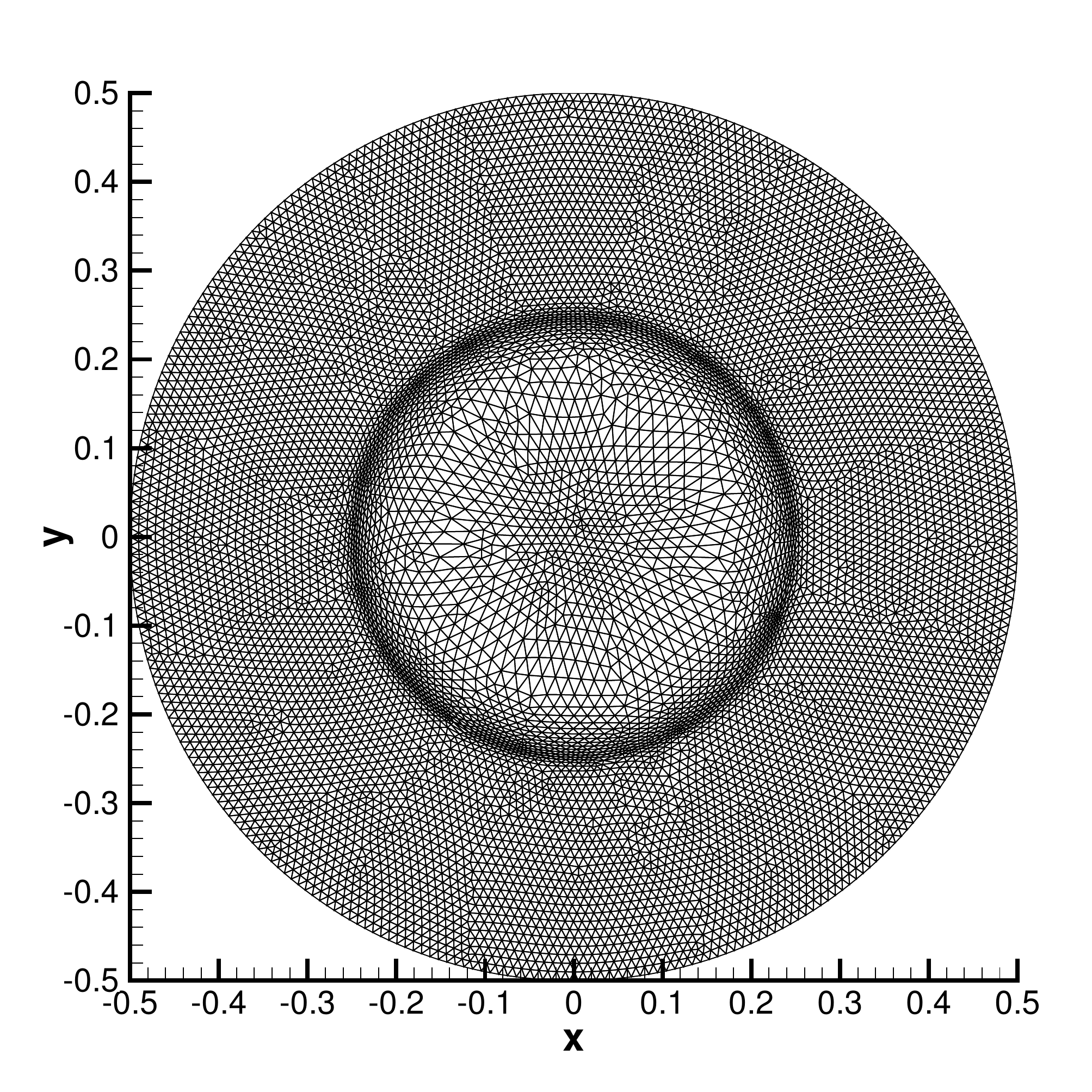}  &           
\includegraphics[width=0.45\textwidth]{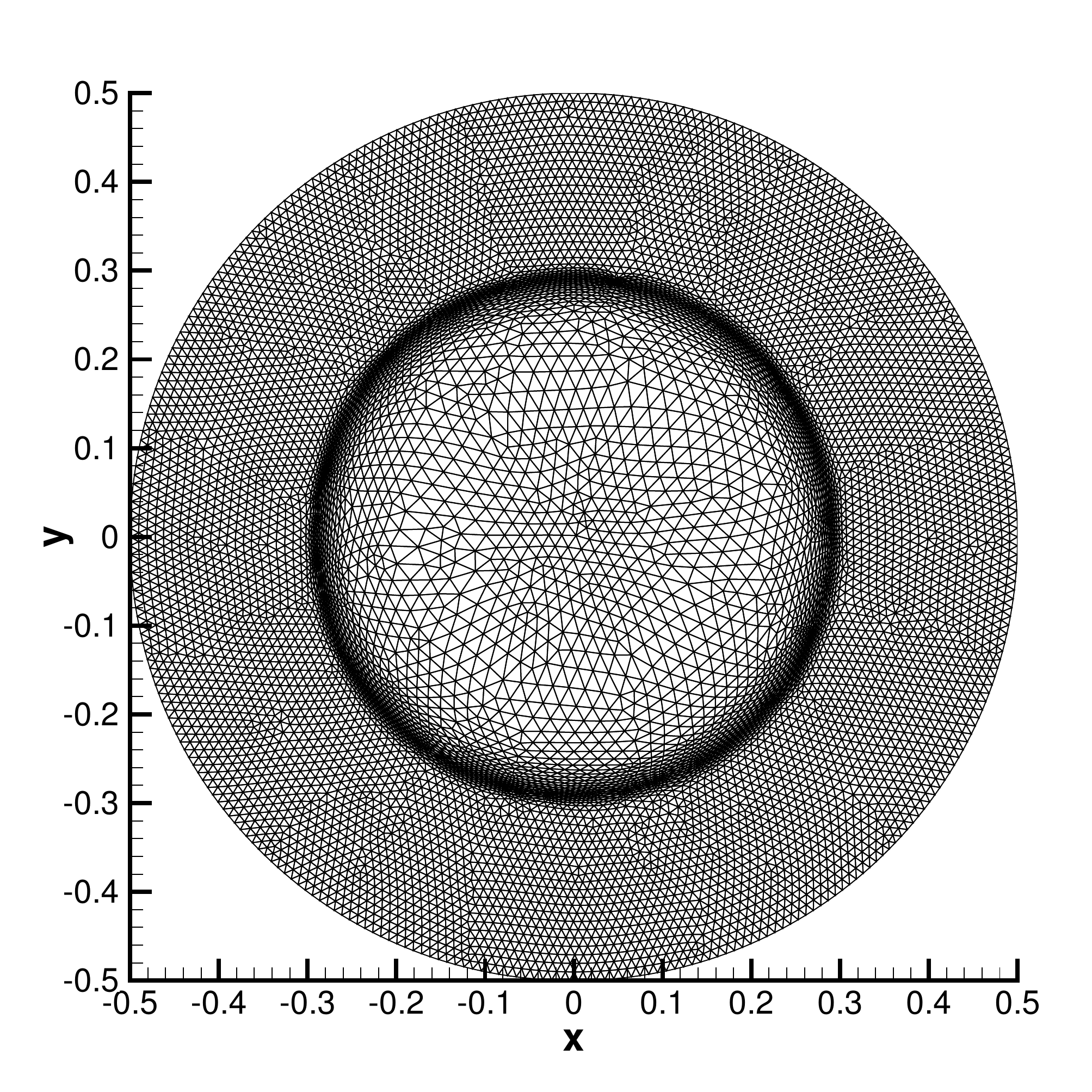} \\
\end{tabular} 
\caption{Evolution of a coarse version of the moving unstructured Lagrangian mesh with rezoning for the RMHD blast wave problem at output times $t=0.20$ and $t=0.30$.} 
\label{fig.RMHDBlastMesh}
\end{center}
\end{figure}


\vspace{-6pt}
\section{Conclusions and Outlook}
\label{sec.concl} 
\vspace{-2pt}
We have presented a family of high--order accurate cell-centered direct ALE ADER-WENO finite volume schemes on two--dimensional unstructured meshes, where three different node solvers have been used for the computation of 
the node velocity in order to move the grid in time. The first solver $\mathcal{NS}_{cs}$ is simply defined as the mass weighted average of the states in the cells surrounding the node, while the solver $\mathcal{NS}_{m}$ 
has been introduced 
in our scheme for hydrodynamics and for the classical MHD equations. Finally, we have used the very recent multidimensional HLL Riemann solver \cite{BalsaraMultiDRS} as a new node solver type for the first time in ALE schemes.  Furthermore we have applied the ALE WENO finite volume algorithm to the Euler equations of compressible gas dynamics, as well as to the classical and relativistic MHD equations. 

Future work will regard the extension of the presented scheme to unstructured tetrahedral meshes together with the three different node solvers. We plan also to use the multidimensional HLL and HLLC Riemann solvers not only as node solver, but also for the numerical flux evaluation, that would probably lead to an increase of the maximum admissible timestep $\Delta t$, hence improving the computational efficiency of Lagrangian schemes, which are usually  characterized by small time steps due to strong mesh deformation.

\vspace{-6pt}
\section*{Acknowledgements}
\label{sec.ack} 
\vspace{-2pt}

W.B. and M.D. have been financed by the European Research Council (ERC) under the European Union's Seventh Framework 
Programme (FP7/2007-2013) with the research project \textit{STiMulUs}, ERC Grant agreement no. 278267. 
D.B. acknowledges support via NSF grants NSF-AST-1009091 and NSF-ACI-1307369 as well as support via NASA 
grants from the Fermi program and the grant NASA-NNX 12A088G.

The authors would like to thank Rapha\"el Loub\`ere for the inspiring discussions on the subject and acknowledge PRACE 
for awarding access to the SuperMUC supercomputer based in Munich, Germany at the Leibniz Rechenzentrum (LRZ). 

\vspace{2pt}

\bibliography{LagrangeMHD}
\bibliographystyle{wileyj}

\end{document}